\documentclass[12pt]{article}
\usepackage{amsmath,amssymb,amscd,graphicx,fullpage}
\usepackage[all]{xy}
\xyoption{import}
\SelectTips{cm}{}

\numberwithin{equation}{section}
\def\Large{\large}

\def\a{\alpha}\def\b{\beta}\def\d{\delta}

\def\o{\omega}\def\s{\sigma}\def\t{\tau}
\def\G{\Gamma}\def\L{\Lambda}\def\O{\Omega}
\def\CB{{\cal B}}
\def\CE{{\cal E}}\def\CF{{\cal F}}\def\CI{{\cal I}}\def\CJ{{\cal J}}
\def\CN{{\cal N}}\def\CO{{\cal O}}
\def\CX{{\cal X}}
\def\CY{{\cal Y}}
\def\BA{\mathbf{A}}\def\BB{\mathbf{B}}\def\BC{\mathbf{C}}
\def\BD{\mathbf{D}}\def\BE{\mathbf{E}}\def\BO{\mathbf{O}}
\def\BX{\mathbf{X}}\def\Bs{\mathbf{s}}\def\Bz{\mathbf{z}}
\def\Balpha{\boldsymbol{\alpha}}\def\Bbeta{\boldsymbol{\beta}}
\def\Bdelta{\boldsymbol{\delta}}
\def\IZ{{\mathbb Z}}
\def\IQ{{\mathbb Q}}\def\IR{{\mathbb R}}\def\IC{{\mathbb C}}
\def\IP{{\mathbb P}}

\def\jbar{{\bar\jmath}}
\def\mbar{{\bar m}}\def\nbar{{\bar n}}
\DeclareMathOperator{\lcm}{lcm}
\DeclareMathOperator{\diag}{diag}
\DeclareMathOperator{\re}{Re}      
\DeclareMathOperator{\im}{Im}

\DeclareMathOperator{\Pic}{Pic}

\DeclareMathOperator{\Span}{Span}

\def\tilde{\widetilde}

\def\w{\wedge}
\def\ha{\frac{1}{2}}
\def\half{\tfrac{1}{2}}

\def\tor{{\rm tor}}
\def\eff{{\rm eff}}
\def\weak{{\rm weak}}
\def\Loop{{\rm loop}} 
\def\Kodaira{{\rm Kodaira}}
\def\NS{{\rm NS}}
\def\RR{{\rm RR}}
\def\MW{{\rm MW}}
\def\SL{{\rm SL}}
\long\def\symbolfootnote[#1]#2{\begingroup%
\def\thefootnote{\fnsymbol{footnote}}\footnote[#1]{#2}\endgroup}
\def\Mtheory{\hbox{M-theory}}
\def\Ftheory{\hbox{F-theory}}

\newcommand{\newsec}{\section}
\newcommand{\subsec}{\subsection}
\newcommand{\nonnumberedsubsec}{\subsubsection*}
\newcommand{\myeqlabel}[1]{\label{eq:#1}
  \expandafter\gdef\csname #1\endcsname{(\ref{eq:#1})}}

\begin{document}
\begin{titlepage}
  \setcounter{page}{0}
  \begin{flushright}
    arXiv:\,0810.5195\\
    UPR 1201-T
  \end{flushright}
  \vspace*{\stretch{1}}
  \begin{center}
    \Large\bf Abelian Fibrations, String Junctions,\\
    and Flux/Geometry Duality
  \end{center}
  \vspace*{\stretch{0.75}}
  \begin{center}
    \large Ron Donagi,$^a$\symbolfootnote[1]{donagi at math.upenn.edu}
    \large Peng Gao,$^b$\symbolfootnote[2]{gaopeng at physics.utoronto.ca}
    \large and Michael B. Schulz$^{c}$\symbolfootnote[3]{mbschulz at brynmawr.edu}
  \end{center}
  \begin{center}
    \textit{$^a$ Department of Mathematics, University of Pennsylvania\\
      Philadelphia, PA 19104, USA}\\[5pt]
    \textit{$^b$ Departments of Physics, University of Toronto\\
      Toronto, Ontario, Canada\ \ M5S 1A7}\\[5pt]
    \textit{$^c$ Department of Physics, Bryn Mawr College\\
      Bryn Mawr, PA 19010, USA} 
  \end{center}
  \vspace*{\stretch{1}}
  \begin{abstract}
    \normalsize In previous work, it was argued that the type IIB $T^6/\IZ_2$
    orientifold with a choice of flux preserving $\CN=2$ supersymmetry is dual
    to a class of purely geometric type IIA compactifications on abelian surface
    ($T^4$) fibered Calabi-Yau threefolds.  We provide two explicit
    constructions of the resulting Calabi-Yau duals.  The first is a monodromy
    based description, analogous to F-theory encoding of Calabi-Yau geometry via
    7-branes and string junctions, except for $T^4$ rather than $T^2$ fibers.
    The second is an explicit algebro-geometric construction in which the $T^4$
    fibers arise as the Jacobian tori of a family of genus-2 curves.  This
    improved description of the duality map will be a useful tool to extend our
    understanding of warped compactifications.  We sketch applications to
    related work to define warped Kaluza-Klein reduction in toroidal
    orientifolds, and to check the modified rules for D-brane instanton zero
    mode counting due to the presence of flux and other D-branes.  The
    nontrivial fundamental groups of the Calabi-Yau manifolds constructed also
    have potential applications to heterotic model building.
  \end{abstract}
  \vspace*{\stretch{5}}
  \flushleft{29 October 2008}
\end{titlepage}


\tableofcontents
\newpage


\newsec{Introduction}

An enduring theme in string theory has been duality---the existence of
dissimilar, but\break nonetheless equivalent discriptions of the same physical
string theory vacuum.  One particularly fruitful avenue has been open/closed
string duality, which relates the physics of D-branes and open strings to purely
geometrical closed string backgrounds.  In this way, we have gained insight into
strongly coupled gauge theories from geometry, and into the geometry of special
holonomy manifolds from consistent D-brane constructions.

Given the prominence of type IIB flux compactifications in string theory model
building, it seems useful to ask whether duality might again be put to
productive use to address shortcomings in our understanding of this class of
vacua.  For example: Flux compactifications are warped compactifications with
D-branes and internal magnetic flux.  The word ``warped'' means that a scale
factor governing the overall size of 4D spacetime varies from point to point in
the 6D compact extra dimensions.  In contrast to standard compactifications,
where there is a well defined Kaluza-Klein (KK) procedure for extracting the 4D
effective field theory from 10D, there is at present only a partial
understanding of the analogous procedure for warped
compactifications~\cite{DeWolfeGiddings,GiddingsMaharana,FreyMaharana,Burgess,STUD}.
Duality is one route through which a definition can be sought.

In the context of $\CN=2$ supersymmetry, it was shown in
Ref.~\cite{CYDuals}\ that the simplest class of IIB flux compactifications, the
type $T^6/\IZ_2$ orientifold with D3 branes and $\CN=2$ flux, is dual to a class
of purely geometric type IIA Calabi-Yau compactifications with no warping, no
\hbox{D-branes} and no flux.\footnote{A similar duality was explored in
  Ref.~\cite{Aspinwall}, where it was shown that the class of
  \Ftheory\ compactifications on ${\rm K3}\times{\rm K3}$ with $\CN=2$ flux is
  again dual to purely geometric Calabi-Yau compactifications of type IIA string
  theory.  In this case, the resulting manifolds are K3 fibered. }  $T^6/\IZ_2$,
while too simplistic to itself form the basis for a realistic model, has a good
track record as a source of insight into flux compactifications.  It was the
setting for the first global analysis of moduli stabilization from
flux~\cite{ModStab,Frey}, and a point of departure for the study of de Sitter
vacua~\cite{EvaDeSitter} and other forms of NS sector
flux~\cite{NewSUSY,SUSYorient,WechtFlux,DbraneNonGeo}, including twists of
topology and nongeometric flux~\cite{GeoNonGeo}.  A deeper understanding of the
duality map between $T^6/\IZ_2$ and conventional type IIA Calabi-Yau
compactifications stands to shed insight into warped compactifications in
general.  However, a prerequisite is a precise understanding of both sides of
the duality map.

The purpose of the work reported here is to explicitly construct the type IIA
Calabi-Yau compactifications dual to $T^6/\IZ_2$, and to make more precise the
encoding of their topology and geometry by the dual configuration of D-branes
and flux.  The Calabi-Yau manifolds $\CX_{m,n}$ that arise are abelian surface
($T^4$) fibrations over $\IP^1$.  Many of their topological properties were
deduced in Ref.~\cite{CYDuals}, however an explicit construction was left for
future work.  We provide two such constructions here.

The first construction is in terms of monodromy matrices of the abelian
fibrations $\CX_{m,n}$.  This construction is largely inspired by the work
\cite{DeWolfeI,DeWolfeII,DeWolfeIII,DeWolfeIV,MWJunction}\ on string junctions
in \Ftheory\ and its relation to the geometry of elliptic surfaces.  Our work
generalizes this formalism to the case of $T^4$ rather than $T^2$ fiber.  We
show that the D3 tadpole condition in $T^6/\IZ_2$ maps to the condition that the
total monodromy about all singular fibers of $\CX_{m,n}$ is unity.  Building on
Ref.~\cite{MWJunction}, we show that the junction formalism is again an
efficient means to compute Mordell-Weil lattice of sections of the abelian
fibration in the $T^4$ fibered case.  For the Calabi-Yau manifold $\CX_{m,n}$,
we show that the free component of the Mordell-Weil lattice is the $D_{M}$ root
lattice, where $M = 16-4mn$ is the number of D3-branes in $T^6/\IZ_2$.  The
generic torsion component of the Mordell-Weil group is $\IZ_m\times \IZ_m$, in
agreement with the isometry group inferred in Ref.~\cite{CYDuals}; at special
points in moduli space it is enhanced to a larger discrete group, characterized
by the lattice of weakly integral null junctions.  We are also able to use the
monodromy description to verify the $\IZ_n\times \IZ_n$ fundamental group
deduced by duality in Ref.~\cite{CYDuals}.  Finally, quotienting by isometries
gives a general way to construct new Calabi-Yau manifolds with nontrivial
fundamental group.

The second construction takes an explicit algebraic geometry approach.  In the
case, we begin with an auxilliary surface $S$ that is fibered by genus-2 curves
over $\IP^1$.  By replacing each genus-2 curve with its Jacobian torus, we
obtain a 3-fold that we show is Calabi-Yau.  Its homology, interection numbers,
Mordell-Weil lattice and second Chern class all agree with those of $\CX_{1,1}$.
Therefore, the two are equivalent up to homotopy type by Wall's
theorem~\cite{Hubsch,Tomasiello}.  Finally, we again explore the enhancement of
Mordell-Weil torsion (i.e., the isometry group) at special loci in moduli space
in this framework, and connect to a subset of the results obtained from the
junction description.

This work, together with Ref.~\cite{CYDuals}, lays the groundwork for the
following applications, to be reported in separate articles, as sketched in
Sec.~\ref{Sec:Conclusions}:
\begin{enumerate}
  \item[i.] to define warped KK reduction for simple warped compactifications by
    duality to conventional Calabi-Yau compactifications~\cite{WarpedKK}.

  \item[ii.] in the context of D3-brane instantons, to check the modified rules
    for instanton zero mode counting due to flux or intersections with other
    localized objects, by duality to worldsheet instantons in type
    IIA~\cite{DInstanton}.

\end{enumerate}

\noindent As a byproduct, the following application arises, as explained in
Sec.~\ref{Sec:Connecting}:
\begin{enumerate}
  \item[iii.] to construct examples of new Calabi-Yau manifolds with nontrivial
    fundamental group of interest for heterotic model
    building~\cite{DonagiSaito,SaitoTalk}.  As highlighted in
    Refs.~\cite{GrossPheno,GrossPopescu,Borisov,Candelas,BDI,BDII}, very few
    Calabi-Yau manifolds with nontrivial fundamental group are explicitly known.

\end{enumerate}

An outline of the paper is as follows:
\medskip

In Sec.~2, we review the monodromy and junction description of elliptic
fibrations.  This establishes the background and point of view in preparation
for an analogous description of abelian surface fibered Calabi-Yau manifolds in
Sec.~3.  For simplicity, we focus on the case of $\IP^1$ base.  The topology of
a generic elliptic fibration over $\IP^1$ is defined by a collection of points
on the $\IP^1$ and the monodromies about the I$_1$ singular fibers at these
points.  The collection of monodromies is unique up to braiding operations and
overall $SL(2,\IZ)$ conjugation.  Via the F-theory, this is the same data that
determines a collection of 7-branes in a type IIB compactification on the base
of the elliptic fibration, and their $(p,q)$ types.  The $W$-bosons of the
spontaneously broken gauge theory on the 7-branes are string junctions
terminating on 7-branes.  Each can be represented by a tree graph on the base,
and encodes a curve in the elliptic fibration with zero intersection with the
fiber and base.  Following Refs.~\cite{DeWolfeI,DeWolfeII,DeWolfeIII,DeWolfeIV},
we explain how various coalescing collections of 7-branes realize unbroken ADE
type gauge symmetries.  Finally, following Ref.~\cite{MWJunction}, we describe
how the junction lattice determines the Mordell-Weil group of rational sections
of the elliptic fibration.\footnote{To be precise, we determine both the
  Mordell-Weil group and the Mordell-Weil lattice.  The Mordell-Weil
  \emph{group} includes the torsion sections, but not the lattice inner product.
  The Mordell-Weil \emph{lattice} includes the lattice inner product, but not
  the torsion sections.}

In Sec.~3, we review the duality map between the type IIB $T^6/\IZ_2$
orientifold with $\CN=2$ flux and type IIA compactified on the Calabi-Yau
manifolds $\CX_{m,n}$.  Then, we generalize the mondoromy and junction
description of the previous sections to be applicable to abelian surface
fibrations, focusing on the $\CX_{m,n}$.  In particular, we determine
collections of $SL(4,\IZ)$ monodromy matrices that define the topology of the
$\CX_{m,n}$.  The condition that the total monodromy be unity reproduces the D3
charge cancellation condition of $T^6/\IZ_2$.  The $\CX_{m,n}$ are abelian
surface fibrations over $\IP^1$, where an abelian surface is a $T^4$ that admits
an embedding in complex projective space.  The latter endows the $T^4$ with a
Hodge form (or equivalently, a theta divisor), which is precisely the additional
ingredient necessary to define an inner product and give the space of junctions
the structure of a lattice.  This lattice again determines the Mordell-Weil
group of sections, including torsion, and the torsion subgroup is an isometry
group of the Calabi-Yau manifold.  We describe its enhancement at singular loci
in moduli space in terms of weakly integral null junctions that become relevant
when I$_1$ fibers coalesce.  Quotienting by these isometries gives new
Calabi-Yau manifolds with nontrivial fundamental group.

In Sec.~4, we provide an algebro-geometric construction of $\CX_{1,1}$ as the
relative Jacobian of a genus-2 fibration over $\IP^1$.  Every smooth principally
polarized abelian surface is the Jacobian of some genus-2 curve.  Therefore, in
the principally polarized cases $m=n$, we might expect to realize the Calabi-Yau
threefold $\CX_{m,n}$ as the relative (i.e., fiberwise) Jacobian of an
auxilliary surface fibered by genus-2 curves.  We show that this is indeed the
case.  Since a genus-2 curve is the double cover of $\IP^1$ with 6 branch
points, we consider a surface $S$ that is the double cover of $\IP^1\times\IP^1$
branched over a $(6,2)$ curve $B$.  The relative Jacobian $\CJ_{S/\IP^1}$ indeed
reproduces the Calabi-Yau manifold $\CX_{1,1}$.  After verifying the Calabi-Yau
condition, we show that the Hodge numbers, intersection numbers, and second
Chern class of $\CJ_{S/\IP^1}$ match those of $\CX_{1,1}$.  These are the
classifying data of a Calabi-Yau threefold up to homotopy type, by Wall's
theorem and its extensions.  Finally, we compute the Mordell-Weil lattice and
show that it matches as well.  At special loci in moduli space, where the branch
curve $B$ factorizes, we compute the Mordell-Weil torsion, and reproduce some of
the results of Sec.~3.

We conclude with summary of results and a discussion of connections to related
and ongoing work.  Ongoing work by the authors include applications to warped KK
reduction, D-brane instanton corrections, heterotic model building, and $SU(2)$
stucture Calabi-Yau compactifications, where the topology of an abelian surface
fibration spontaneously breaks extended supersymmetry.  We also outline
connections to recent work on D(imensional) duality~\cite{Dduality} and
semi-flat T-fold compactifications~\cite{Hull,McGreevy}.

Derivation of key results and relevant mathematical and physical background can
be found in the appendices.  App.~A describes how the monodromies of branes or
singular fibers are transformed under braiding motions of the locations of these
objects.  The duality derivation of the monodromy matrices of $\CX_{m,n}$ is
given in App.~D, and the lattice vectors of null loop junctions are computed in
App.~E\@.  Apps.~B and C contain background on abelian varieties and the
Mordell-Weil lattice.  App.~F is an introduction to complex curves, their
Jacobians, and line bundles.  App.~H gives background on direct images.
Finally, Apps.~G, I and J contain the derivations of mathematical results used
in Sec.~4.


\section{Monodromy and junction description of elliptic fibrations}

Given an elliptically fibered\footnote{Elliptic and abelian surface fibrations
  always refer to fibrations with section in this paper.}  Calabi-Yau manifold
$\CX$ with base $\CB$, \Ftheory\ provides a nonperturbative definition of a
family of type IIB string theory vacua with spatially varying dilaton-axion $\t$
\cite{VafaF}.  The type IIB vacua are defined by ``Newton's Law,'' $F =
M|_{A_{T^2}\to0}$ \cite{VafaF,BanksFMA}.  That is, we consider \Mtheory\ on
$\CX$ in the limit that the area of the elliptic fiber goes to zero, while
holding the complex structure fixed.  The result is type IIB compactified on
$\CB$, with $\tau$ identified with the complex structure modulus of the elliptic
fiber at each point on $\CB$.  The codimension 1 singular locus of the elliptic
fibration on $\CB$ is wrapped by 7-branes in the IIB description, where the type
of each 7-brane is determined by the monodromy of $\tau$ about the 7-brane
worldvolume.

Conversely, the nonperturbative type IIB description provides a useful encoding
of the geometry of $\CX$.  The collection of $(p,q)$ 7-branes determines the
degenerations of the elliptic fibration and the corresponding monodromy matrices
about singular fibers.  These data determine the topology of $\CX$.  Additional
geometric information is efficiently encoded by string junctions terminating on
the 7-branes.  These string junctions are the $W$-bosons of the 7-brane gauge
theory.  Their equivalence classes form a charge lattice known as the junction
lattice.  The string junctions lift to M2-branes wrapped on 2-cycles of $\CX$,
so they encode information about the geometry of 2-cycles.  For example,
enhanced gauge symmetry corresponds to coalescing groups of 7-branes in IIB\@.
In this case, the massless $W$-bosons are string junctions contractible to zero
length, which lift to 2-cycles contractible to zero volume.  Roughly speaking,
in the case of $\IP^1$ base, $H_2(\CX)$ comes from the generic fiber, extra
components of singular fibers, and sections of the elliptic fibration $\pi\colon
\CX\to\IP^1$.  The string junctions are related to the latter.  As we will see
in Sec.~2.5.2, the junction lattice determines the Mordell-Weil lattice of
rational sections of the elliptic fibration~\cite{MWJunction}.

To describe the geometry of the Calabi-Yau duals of $T^6/\IZ_2$ in Sec.~3, we
apply a similar monodromy and junction based description to the case of $T^4$
rather than $T^2$ fiber.  With this goal in mind, the remainder of this section
is devoted to laying the groundwork for the generalization by analyzing the
simpler elliptically fibered case in more detail.


\subsection{\Ftheory}

\subsubsection{The \Ftheory\ limit}

Let us briefly review the duality chain and limit that relates the initial
\Mtheory\ background to the final type IIB background.  We begin with \Mtheory\
compactified on an elliptic fibration $\CX$ over base $\CB$.  The generic fiber has
two nontrivial \hbox{1-cycles} $\a$ and $\b$\@.  In the limit of small $R_\a$,
the background is described by perturbative type IIA string theory
with\footnote{Here, we assume unit periodicity $x\cong x+1$ for toroidal
  coordinates and set $2\pi\sqrt{\a'}=1$ for simplicity.}  $g^\textrm{IIA}_s =
R_\a\ll 1$.  The IIA compactification manifold is the base of the $S^1_\a$
fibration.  If $R_\b$ is also small, then it is appropriate to \hbox{T-dualize}
to type IIB.  In the $R_\b\to0$ limit, the type~IIB $\b$ cycle decompactifies,
leaving type~IIB compactified on the base~$\CB$.

For the special case of a rectangular torus, the type~IIB dilaton-axion is
purely imaginary and can be identified with the complex structure modulus of the
elliptic fiber: $i/g_s^\mathrm{IIB} = iR_\b^\mathrm{IIA}/g_s^\mathrm{IIA} =
iR_\b/R_\a = \t$.  For a nonrectangular torus, the identification remains valid
and the real part of the complex structure modulus gives nonzero type IIB axion
$C_{(0)}$.

\subsubsection{Singular fibers, $(p,q)$ 7-branes, and $(p,q)$ strings}

\begin{figure}[ht]
  \def\BraneFiber{\mbox{\includegraphics{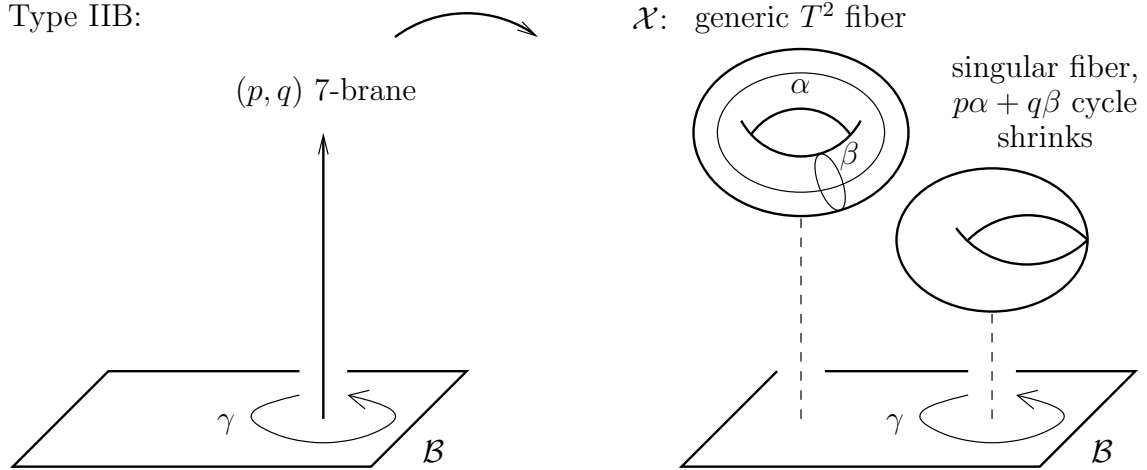}}}
  \begin{equation*}
    \begin{xy}
      \xyimport(6,2.5){\BraneFiber}
      (0.35,2.45)*\txt{Type IIB:} ; (3.4,2.45)*\txt{$\CX$:} ;
      (2.25,0.1)*{\CB} ; (5.8,0.1)*{\CB} ;
      (1.15,0.25)*{\gamma} ; (4.7,0.25)*{\gamma} ;
      (1.679,2.05)*\txt{$(p,q)$ 7-brane} ;
      (4.2,2.45)*\txt{generic $T^2$ fiber} ;
      (5.5,2,0)*\txt{singular fiber,\\ $p\alpha+q\beta$ cycle\\ shrinks} ;
      (4.2,2.08)*{\alpha} ; (4.46,1.7)*{\beta};
    \end{xy}
  \end{equation*}
  \caption{F-theory relates a $(p,q)$ 7-brane in type IIB to an elliptic fiber with
    vanishing $(p,q)$-cycle in $\CX$.}
  \label{fig:BraneFib}
\end{figure}

For now we assume that degenerations of the elliptic fibration $\CX$ are of
Kodaira type I$_1$: a $(p,q)$ 1-cycle $p\a + q\b \subset T^2$ vanishes over a
codimension~1 locus $D$ in the base.  In the corresponding type IIB
interpretation, a $(p,q)$ 7-brane wraps the divisor $D\subset \CB$ and spans the
noncompact dimensions of spacetime (see Fig.~\ref{fig:BraneFib}).  Here, a
$(p,q)$ 7-brane is an object on which a $(p,q)$ string can end.\footnote{A
  $(p,q)$ string is the bound state of $p$ fundamental and $q$ D-strings.  The
  space of all $(p,q)$ strings, with $p$ and $q$ relatively prime, is the
  $SL(2,\IZ)$ S-duality orbit of a fundamental string.}  Thus, a $(1,0)$ 7-brane
is a D7-brane.

The \Ftheory\ limit relates the dual interpretations of the integers $(p,q)$ as
string charge and homology vector.  A $(p,q)$ string in type IIB lifts to an
\Mtheory\ membrane wrapped on a $p\a+q\b$ cycle in the fiber of $\CX$.

\subsection{\Ftheory\ on K3}

The simplest \Ftheory\ compactification manifold with nontrivial base is a K3
surface.  A generic elliptic K3 surface is an elliptic fibration over a $\IP^1$
base, with 24 I$_1$ singular fibers over points on $\IP^1$.  The corresponding
IIB background is a compactification of type IIB string theory on $\IP^1$ with
24 $(p,q)$ 7-branes, each located at a point on $\IP^1$ and filling the 7+1
noncompact dimensions of spacetime.

The weak coupling perturbative interpretation of this background was well known
long before the formulation of \Ftheory.  It is the $T^2/\IZ_2$ orientifold,
dual to type I string theory via T-duality in the two $T^2$ directions.  There
are 16 D7-branes at arbitrary points on $T^2/\IZ_2\cong \IP^1$ and 4 O7-planes
located at the fixed points of the $\IZ_2$ involution.  A total of 16+4=20
objects is too few to correspond to the 24 $(p,q)$ 7-branes required by
\Ftheory.  However, it is already clear from the perturbative description that
the O7-planes are poorly described by leading order supergravity and thus need
not represent fundamental objects~\cite{Sen:1997kz,SUSYorient,CYDuals}.  At
finite distance from the O7 planes, the string coupling diverges.  The geometry
is a warped product $\IR^{7,1}\rtimes \IP^1$ in which a $\IP^1$ dependent scale
factor multiplies the flat metric on $\IR^{7,1}$, and this scale factor diverges
at finite distance from the O7-planes.

F-theory provides an elegant nonperturbative resolution of this pathology.  Each
O7-plane resolves to a pair of $(p,q)$ 7-branes.  Up to braiding operations on
the 7-branes described below, the pair is $(p,q)=(1,-1)$ and $(1,1)$.  The
separation between the two 7-branes depends on the string coupling as
$\exp(-1/g_s)$, so it is invisible in perturbation theory.\footnote{From the
  \Mtheory\ perspective, the corrections come from Kaluza-Klein modes along the
  circles in the elliptic fiber.  From the IIB perspective, the corrections come
  from $(p,q)$-strings, whereas the leading supergravity description only
  contains the effective field theory of a fundamental string.}  This phenomenon
can also be described by $SU(2)$ Seiberg-Witten theory: $\CN=2$, $SU(2)$
super-Yang-Mills theory is the theory on D3-brane probe in the background of an
O7-plane~\cite{BanksDouglas}.  In the classical gauge theory moduli space, there
is an enhanced $SU(2)$ symmetry point where the D3-brane coincides with the
O7-plane.  Quantum mechanically (due to instanton corrections in the gauge
theory), the $SU(2)$ point is lifted and replaced by two massless dyon points
separated by a distance of order
$\exp(-4\pi/g_\mathrm{YM}^2)$~\cite{SeibergWitten}.  The two dyon
hypermultiplets are $(p,q)$ strings stretched from the D3 probe to the
$(p,q)=(1,\pm1)$ 7-branes.  A dyon becomes massless when the D3-brane is
coincident with the corresponding 7-brane.


\subsection{Monodromies and braiding}\label{Sec:MonodBraid}

Three types of $(p,q)$ 7-branes appeared in the type IIB description of
\Ftheory\ on K3 given in the previous section.  Their $(p,q)$ charges are $\BA =
(1,0)$, $\BB=(1,-1)$ and $\BC=(1,1)$.  In this notation, a D7-brane is an $\BA$
brane located at a point on $\IP^1$ and an O7-plane resolves to a $\BB,\BC$
pair.
\begin{figure}[ht]
  \begin{equation*}
    \xygraph{
      !~:{@{-}|@{>}}
      !~-{@{--}}
      {\BA}="A1"-[dd] [ruu]{\dots}="dots"  [r]
      {\BA}="AN"-[dd] [ruu]
      {\BB}="B1"-[dd] [ruu]{\BC}="C1"-[dd] [ruu]
      {\BB}="B2"-[dd] [ruu]{\BC}="C2"-[dd] [ruu]
      {\BB}="B3"-[dd] [ruu]{\BC}="C3"-[dd] [ruu]
      {\BB}="B4"-[dd] [ruu]{\BC}="C4"-[dd]
    }
  \end{equation*}
  \caption{A collection $\BA,\BB,\BC$ points in $\IP^1$, and their branch cuts.
    The collection consists of 16 $\BA$ points and 4 $\BB,\BC$ pairs.}
  \label{fig:ABCdiagram}
\end{figure}

We can represent the IIB background, or equivalently the K3 surface, by
arranging the $\BA$, $\BB$ and $\BC$ points on the projective plane $\IP^1$,
keeping track of the locations of branch cuts (see
Fig.~\ref{fig:ABCdiagram}).  Here, the branch cuts denote discontinuities in
$\tau$.  If we do not introduce branch cuts, then $\tau$ is multiple-valued.
For example, $\tau\to\tau+1$ after circuiting a D7-brane counterclockwise, since
$\re\tau= C_{(0)}$ and since a D7-brane is a source of Ramond-Ramond (RR) flux
$F_{(1)}=d(C_{(0)})$.  Alternatively, we can opt for a single valued RR
potential $C_{(0)}$ plus a Dirac string, that is, a branch cut.  Then, the
discontinuity in crossing the branch cut of a D7-brane counterclockwise is
$\tau\to\tau-1$.  There are similar, $SL(2,\IZ)$ dual, discontinuities along the
branch cuts of other $(p,q)$ 7-branes.  Let us agree to draw all branch cuts as
vertical lines intersecting at the point at infinity on $\IP^1$.  This
determines an ordering (left to right in Figs.~\ref{fig:ABCdiagram}
and~\ref{fig:Xdiagram}) of the 7-branes, which we summarize in the diagram of
Fig.~\ref{fig:ABCdiagram}, or more compactly as
\begin{equation}\label{eq:ABCcollection}
  \BA^{16}\,\BB\BC\,\BB\BC\,\BB\BC\,\BB\BC.
\end{equation}
We denote an arbitrary $(p,q)$ 7-brane (or singular fiber) by $\BX_{[p,q]}$, so
for a more general collection of $(p,q)$ 7-branes, 
\begin{equation}\label{eq:Xcollection}
  \BX_{[p_1,q_1]}\BX_{[p_2,q_2]}\ldots\BX_{[p_n,q_n]},
\end{equation}
we have a diagram of the form shown in
Fig.~\ref{fig:Xdiagram}.
\begin{figure}[ht]
  \begin{equation*}
    \xygraph{
      !~-{@{--}}
      {\BX_{[p_1,q_1]}}="X1"-[dd] [r(1.5)uu]
      {\BX_{[p_2,q_2]}}="X2"-[dd] [r(1.5)uu]{\dots}="dots" [r(1.5)]
      {\BX_{[p_n,q_n]}}="XN"-[dd]
    }
  \end{equation*}
  \caption{A collection of $\BX_{[p,q]}$ points in $\IP^1$, and their branch
    cuts.}
  \label{fig:Xdiagram}
\end{figure}
\bigskip

\begin{figure}[ht]
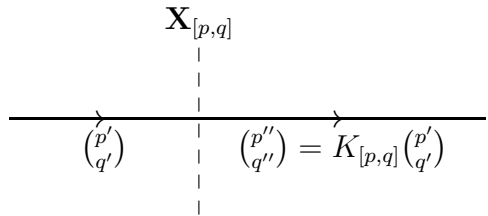

  \begin{equation*}
    \xygraph{
      !~:{@{-}|@{>}}
      !~-{@{--}}
      {\BX_{[p,q]}}="X1"-[dd] [ull]
      :[rr]_{\textstyle\binom{p'}{q'}}
      :[rrr]_{\textstyle\binom{p''}{q''}=K_{[p,q]}\binom{p'}{q'}} [ur]
    }
  \end{equation*}
  \caption{A string crossing a branch cut.}
  \label{fig:StringCrossingCut}
\end{figure}

We have already observed that a $(p,q)$ string can end at $(p,q)$
7-brane.  When a $(p',q')$ string crosses a branch cut of a $(p,q)$
7-brane in the counterclockwise direction about the branch point, the
charges of the string are transformed to new charges $(p'',q'')$, as
shown in Fig.~\ref{fig:StringCrossingCut}.

\noindent Let $\Bz$ denote the column vector $\binom{p}{q}$, with
$\Bz'$ and $\Bz''$ defined analogously.  Then, the transformation is
$\Bz'' = K_{[p,q]}\,\Bz'$, where $K_{[p,q]}$ is the monodromy matrix
\begin{equation}\label{eq:pqMonodromy}
  K_{[p,q]}= 
  \begin{pmatrix}
    1+pq & -p^2\\
    q^2 & 1-pq
  \end{pmatrix} \in \SL(2,\IZ).
\end{equation}
In particular, the monodromy matrices for $\BA$, $\BB$, $\BC$, and the pair $\BO
= \BB\BC$ are
\begin{gather}\label{eq:Kmonod}
  K_\BA = \begin{pmatrix} 1 & -1\\ 0 & 1\end{pmatrix},\quad
  K_\BB = \begin{pmatrix} 0 & -1\\ 1 & 2\end{pmatrix},\quad
  K_\BC = \begin{pmatrix} 2 & -1\\ 1 & 0\end{pmatrix},\\
  \mathrm{and}\quad
  K_\BO = K_\BC K_\BB = -\begin{pmatrix} 1 & 4\\ 0 & 1\end{pmatrix}.
\end{gather}
The fact that $K_\BO$ equals $K_\BA^{-4}$ up to a minus sign indicates that an
O7-plane has RR charge $-4$ times the charge of a D7-brane.  The overall minus
sign is due to the orientifold involution.

The monodromy matrices also have a topological interpretation on the K3 surface.
Here, $p$ and $q$ give the components of a 1-cycle $p\a+q\b$ in an elliptic
fiber, and $K_{[p,q]}$ determines how these components transform when the cycle
crosses the branch cut.

In the case of $\IP^1$ base, a loop that encloses all 7-branes (or singular
elliptic fibers) is contractible to the (smooth) point at infinity on $\IP^1$.
Therefore, the associated monodromy must be trivial.  Indeed, for the
$\BA,\BB,\BC$ description of K3,
\begin{equation}
  \bigl(K_\BC K_\BB\bigr)^4 {K_{\BA}}^{16} = 1.
\end{equation}

Finally, since the $(p,q)$ charges of strings are tranformed when crossing
branch cuts, the charges of 7-branes (on which they can end) are also
transformed.  Thus, the collection of \hbox{7-branes} corresponding to a given
\Ftheory\ compactification is unique only up to: (i)~braiding\break operations
in which 7-branes are are successively transported through the branch cuts of
other 7-branes and (ii)~an overall $SL(2,\IZ)$ conjugation of all
monodromies.\footnote{An overall $SL(2,\IZ)$ conjugation can often be achieved
  by braiding.  In this case, the equivalence (ii) is redundent.}  Examples of
braiding operations can be found in App.~\ref{App:Braiding}.  The collection of
7-branes~\eqref{eq:ABCcollection} describing F-theory on K3 is unique up to
these equivalences.


\subsection{String junctions and gauge symmetry}
\label{Sec:JunctionsGauge}

Given a collection of $N$ parallel D7-branes, no two coincident, the massive
$W$-bosons for the spontaneously broken $SU(N)$ worldvolume gauge symmetry are
$(1,0)$ fundamental strings stretched between the $(1,0)$ D7-branes.  In the
presence of an O7-plane, an enhancement to $SO(2N)$ occurs when the D7-branes
are all coincident with the O7-plane.  In the perturbative description, the
additional $W$-bosons associated with the spontaneous breaking of $SO(2N)$ (over
those of $SU(N)$) are strings stretched between the D7-branes and the O7-plane.
(On the covering space of the orientifold, these are strings connecting
D7-branes with their $\IZ_2$ images.)  In all cases, the masses of the
$W$-bosons can be attributed to finite string lengths $\times$ finite tensions.

Nonperturbatively, an O7-plane resolves to a $\BB,\BC$ pair of $(p,q)$-branes,
so the additional $W$-bosons of $SO(2N)$ should resolve to \emph{string
  junctions} stretched between the D7-branes and a $\BB,\BC$ pair.  Here, a
string junction is a collection of $(p,q)$ string segments, such that each
segment terminates at either a 7-brane or a vertex.  At a vertex, an arbitrary
number of strings can meet; the only requirement is that the total $(p,q)$
charge of the (oriented) strings entering the vertex equals the total $(p,q)$
charge leaving the vertex.

\begin{figure}[ht]
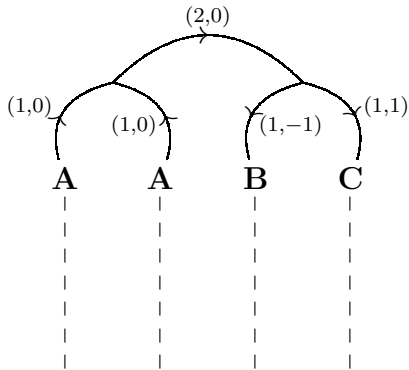

  \begin{equation*}
    \xygraph{
      !~:{@{-}|@{>}}
      !~-{@{--}}
      {\BA}="A1"-[dd] [ruu]
      {\BA}="A2"-[dd] [ruu]
      {\BB}="B"-[dd] [ruu]{\BC}="C"-[dd] [ruu]
      ,"A1":@/^1pc/[u(1)r(0.5)]^{(1,0)}
      ,"A2":@/_1pc/[u(1)l(0.5)]^{(1,0)}
           :@/^1.5pc/[rr]^{(2,0)}
          (:@/^1pc/"C"^{(1,1)}):@/_1pc/"B"^{(1,-1)}
    }
  \end{equation*}
  \caption{Massive string junction for the breaking of $SO(2N)$ to $SU(N)$.}
  \label{fig:DNroot}
\end{figure}

From the perturbative description, the natural guess is that an $SO(2N)$
$W$-boson realized as a string connecting a single D7-brane to an O7-plane
should resolve to a string junction connecting a single D7 brane to a $\BB,\BC$
pair.  But, such a junction is impossible with integer $(p,q)$ and charge
conservation at the trivalent vertex.  In fact, a perturbative
$\mathrm{D7}\to\mathrm{O7}$ string represents \emph{half of a root} of $SO(2N)$.
The $\textrm{D7}_i\to\textrm{D7}_{i+1}$ strings span the root lattice of
$SU(N)$.  Here, $i=1,2,\ldots,N$ denotes an ordering of the D7-branes and an
arrow denotes an oriented string stretch between the two objects.  Adding the
combination $\bigl(\mathrm{D7}_{N-1}\to\mathrm{O7}\bigr) \oplus
\bigl(\mathrm{D7}_N\to\mathrm{O7}\bigr)$ enlarges the lattice to the root
lattice of $SO(2N)$~\cite{Sen:1997kz}.  The nonperturbative resolution of the
last root is a string junction connecting two D7-branes to a $\BB$ and $\BC$
brane.  As shown in Fig.~\ref{fig:DNroot}, the two $(1,0)$ strings emanating
from the D7 branes join to form a $(2,0)$ string; then, the $(2,0)$ string
splits to form a $(1,-1)$ plus a $(1,1)$ string, which terminate on the $\BB$
and $\BC$ branes.


\subsection{Junction lattice}
\label{Sec:JunctionLattice}

The general formalism was worked out in a series of papers by DeWolfe et
al.~\cite{DeWolfeI,DeWolfeII,DeWolfeIII,DeWolfeIV}.  Given a collection of
7-branes, ordered as in Fig.~\ref{fig:Xdiagram}, we define a lattice of
equivalence classes of string junctions as follows.  

First, to each string junction, we associate a lattice vector
  \begin{equation}\label{eq:Qvector}
    Q = \sum_i Q_i \Bs_i,
  \end{equation}
where $Q_i\in\IZ$ is the net number of strings leaving (minus entering) the
$i$th 7-brane $\BX_{[p_i,q_i]}$, and the $\Bs_i$ for $i=1,\dots,N$ are a formal
basis for a rank $N$ lattice, where $N$ is the number of 7-branes.  We can think
of $\Bs_i$ as an outward oriented $(p_i,q_i)$ half-string emanating from
$\BX_{[p_i,q_i]}$.  Two strings junctions are equivalent if they have the same
$Q$.

\begin{figure}[ht]
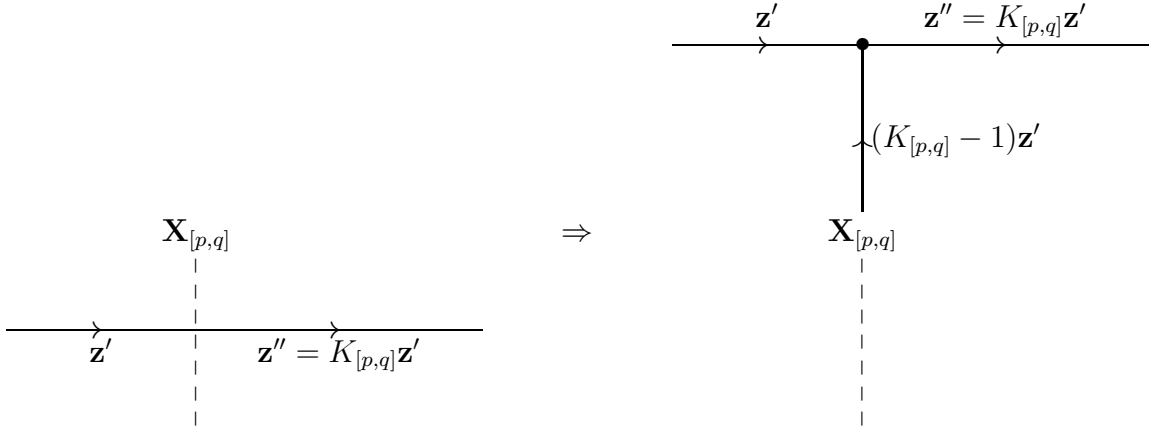

  \begin{equation*}
    \xygraph{
      !~:{@{-}|@{>}}
      !~-{@{--}}
      {\BX_{[p,q]}}="X1"-[dd] [ull]
      :[rr]_{\textstyle\Bz'}
      :[rrr]_{\textstyle\Bz''=K_{[p,q]}\Bz'} [ur]
      {\Rightarrow} [rrr]
      {\BX_{[p,q]}}(-[dd])
      :[uu]_{\textstyle(K_{[p,q]}-1)\Bz'} [ll]
      :[rr]="Vertex"^{\textstyle\Bz'\vphantom{K_{[p,q]}}}
      :[rrr]^{\textstyle\Bz''=K_{[p,q]}\Bz'} [ur]
      ,"Vertex"{\bullet}
    }
  \end{equation*}
  \caption{Pushing a string through a 7-brane: the Hanany-Witten effect.}
  \label{fig:HananyWitten}
\end{figure}

Each equivalence class $Q$ can be represented by a junction in \emph{standard
  presentation}, that is, by a tree-graph with trivalent vertices.  The
nontrivial step in converting a given representative to standard presentation is
the operation of pushing a string through a \hbox{7-brane $\BX_{[p,q]}$}.  This
operation is illustrated in Fig.~\ref{fig:HananyWitten}.  Below the 7-brane,
string charge conservation requires a discontinuity from $\Bz'=\binom{p'}{q'}$
to $\Bz''=K_{[p,q]}\,\Bz'$ across the branch cut.  Above the 7-brane, the
discontinuity can only be accounted for by the appearance of a new string that
connects the 7-brane $\BX_{[p,q]}$ to the point of discontinuity.  This is an
example of the Hanany-Witten effect~\cite{HananyWitten}.

As required, the new string emanating from $\BX_{[p,q]}$ has string
charge proportional to $\Bz=\binom{p}{q}$:
\begin{equation}\label{eq:HWrule}
  (K_{[p,q]}-1)\Bz' = (\Bz'\cdot\Bz)\, \binom{p}{q},
  \quad\text{where}\quad
  \Bz'\cdot\Bz = \begin{vmatrix} p' & p\\ q' & q\end{vmatrix}
                = p'q-q'p.
\end{equation}

The \emph{junction lattice} $J$ is defined to be the lattice of equivalence
classes $Q$ of junctions together with the inner product,
\begin{equation}\label{eq:innerproduct}
  \begin{split}
  (s_i,s_i) &= -1,\\
  (s_i,s_j) &= (s_j,s_i) = \ha(p_i q_j - p_j q_i),\quad i < j.
  \end{split}
\end{equation}
A good way to think about this inner product and the tranformation rule
\eqref{eq:HWrule} is that they are both related to the antisymmetric
intersection pairing $(p',q')\cdot(p,q) = p'q-q'p\,$ of \hbox{1-cycles} on the
elliptic fiber.  To define the symmetric inner product \eqref{eq:innerproduct},
we use the additional structure provided by the ordering of the branch cuts.

On $\IC^1$ or an open subset of $\IP^1$, the junction lattice $J$ is the full
rank $N$ lattice generated by the $\Bs_i$.  However, some of the lattice vectors
correspond to junctions with asymptotic $(p,q)$ charge at infinity.
We define the \emph{proper junction lattice} $J_\text{proper}$ to be the
sublattice of \emph{proper string junctions} in which no net charge is carried
by strings that run off to infinity,\footnote{Eq.~\eqref{eq:properlattice} gives
  a homomophism from $\Span(\{\Bs_i\})\cong\IZ^N$ to the $\IZ^2$ of $(p,q)$
  string charges.  The proper junction lattice is the kernel of this
  homomorphism.}
\begin{equation}\label{eq:properlattice}
  \sum_i Q_i\binom{p_i}{q_i} = \binom{0}{0}
  \quad\text{(proper junction lattice).}
\end{equation}
This constraint means that the rank of $J_\text{proper}$ is less than the number
of 7-branes by 1 or 2.\footnote{For example, for a collection of D7 branes only,
  Eq.~\eqref{eq:properlattice} gives only 1 constraint since $q_i=0$ for all
  $i$.  For a collection that spans the $\IZ^2$ of possible $(p,q)$, there are 2
  constraints.}

On $\IP^1$, there is no distinction between the junction lattice and proper
junction lattice.  The point at infinity is contained in $\IP^1$, and a string
cannot terminate there unless it is the location of a 7-brane or vertex.  Thus,
on $\IP^1$, we have $J = J_\text{proper}$, of rank less than the number of
7-branes $N$.

The root lattices of the A-D-E Lie algebras can each be realized as the proper
junction lattice of collections of $\BA$, $\BB$ and $\BC$ type 7-branes, as
indicated in Table~\ref{tbl:ADE}.  (See
Refs.~\cite{DeWolfeI,DeWolfeII,DeWolfeIII,DeWolfeIV} for further details.)
\begin{table}[ht]
  \centering
  \begin{equation*}
    \begin{split}
      A_N\quad & \BA^{N+1}\quad (N\ge1),\\
      D_N\quad& \BA^N \BB \BC\quad (N\ge4),\\
      E_N\quad & \BA^{N-1} \BB \BC \BC\quad (N=6,7,8),\\
      H_N\quad & \BA^{N+1} \BC\quad (N=0,1,2).
    \end{split}
  \end{equation*}
  \caption{A-D-E Lie algebras and the corresponding 7-brane collections.}
  \label{tbl:ADE}
\end{table}

The proper string junctions of each of these collections are the $W$-bosons of
the corresponding gauge symmetry.  When the 7-branes coalesce to a point, the
$W$-bosons are massless and the gauge symmetry is unbroken.  In general, a
collection of 7-branes can coalesce if and only if the inner product of the
proper junction lattice of the collection is negative definite.  This condition
is satisfied for the classical A-D-E Lie algebras with $N$ in the ranges given
in Table~\ref{tbl:ADE}, but not for the more exotic algebras like $E_N$ with
$N>8$.

The $H_N$ row of Table~\ref{tbl:ADE} provides a second way to realize $A_N$
gauge symmetry.  In contrast to the perturbative realization via $N+1$ $\BA$
branes (D7-branes), the $H_N$ realization is strongly coupled.  Likewise, in the
moduli space of the $\CN=2$ worldvolume theory on a D3-brane probe in the
presence of a coalesced $H_{N}$ collection of 7-branes, there is a strongly
coupled Argyres-Douglas point~\cite{ArgyresDouglas}, at which $N+1$
hypermultiplets (of two mutually nonlocal electromagnetic charges) become
massless.  The hypermultiplets come from a $(1,0)$ string or $(1,1)$ string
stretched between the D3 brane and an $\BA$ or $\BC$ brane, respectively.

\begin{figure}[ht]
\def\SpheresFromJunctions{\mbox{\includegraphics{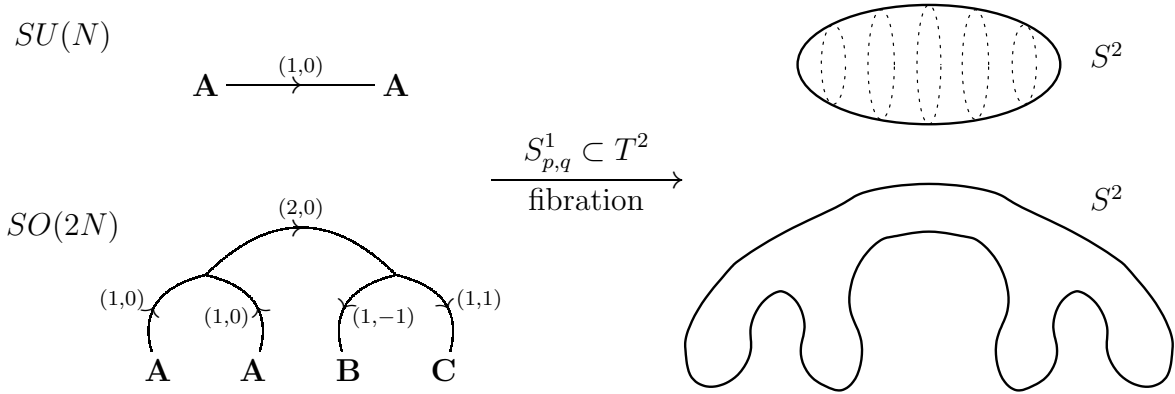}}}
  \begin{equation*}
      \xygraph{
        !~:{@{-}|@{>}}
        !~-{@{>}}
        ="Here" [u(3.75)]{SU(N)} [d(2)]{SO(2N)} [u(1.5)r(1.5)]
        {\BA}="A1":[rr]{\BA}="A2"^{(1,0)} [d(3)l(2.5)]
        {\BA}="A3" [r]
        {\BA}="A4" [r]
        {\BB}="B"  [r]{\BC}="C" [r]
        ,"A3":@/^1pc/[u(1)r(0.5)]^{(1,0)}
        ,"A4":@/_1pc/[u(1)l(0.5)]^{(1,0)}
        :@/^1.5pc/[rr]^{(2,0)}
        (:@/^1pc/"C"^{(1,1)}):@/_1pc/"B"^{(1,-1)}
        ,"A2" [dr]-[rr]^{\displaystyle S^1_{p,q}\subset T^2}_{\txt{fibration}}
      }
    \begin{xy}
     \xyimport(2.8,2.2){\SpheresFromJunctions}
     (2.4,1.9)*{S^2} ;
     (2.4,1.1)*{S^2}
    \end{xy}
  \end{equation*}
  \caption{Lift of $SU(N)$ roots and the additional roots of $SO(2N)$ from
    string junctions stretched between 7-branes in type IIB to 2-cycles in $\CX$.}
  \label{fig:JunctionLift}
\end{figure}

In terms of the elliptic fibration, string junctions lift to 2-cycles in $\CX$.
A junction in standard presentation (i.e., a tree graph) lifts to a 2-cycle that
is topologically $S^2$.  As mentioned earlier, the 2-cycle is obtained by
fibering the circle $S^1_{p,q}\subset T^2$ over each $(p,q)$ string segment in
$\IP^1$ (see Fig.~\ref{fig:JunctionLift}).  The 2-cycle smoothly pinches off at
the locations of the singular fibers, at which an $S^1$ shrinks to zero size.
The junction inner product \eqref{eq:innerproduct} reproduces the standard
intersection pairing on $H_2(\CX)$.  The statement that a collection of 7-branes
can coalesce to a point on $\IP^1$ only for negative definite inner product
reproduces the standard result that a collection of 2-cycles can be collapse to
zero size only for negative definite intersection matrix.

Finally, coalescing $(p,q)$ 7-branes in type IIB correspond to coalescing I$_1$
singular fibers in~$\CX$.  The relation between the choice of coalescing
collection (in terms of $\BA$, $\BB$ and $\BC$ type I$_1$~fibers) and the
Kodaira type of the resulting singular fiber is shown in
Table~\ref{tbl:Kodaira}~\cite{Wikipedia,Barth}.  The table also indicates the
number of irreducible components of the resulting singular fiber and the
intersection matrix of these components.
\begin{table}[ht]
  \centering
  \begin{tabular}[ht]{lllll}
    \multicolumn{2}{l}{Coalescing collection}  & Kod.\ type &  
    Number of components & Intersection matrix\\
    \ --- &\quad --- & I$_0$ & 1 (elliptic) & 0\\
    $A_{N-1}$ & ($\BA^N$) & I$_N$  & $N$ ($N$ distinct intersect.\ pts.)
    & affine $A_{N-1}$\\
    $D_{N+4}$ & ($\BA^{N+4}\,\BB \BC$)  & I$^*_N$ & $N+5$ & affine $D_{N+4}$\\
    $E_6$ & ($\BA^5 \BB \BC \BC$) & IV$^*$ & 7 & affine $E_6$\\
    $E_7$ & ($\BA^6 \BB \BC \BC$) & III$^*$ & 8 & affine $E_7$\\
    $E_8$ & ($\BA^7 \BB \BC \BC$) & II$^*$ & 9 & affine $E_8$\\
    $H_0$ & ($\BA \BC$) & II & 1 (with cusp) & 0\\
    $H_1$ & ($\BA^2 \BC$) & III & 2 (meet in one pt.\ of order 2) & affine $A_1$\\
    $H_2$ & ($\BA^3 \BC$) & IV & 3 (all meet in 1 pt.) & affine $A_2$
  \end{tabular}
  \caption{Coalescing collections of I$_1$ fibers and Kodaira type of
    the resulting singular fiber.}
  \label{tbl:Kodaira}
\end{table}


\subsubsection{Junction lattice of $\half \mathrm{K3} = \mathrm{dP}_9$}

The simplest nontrivial choice of the Calabi-Yau manifold $\CX$ is an elliptic K3
surface.  In this case, we can simplify matters further by considering the
stable degeneration limit, in which the base degenerates to two $\IP^1$s meeting
at a point, and K3 factorizes into two dP$_9$ surfaces\footnote{A dP$_9$ is a
  rational elliptic surface: rational, since it is the blow-up of $\IP^2$ in
  nine points, and elliptic, since it admits an elliptic fibration over $\IP^1$.
  The sections are the nine blow-up $\IP^1$s and the elliptic fiber is
  represented by the canonical class $K=-3H+\sum_{i=1}^9\CE_i$, with $K^2=0$.
  Here, $H$ is the hyperplane class of $\IP^2$ and $\CE_i$ is the class of the
  $i$th exceptional $\IP^1$.}  meeting in an elliptic curve.  Via duality to the
$E_8\times E_8$ heterotic string, each dP$_9$ corresponds to a single $E_8$
factor, so we can focus on a single dP$_9$.

As an elliptic fibration, a generic smooth dP$_9$ has 12 singular fibers of type
I$_1$.  Up to the equivalences discussed in App.~\ref{App:Braiding} (braiding
and $SL(2,\IZ)$ conjugation), the collection of singular fibers is
$\BA^8\BB\BC\BB\BC$---exactly half of the corresponding
collection~\eqref{eq:ABCcollection} for K3.  The total monodromy about all
singular fibers is again unity, as required,
\begin{equation}
  \bigl(K_\BC K_\BB\bigr)^2 {K_{\BA}}^8 = 1.
\end{equation}

The junction lattice of dP$_9$ was analyzed in great detail in
Ref.~\cite{MWJunction}.  For a smooth dP$_9$ with 12 I$_1$ fibers, the junction
lattice is the semidefinite lattice
\begin{equation}
  J = E_8^-\oplus\IZ\Bdelta_1\oplus\IZ\Bdelta_2,
\end{equation}
the direct sum of the $E_8^-$ lattice (where the minus indicates that
the inner product is minus that of the $E_8$ root lattice) and a 2D
null lattice generated by the charge vectors
\begin{equation}\label{eq:dPQ}
  \begin{split}
    \Bdelta_1 &= (0,0,0,0,0,0,0,0,-1,-1,1,1),\\
    \Bdelta_2 &= (-1,-1,-1,-1,-1,-1,-1,-1,7,5,-3,-1).
  \end{split}
\end{equation}
By brane motions (the braiding operations of App.~\ref{App:Braiding}),
$\BA^8\BB\BC\BB\BC\cong\BA^7\BB\BC\BC\BX_{[3,1]}\BA$.  The $E_8^-$ is then
generated by the proper junctions of $\BA^7\BB\BC\BC$.  The null vectors
$\Bdelta_1$ and $\Bdelta_2$ have the following interpretation.  They can each be
represented by a \emph{loop junction}, a counterclockwise loop of $(p,q)$
strings circling all of the the 7-branes.  (See Fig.~\ref{fig:LoopJunctions}
below.)  Since the total monodromy is unity, a $(p,q)$ string starting above all
of the branch cuts comes back again to a $(p,q)$ string after traversing the
complete loop.  So, it can close.  Such a loop is contractible to the point at
infinity and is in this sense trivial.\footnote{Likewise, the 2-cycle in $\CX$
  obtained by fibering an $S^1$ over this loop is a homologically trivial $T^2$
  in $\CX$.}  For $(p,q) = (1,0)$ and $(0,1)$, we obtain the charge vectors
$\Bdelta_1$ and $\Bdelta_2$, respectively.  In order to obtain the charge
vectors quoted in Eq.~\eqref{eq:dPQ}, it is necessary to convert the loop
junction to standard tree presentation by ``pushing strings through vertices''
using the Hanany-Witten effect, as described in Sec.~2.5.

\begin{figure}[ht]
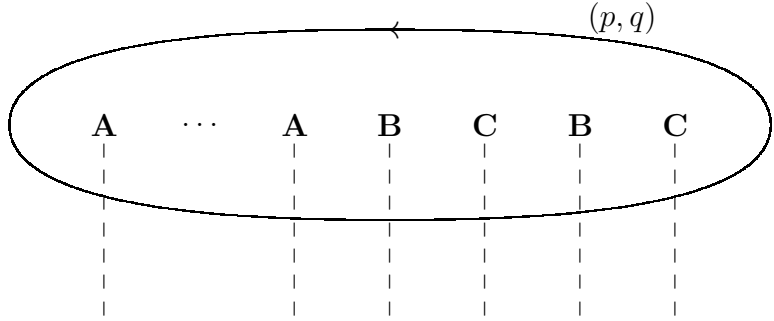

  \begin{equation*}
    \xygraph{
      !~:{@{-}}
      !~-{@{--}}
      {\BA}="A1"-[dd] [ruu]{\dots}="dots"  [r]
      {\BA}="A8"-[dd] [ruu] 
      {\BB}="B1"-[dd] [ruu]{\BC}="C1"-[dd] [ruu]
      {\BB}="B2"-[dd] [ruu]{\BC}="C2"-[dd]
      ,"B1"[u] ="N"
      ,"A1"[l] ="W"
      ,"B1"[d] ="S"
      ,"C2"[r] ="E"
      ,"N":@(l,u)"W" :@(d,l)"S" :@(r,d)"E" :@{->}@(u,r)"N"_{\displaystyle(p,q)}
    }
  \end{equation*}
  \caption{A loop junction, contractible to the point at infinity.  The null
    generators $\Bdelta_1$ and~$\Bdelta_2$ are obtained for $(p,q)=(1,0)$ and
    $(0,1)$, respectively.}
  \label{fig:LoopJunctions}
\end{figure}


\subsubsection{Mathematical interpretation of the junction lattice}
\label{Sec:MathInterp}

In this section, we establish terminology for variety of lattices related to the
junction lattice~$J$, and then relate these lattices to the homology of the
elliptic fibration.  (See Eqs.~\eqref{eq:Qvector} and \eqref{eq:innerproduct}
for the definition of the junction lattice.)
\begin{enumerate}
  
\item The \emph{null junction lattice} $J_\text{null}\subset J$ is the null
  sublattice of the junction lattice.

\item The \emph{loop junction lattice} $J_\Loop\subset J_\text{null}$ is the
  null sublattice generated by loop junctions circling all 7-brane locations
  (locations of singular fibers) on $\IP^1$ and contractible to the point at
  infinity.\footnote{For a compact elliptic surface, there is no torsion in the
    N\'eron-Severi lattice (the algebraic part of $H_2(\CX,\IZ)$), and no
    distinction between $J_\Loop$ and $J_\text{null}$.  However, it is useful to
    introduce the extra terminology here, so that it will carry over to abelian
    surface fibrations without alteration.  As we will see, $J_\Loop\ne
    J_\text{null}$ for the abelian surface fibrations $\CX_{m,n}$ with $m\ne 1$
    discussed in Sec.~3.}

\item The \emph{effective junction lattice} is the quotient
  \begin{equation}
        J_\eff = J/J_\Loop.
  \end{equation}
  We have just seen that the effective junction lattice of a smooth dP$_9$ is
  $J_\eff({\rm dP}_9) = E_8^-$.
    
\item The \emph{Kodaira junction lattice} $J_{\rm Kodaira}$ is the sum of the
  proper junction sublattices associated to each of the collections of
  coalescing 7-branes.  It is trivial for a smooth surface, where all singular
  fibers are of type I$_1$, but nontrivial when a collection of 7-branes has
  coalesced and the elliptic fibration acquires a multicomponent Kodaira fiber,
  as in Table~\ref{tbl:Kodaira}.

\item The \emph{tadpole junction lattice} is
  \begin{equation}
    J_0 = J^\perp_{\rm Kodaira}\equiv \text{orthogonal complement of $J_{\rm
        Kodaira}$ in $J_\eff$.}
  \end{equation}
  For a smooth surface, $J_0 = J_\eff$.  The reason for the terminology is that
  the sublattice orthogonal to the proper junctions associated to a
  subcollection of coalescing branes can be represented by junctions that are
  tree graphs away from the collection, with a possible termination in a
  \emph{tadpole loop} circling the collection
  (cf.~Figs.~\ref{fig:NullTadpoleJunction} and \ref{fig:TadpoleJunctions}).

\end{enumerate}
In Sec.~\ref{Sec:WeaklyIntegral}, we will also define \emph{weakly integral}
analogs of these lattices.

Let us describe the homological interpretation of the various lattices just
defined.  First consider an arbitrary elliptic fibration $\pi\colon\,\CX\to
\CB$.  The existence of the projection $\pi$ induces a filtration on the
cohomology that allows us to write \cite{Aspinwall,GandH}.\footnote{To be
  precise, Eq.~\eqref{eq:CohomSplit} is not literally correct as written.  There
  exists a filtration, but additional assumptions are necessary for the sequence
  to split as shown.  For a further discussion, see App.~\ref{App:DirIm}.}
\begin{equation}\label{eq:CohomSplit}
  H^2(\CX,\IZ) = H^2(\CB,\IZ)\oplus H^1(\CB,R^1\pi_*\IZ)\oplus H^0(\CB,R^2\pi_*\IZ).
\end{equation}
Here, $H^p(\CB,R^q\pi_*\IZ)$ can roughly be thought of as the cohomology of
degree $p$ along the base and $q$ along the fiber.\footnote{The sheaf
$R^q\pi_*\IZ$ on $\CB$ associates the group $H^q(\pi^{-1}(U),\IZ)$ to each open
set $U\subset \CB$.  For $U$ a neighborhood of a generic point $x\in\IP^1$ this
becomes the cohomology group $H^q(f_x,\IZ)$ along the elliptic fiber
$f_x=\pi^{-1}(x)$, modulo monodromy equivalences.  See App.~\ref{App:DirIm} for
further background on $R^q\pi_*$.}  We have contributions:

\begin{enumerate}
  \item $H^2(\CB,\IZ)$ from the base,

  \item $H^0(\CB,R^2\pi_*\IZ)$ from the generic fiber and extra components of
    reducible fibers,
    
  \item $H^1(\CB,R^1\pi_*\IZ)$ from everything else.
\end{enumerate}

Now restrict to an elliptic surface $\CX$.  The string junctions of the previous
section are real 1D graphs on $\CB$ that lift to 2-cycles in $\CX$ by fibering an
$S^1\subset T^2$ over each segment of the graph.  Thus we might expect that they
are related to $H^1(\CB,R^1\pi_*\IZ)$.  Indeed, this is the interpretation of
the tadpole junction lattice given in Ref.~\cite{MWJunction},
\begin{equation}
  J_0 = H^1(\CB,R^1\pi_*\IZ).
\end{equation}
Likewise,
\begin{equation}
  H^0(\CB,R^2\pi_*\IZ) = J_{\rm Kodaira}\oplus F\,\IZ,
\end{equation}
where $F$ is the generic fiber.

The group of rational sections of an elliptic fibration is known as the
\emph{Mordell-Weil} group $\MW$\@.  The \emph{narrow Mordell-Weil group} $\MW_0$
is the subgroup of sections that intersect the same components of singular
fibers as the zero section $\s_0$.  In general
\cite{CoxZucker,MWJunction,Aspinwall},
\begin{equation}
  \MW_0(\CX) = H^1(\CB,R^1\pi_*\IZ)\cap H^{1,1}(\CX,\IC).
\end{equation}
For dP$_9$, the intersection removes nothing, and $\MW_0 = H^1(\CB,R^1\pi_*\IZ)
= J_0 $, where $J_0=E_8^-$ from the previous section.  This is the situation to
keep in mind for the generalization to abelian surface fibered Calabi-Yau
manifolds in Sec.~3, where again $H^{2,0}(\CX)=0$.  On the other hand, for $\CX$
an elliptic K3, $H^{2,0}\ne0$, and the intersection with $H^{1,1}$ depends on
the choice of complex structure moduli.\footnote{For an elliptic K3,
  $H^1(\CB,R^1\pi_*\IZ) = E_{10}^-\oplus E_{10}^-$ is a signature $(2,18)$
  sublattice of $H^2({\rm K3},\IZ)$, which when combined with $\s_0,f$ gives the
  full $H^2$ lattice \cite{DeWolfeIV}. On the other hand $\MW_0$ varies from
  rank $0$ to $16$ depending on complex structure.}

The junction description of the full Mordell-Weil group $\MW$ and its torsion
subgroup $MW_\tor$ will be given in Sec.~\ref{Sec:WeaklyIntegral}, once we have
introduced the notion of weak integrality.


\subsubsection{An example with coalesced 7-branes}
\label{Sec:EllipticExample}

We now turn to an example in which some of the 7-branes (I$_1$ fibers of $\CX$)
have coalesced.  For simplicity, we start again with dP$_9$.  From the braiding
operations discussed in App.~\ref{App:Braiding}, we have
\begin{equation}
  \BA^8\, \BB \BC\, \BB \BC \cong
  \BA^4\, \BB \BC \BA^2\, \BB \BC \BA^2\cong
  \BA^4\, \BB \BA^2 \BB\, \BB \BA^2 \BB\cong
  \BA^4\, \BB^2 \BD^2\, \BB^2 \BD^2,
\end{equation}
where $\BD = \BX_{[0,1]}$.  We adopt the last collection $\BA^4 \BB^2\BD^2
\BB^2 \BD^2$ for this example.  

In the basis corresponding to this collection, the null loop junctions of
Sec.~\ref{Sec:JunctionLattice} become
\begin{equation}\label{eq:ExNull}
  \begin{split}
    \Bdelta_1 &= (0,0,0,0,-1,-1,-1,-1,1,1,1,1),\\
    \Bdelta_2 &= (-1,-1,-1,-1,3,3,-2,-2,-1,-1,0,0),
  \end{split}
\end{equation}
from loops with $(p,q) = (1,0)$ and $(0,1)$, respectively.  In our collection,
adjacent branes of the same type can coalesce.  We will use surrounding
parentheses to denote subcollections of coalesced branes.  Thus,
\begin{equation}\label{eq:ExJKodaira}
  (\BA^2) (\BA^2) (\BB^2) (\BD^2) (\BB^2) (\BD^2),
  \quad\text{with}\quad J_\Kodaira = A_1^-\,{}^{\oplus6},
\end{equation}
denotes a brane collection in which all branes have coalesced pairwise.  In the
elliptic fibration of the dP$_9$, the twelve I$_1$ fibers have coalesced
pairwise to become six I$_2$ fibers, each giving an $A_1$ surface singularity.
The Kodaira junction lattice is $A_1^-\,{}^{\oplus6}$, generated by $(p,q)$
strings that connect the two 7-branes in each pair:
\begin{equation}
  \begin{split}
    \Balpha_1 &= \Bs_1-\Bs_2,\quad
    \Balpha_2 = \Bs_3-\Bs_4,\quad
    \Balpha_3 = \Bs_5-\Bs_6,\\
    \Balpha_4 &= \Bs_7-\Bs_8,\quad
    \Balpha_5 = \Bs_9-\Bs_{10},\quad
    \Balpha_6 = \Bs_{11}-\Bs_{12}.
  \end{split}
\end{equation}
Here, $A_1^-$ denotes a lattice whose inner product is minus that of the $A_1$
root lattice.  

\begin{figure}[ht]
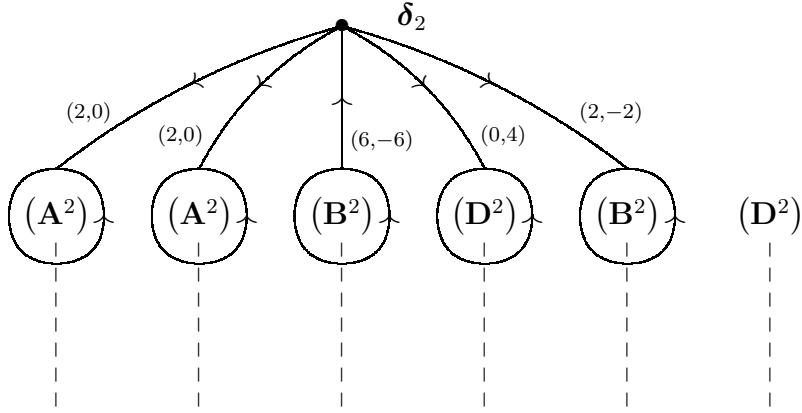

  \begin{equation*}
    \xygraph{
      !~:{@{-}}
      !~-{@{--}}
      {\bigl(\BA^2\bigr)}="A1"-[dd] [r(1.5)uu]
      {\bigl(\BA^2\bigr)}="A2"-[dd] [r(1.5)uu]
      {\bigl(\BB^2\bigr)}="B1"-[dd] [r(1.5)uu]
      {\bigl(\BD^2\bigr)}="D1"-[dd] [r(1.5)uu]
      {\bigl(\BB^2\bigr)}="B2"-[dd] [r(1.5)uu]
      {\bigl(\BD^2\bigr)}="D2"-[dd]
      ,"A1" [u(0.5)] ="A1N" [dl(0.5)] ="A1W" [dr(0.5)]  ="A1S" [ur(0.5)] ="A1E"
      ,"A1N":@(l,u)@/_/"A1W":@(d,l)@/_/"A1S":@{->}@(r,d)@/_/"A1E":@(u,r)@/_/"A1N"
      ,"A2" [u(0.5)] ="A2N" [dl(0.5)] ="A2W" [dr(0.5)]  ="A2S" [ur(0.5)] ="A2E"
      ,"A2N":@(l,u)@/_/"A2W":@(d,l)@/_/"A2S":@{->}@(r,d)@/_/"A2E":@(u,r)@/_/"A2N"
      ,"B1" [u(0.5)] ="B1N" [dl(0.5)] ="B1W" [dr(0.5)]  ="B1S" [ur(0.5)] ="B1E"
      ,"B1N":@(l,u)@/_/"B1W":@(d,l)@/_/"B1S":@{->}@(r,d)@/_/"B1E":@(u,r)@/_/"B1N"
      ,"D1" [u(0.5)] ="D1N" [dl(0.5)] ="D1W" [dr(0.5)]  ="D1S" [ur(0.5)] ="D1E"
      ,"D1N":@(l,u)@/_/"D1W":@(d,l)@/_/"D1S":@{->}@(r,d)@/_/"D1E":@(u,r)@/_/"D1N"
      ,"B2" [u(0.5)] ="B2N" [dl(0.5)] ="B2W" [dr(0.5)]  ="B2S" [ur(0.5)] ="B2E"
      ,"B2N":@(l,u)@/_/"B2W":@(d,l)@/_/"B2S":@{->}@(r,d)@/_/"B2E":@(u,r)@/_/"B2N"
      ,"B1" [uu]="Vertex" {\bullet}
      !~:{@{-}|@{>}}
      ,"Vertex":@/_/"A1N"_(0.8){(2,0)}
      ,"Vertex":@/_/"A2N"_(0.88){(2,0)}
      ,"B1N":"Vertex"_(0.2){(6,-6)}
      ,"Vertex":@/^/"D1N"^(0.88){(0,4)}
      ,"Vertex":@/^/"B2N"^(0.8){\;\;(2,-2)}^(0.2){\strut\displaystyle\Bdelta_2}
    }
  \end{equation*}
  \caption{The null junction $\Bdelta_2$ represented as a tadpole junction.}
  \label{fig:NullTadpoleJunction}
\end{figure}

\begin{figure}[ht]
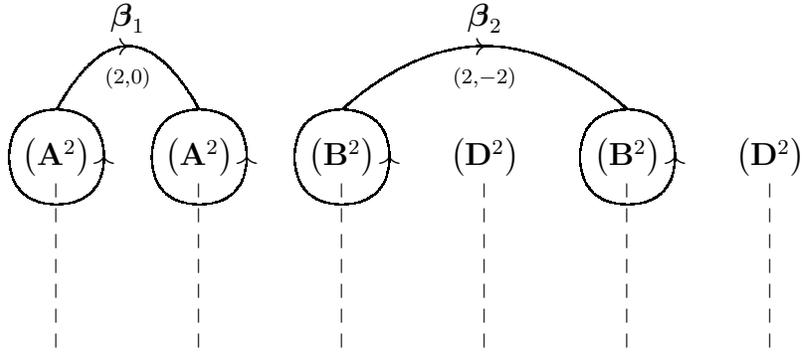

  \begin{equation*}
    \xygraph{
      !~:{@{-}}
      !~-{@{--}}
      {\bigl(\BA^2\bigr)}="A1"-[dd] [r(1.5)uu]
      {\bigl(\BA^2\bigr)}="A2"-[dd] [r(1.5)uu]
      {\bigl(\BB^2\bigr)}="B1"-[dd] [r(1.5)uu]
      {\bigl(\BD^2\bigr)}="D1"-[dd] [r(1.5)uu]
      {\bigl(\BB^2\bigr)}="B2"-[dd] [r(1.5)uu]
      {\bigl(\BD^2\bigr)}="D2"-[dd]
      ,"A1" [u(0.5)] ="A1N" [dl(0.5)] ="A1W" [dr(0.5)]  ="A1S" [ur(0.5)] ="A1E"
      ,"A1N":@(l,u)@/_/"A1W":@(d,l)@/_/"A1S":@{->}@(r,d)@/_/"A1E":@(u,r)@/_/"A1N"
      ,"A2" [u(0.5)] ="A2N" [dl(0.5)] ="A2W" [dr(0.5)]  ="A2S" [ur(0.5)] ="A2E"
      ,"A2N":@(l,u)@/_/"A2W":@(d,l)@/_/"A2S":@{->}@(r,d)@/_/"A2E":@(u,r)@/_/"A2N"
      ,"B1" [u(0.5)] ="B1N" [dl(0.5)] ="B1W" [dr(0.5)]  ="B1S" [ur(0.5)] ="B1E"
      ,"B1N":@(l,u)@/_/"B1W":@(d,l)@/_/"B1S":@{->}@(r,d)@/_/"B1E":@(u,r)@/_/"B1N"
      ,"B2" [u(0.5)] ="B2N" [dl(0.5)] ="B2W" [dr(0.5)]  ="B2S" [ur(0.5)] ="B2E"
      ,"B2N":@(l,u)@/_/"B2W":@(d,l)@/_/"B2S":@{->}@(r,d)@/_/"B2E":@(u,r)@/_/"B2N"
      !~:{@{-}|@{>}}
      ,"A1N":@/^2pc/"A2N"^{\strut\displaystyle\Bbeta_1}_{\strut (2,0)}
      ,"B1N":@/^2pc/"B2N"^{\strut\displaystyle\Bbeta_2}_{\strut (2,-2)}
    }
  \end{equation*}
  \caption{Tadpole junctions $\Bbeta_1$ and $\Bbeta_2$.}
  \label{fig:TadpoleJunctions}
\end{figure}

The tadpole junction lattice $J_0=J_\Kodaira^\perp\subset J_\eff$ is generated
by tadpole junctions modulo null loops.  Recall that tadpole junctions terminate
at noncoalesced branes and/or at tadpole loops circling coalesced
branes.\footnote{Instead of the tadpole terminations, we can alternatively push
the tadpole loops through branch points to obtain terminations at coalesced
branes.  Note however, that in the full junction lattice, the charge vector of
the resulting junction diagram is ambiguous.  For example, a (2,0) string
terminating on a coalesced $(\BA^2)$ pair might have the junction charges of a
(1,0) string terminating on each $\BA$, or a $(2,0)$ string terminating on one
of them.  A tadpole termination makes it unambigous that we mean the former (by
a small deformation of the coalesced brane locations).  This ambiguity
disappears once we restrict to the tadpole junction lattice, where the only
charges permitted are those corresponding to the tadpole terminations.}  It is
meaningful to quotient by null loops since the null junctions \eqref{eq:ExNull}
can also be represented by tadpole junctions.  For example, $\Bdelta_2$ is shown
in Fig.~\ref{fig:NullTadpoleJunction}.  Tadpole junctions representing the
lattice vectors
\begin{equation}
  \begin{split}
    \Bbeta_1 &= \Bs_1+\Bs_2-\Bs_3-\Bs_4,\\
    \Bbeta_2 &= \Bs_5+\Bs_6-\Bs_9-\Bs_{10}, 
  \end{split}
\end{equation}
are shown in Fig.~\ref{fig:TadpoleJunctions}. Together, $\Bbeta_1$ and
$\Bbeta_2$ span the tadpole junction lattice of the collection
\eqref{eq:ExJKodaira}:\footnote{A similar junction joining the two $(\BD^2)$'s
  in Fig.~\ref{fig:TadpoleJunctions} differs from $\Bbeta_2$ by $\Bdelta_1$, so
  it is not linearly independent.}
\begin{equation}
  J_0 = \Bbeta_1\IZ\oplus\Bbeta_2\IZ = A_1^-\,{}^{\oplus2}.
\end{equation}
This lattice is isomorphic to the narrow Mordell-Weil lattice $\MW_0$ of the
corresponding singular elliptic dP$_9$ with six I$_2$ fibers.


\subsection{Weakly integral junction lattice and torsion sections} 
\label{Sec:WeaklyIntegral}

\begin{figure}[ht]
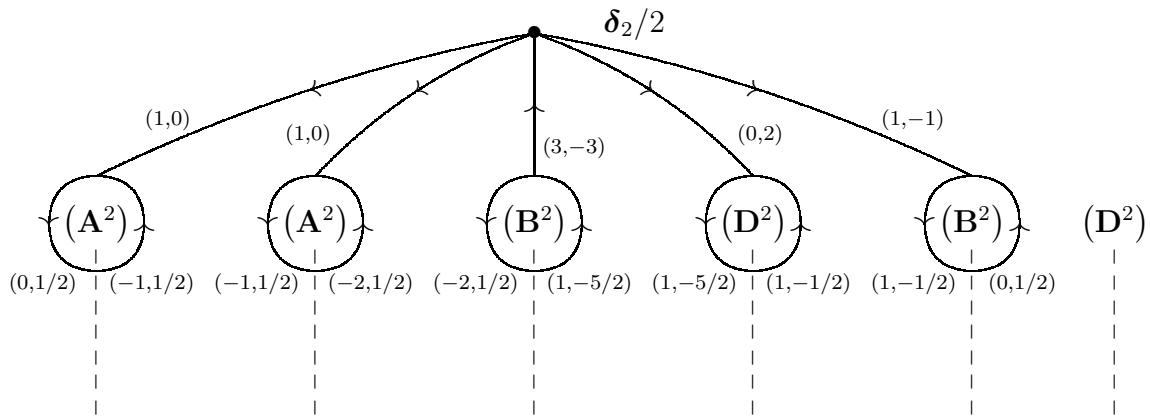

  \begin{equation*}
    \xygraph{
      !~:{@{-}}
      !~-{@{--}}
      {\bigl(\BA^2\bigr)}="A1"-[dd] [r(2.3)uu]
      {\bigl(\BA^2\bigr)}="A2"-[dd] [r(2.3)uu]
      {\bigl(\BB^2\bigr)}="B1"-[dd] [r(2.3)uu]
      {\bigl(\BD^2\bigr)}="D1"-[dd] [r(2.3)uu]
      {\bigl(\BB^2\bigr)}="B2"-[dd] [r(1.5)uu]
      {\bigl(\BD^2\bigr)}="D2"-[dd]
      ,"A1" [u(0.5)] ="A1N" [dl(0.5)] ="A1W" [dr(0.5)]  ="A1S" [ur(0.5)] ="A1E"
      ,"A1N":@{->}@(l,u)@/_/"A1W":@(d,l)@/_/"A1S"_(0.65){(0,1/2)\,}
            :@{->}@(r,d)@/_/"A1E"_(0.35){(-1,1/2)}:@(u,r)@/_/"A1N"
      ,"A2" [u(0.5)] ="A2N" [dl(0.5)] ="A2W" [dr(0.5)]  ="A2S" [ur(0.5)] ="A2E"
      ,"A2N":@{->}@(l,u)@/_/"A2W":@(d,l)@/_/"A2S"_(0.65){(-1,1/2)\,}
            :@{->}@(r,d)@/_/"A2E"_(0.35){(-2,1/2)}:@(u,r)@/_/"A2N"
      ,"B1" [u(0.5)] ="B1N" [dl(0.5)] ="B1W" [dr(0.5)]  ="B1S" [ur(0.5)] ="B1E"
      ,"B1N":@{->}@(l,u)@/_/"B1W":@(d,l)@/_/"B1S"_(0.65){(-2,1/2)\,}
            :@{->}@(r,d)@/_/"B1E"_(0.35){(1,-5/2)}:@(u,r)@/_/"B1N"
      ,"D1" [u(0.5)] ="D1N" [dl(0.5)] ="D1W" [dr(0.5)]  ="D1S" [ur(0.5)] ="D1E"
      ,"D1N":@{->}@(l,u)@/_/"D1W":@(d,l)@/_/"D1S"_(0.65){(1,-5/2)\,}
            :@{->}@(r,d)@/_/"D1E"_(0.35){(1,-1/2)}:@(u,r)@/_/"D1N"
      ,"B2" [u(0.5)] ="B2N" [dl(0.5)] ="B2W" [dr(0.5)]  ="B2S" [ur(0.5)] ="B2E"
      ,"B2N":@{->}@(l,u)@/_/"B2W":@(d,l)@/_/"B2S"_(0.65){(1,-1/2)\,}
            :@{->}@(r,d)@/_/"B2E"_(0.35){(0,1/2)}:@(u,r)@/_/"B2N"
      ,"B1" [uu]="Vertex" {\bullet}
      !~:{@{-}|@{>}}
      ,"Vertex":@/_/"A1N"_(0.8){(1,0)}
      ,"Vertex":@/_/"A2N"_(0.88){(1,0)}
      ,"B1N":"Vertex"_(0.2){(3,-3)}
      ,"Vertex":@/^/"D1N"^(0.88){(0,2)}
      ,"Vertex":@/^/"B2N"^(0.8){\;\;(1,-1)}^(0.2){\strut\displaystyle\Bdelta_2/2}
    }
  \end{equation*}
  \caption{The weakly integral null junction $\Bdelta_2/2$.}
  \label{fig:WeakNullTadpole}
\end{figure}

\begin{figure}[ht]
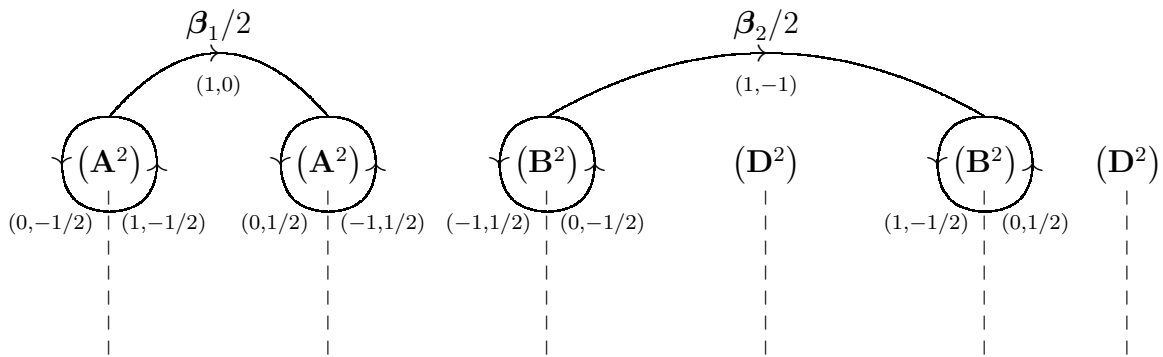

  \begin{equation*}
    \xygraph{
      !~:{@{-}}
      !~-{@{--}}
      {\bigl(\BA^2\bigr)}="A1"-[dd] [r(2.3)uu]
      {\bigl(\BA^2\bigr)}="A2"-[dd] [r(2.3)uu]
      {\bigl(\BB^2\bigr)}="B1"-[dd] [r(2.3)uu]
      {\bigl(\BD^2\bigr)}="D1"-[dd] [r(2.3)uu]
      {\bigl(\BB^2\bigr)}="B2"-[dd] [r(1.5)uu]
      {\bigl(\BD^2\bigr)}="D2"-[dd]
      ,"A1" [u(0.5)] ="A1N" [dl(0.5)] ="A1W" [dr(0.5)]  ="A1S" [ur(0.5)] ="A1E"
      ,"A1N":@{->}@(l,u)@/_/"A1W":@(d,l)@/_/"A1S"_(0.65){(0,-1/2)\,}
            :@{->}@(r,d)@/_/"A1E"_(0.35){(1,-1/2)\,}:@(u,r)@/_/"A1N"
      ,"A2" [u(0.5)] ="A2N" [dl(0.5)] ="A2W" [dr(0.5)]  ="A2S" [ur(0.5)] ="A2E"
      ,"A2N":@{->}@(l,u)@/_/"A2W":@(d,l)@/_/"A2S"_(0.65){(0,1/2)\,}
            :@{->}@(r,d)@/_/"A2E"_(0.35){(-1,1/2)\,}:@(u,r)@/_/"A2N"
      ,"B1" [u(0.5)] ="B1N" [dl(0.5)] ="B1W" [dr(0.5)]  ="B1S" [ur(0.5)] ="B1E"
      ,"B1N":@{->}@(l,u)@/_/"B1W":@(d,l)@/_/"B1S"_(0.65){(-1,1/2)\,}
            :@{->}@(r,d)@/_/"B1E"_(0.35){(0,-1/2)}:@(u,r)@/_/"B1N"
      ,"B2" [u(0.5)] ="B2N" [dl(0.5)] ="B2W" [dr(0.5)]  ="B2S" [ur(0.5)] ="B2E"
      ,"B2N":@{->}@(l,u)@/_/"B2W":@(d,l)@/_/"B2S"_(0.65){(1,-1/2)\,}
            :@{->}@(r,d)@/_/"B2E"_(0.35){(0,1/2)}:@(u,r)@/_/"B2N"
      !~:{@{-}|@{>}}
      ,"A1N":@/^2pc/"A2N"^{\strut\displaystyle\Bbeta_1/2}_{\strut (1,0)}
      ,"B1N":@/^2pc/"B2N"^{\strut\displaystyle\Bbeta_2/2}_{\strut (1,-1)}
    }
  \end{equation*}
  \caption{Weakly integral tadpole junctions $\Bbeta_1/2$ and $\Bbeta_2/2$.}
  \label{fig:WeakTadpole}
\end{figure}

In the presence of coalesced branes, there exists a natural extension of the
junction lattice $J$ known as the \emph{weakly integral junction lattice}
$J^\weak$ \cite{MWJunction}.  So far we have considered only physical string
junctions in which the string charges $(p,q)$ of each segment are integral.  A
weakly integral string junction is a string junction in which we relax the
integrality requirement on the strings in tadpole loops, requiring only that the
string charge entering or leaving a tadpole termination be integral.  For
example, consider a dP$_9$ surface.  From the junctions in
Figs.~\ref{fig:NullTadpoleJunction} and~\ref{fig:TadpoleJunctions}, we obtain
the weakly integral junctions $\Bdelta_2/2$, $\Bbeta_1/2$, and $\Bbeta_2/2$
shown in Figs.~\ref{fig:WeakNullTadpole} and~\ref{fig:WeakTadpole}.

Now return to the case of a general elliptic surface.  In addition to $J^\weak$,
we define the related junction lattices $J^\weak_\text{null}$, $J^\weak_\eff$,
$J^\weak_\Kodaira$ and $J^\weak_0$.  The definitions are straightforward analogs
of those in Sec.~\ref{Sec:MathInterp}, with the possible exception of loop
junction sublattice, which is defined to be the same in either case:
$J^\weak_\Loop = J_\Loop$.

The full Mordell-Weil group and its torsion subgroup are determined by the
various junction lattices as follows \cite{MWJunction}:
\begin{align}
  \MW &= J^\weak_0 && \text{(weakly integral tadpole junctions mod loops),}\\
  \MW_\tor &= J^\weak_\text{null}/J_\Loop
           && \text{(weakly integral null junctions mod loops)}.
\end{align}
When the N\'eron-Severi lattice\footnote{The N\'eron-Severi lattice is the
  lattice of algebraic divisors modulo algebraic equivalence.  For a compact
  elliptic surface $\CX$, the N\'eron-Severi lattice is torsion free; under the
  isomorphism $H_2(\CX,\IZ)\cong H^2(\CX,\IZ)$, it is the sublattice of
  $H_2(\CX,\IZ)$ identified with $H^2(\CX,\IZ)\cap H^{1,1}(\CX,\IC)$.}
$\NS(\CX)$ of the elliptic surface is unimodular, the Mordell-Weil lattice
$\MW/\MW_\tor$ is $MW_0^*=J_0^*$, the dual lattice of the narrow Mordell-Weil
lattice and tadpole junction lattice.  In particular, this is the case for
dP$_9$.

In the singular dP$_9$ example above, $J_0 = A_1\,{}^{\oplus2}$, so
$\MW/\MW_\tor = A_1^*\,{}^{\oplus2}$.  The weakly integral null lattice is
\begin{equation}
  J^\weak_\text{null} = (\Bdelta_1/2)\IZ\oplus (\Bdelta_2/2)\IZ,
\end{equation}
so $\MW_\tor = J^\weak_\text{null}/J_\Loop = \IZ_2\oplus\IZ_2$, generated by 
$\Bdelta_1/2$ and $\Bdelta_2/2$.

For a more complete and very readable exposition on the relation between string
junctions and the Mordell-Weil lattice, the reader is referred to
Ref.~\cite{MWJunction}, in which the Oguiso-Shioda classification of the
Mordell-Weil lattices of dP$_9$ is reproduced entirely in the junction
framework.\footnote{Indeed, Ref.~\cite{MWJunction} even caught a two errors in
  Oguiso and Shioda's list of Mordell-Weil groups of dP$_9$
  \cite{OguisoShioda}.}  As emphasized in Ref.~\cite{MWJunction}, $\MW_\tor$ has
physical consequences.  Whereas the gauge \emph{algebra} on a collection of
coalescing 7-branes is determined by its root lattice, and hence by
$J_\Kodaira$, the group $\MW_\tor$ determines $\pi_1$ of the global gauge
\emph{group}.


\section{Monodromy and junction description of CY duals of $T^6/\IZ_2$}

\subsection{The duality map}
\label{Sec:DualityMap}

In Ref.~\cite{CYDuals}, it was shown that type IIB $T^6/\IZ_2$ orientifold with
a choice of flux preserving $\CN=2$ supersymmetry is dual to a class of purely
geometrical type IIA Calabi-Yau compactifications with no flux.  The dual
Calabi-Yau manifolds $\CX_{m,n}$ are abelian surface fibrations over $\IP^1$.
They bear a number of similarities to the elliptic fibrations over $\IP^1$
described in Sec.~2.  In this section, we will provide a similar monodromy and
junction based description of the topology of $\CX_{m,n}$ and its geometry of
curves.

To see why the IIA dual geometry is a $T^4$ fibration, first recall that in the
absence of flux, we have the $\CN=4$ duality,
\begin{equation}\label{eq:DualityNoFlux}
  \begin{matrix} 
    T^6/\IZ_2 \text{ orientifold} & \leftrightarrow & \text{IIA on K3}\times
    T^2\hfill\\ & & (\text{where K3 $= T^2$ fibration over $P^1$}).\hfill
  \end{matrix}
\end{equation}
Here, note that K3$\times T^2$ can be viewed as $T^4$ fibration over $\IP^1$,
where a $T^2\subset T^4$ trivially factorizes.  This duality can be understood
in several ways.  Two are as follows.

\begin{enumerate}

\item \emph{Via heterotic/IIA duality:} The IIB $T^6/\IZ_2$ orientifold
  $\leftrightarrow$ type I on $T^6$ (by T-duality) $\leftrightarrow$ heterotic
  $SO(32)$ on $T^6$ (by S-duality) $\leftrightarrow$ heterotic $E_8\times E_8$
  on $T^6$ (by T-duality) $\leftrightarrow$ IIA on K3$\times T^2$ (by
  heterotic/IIA duality).

\item \emph{Via \Mtheory:} First T-dualize $T^6/\IZ_2$ on a $T^3$ to obtain the
  IIA $T^3/\IZ_2\times T^3$ D6/O6 orientifold.  This is dual to type IIA on
  K3$\times T^2$ via a circle swap: The IIA orientifold lifts to \Mtheory\ on
  K3$\times T^3$, where the K$3\cong T^4/\IZ_2$ comes from the lift of the
  $T^3/\IZ_2$ factor.  Then compactifying on an $S^1$ in the $T^3$ factor gives
  IIA on K3$\times T^2$.
\end{enumerate}
The modification of the duality \eqref{eq:DualityNoFlux} due to $\CN=2$ flux is
as follows:
\begin{equation}\label{eq:DualityWithFlux}
  \begin{matrix} 
    T^6/\IZ_2 \text{ orientifold} & \leftrightarrow & \text{IIA on a Calabi-Yau }
    \CX_{m,n}\hfill\\ & & (\text{where $\CX_{m,n} = T^4$ fibration over $P^1$}).\hfill
  \end{matrix}
\end{equation}
The flux dualizes to twists of the topology that: (i) mix the previous $T^2$
factor with the $T^2$ fiber of K3, and (ii) require a reduction in the number of
exceptional divisors from the 16 that would be associated with K3 to a smaller
number $M<16$.  On the $T^6/\IZ_2$ side of the duality, the integers $m$ and $n$
parametrize the choice of flux
\begin{equation}
  \begin{split}
    F_\RR &= 2m(dx^1\w dy^1 + dx^2\w dy^2)\w dy^3,\\
    H_\NS &= 2n(dx^1\w dy^1 + dx^2\w dy^2)\w dx^3,
  \end{split}
\end{equation}
and $M$ is the number of D3-branes.

In Ref.~\cite{CYDuals}, this duality was studied by mapping the family of
classical 10D type IIB supergravity solutions through the duality chain 2 above.
The resulting description of $\CX_{m,n}$ includes an explicit metric that is
valid to leading order in the relative K\"ahler modulus $h/s$ (fiber size/base
size).  The harmonic forms in the metric can also be given explicitly.  For
$h/s\ll1$ the metric is a good approximation everywhere except near the singular
loci of a subset of the singular fibers (the $\BB_i,\BC_i$ fibers of
Sec.~\ref{Sec:CYmonodromy}).


\subsection{Known properties of type IIA Calabi-Yau duals $\CX_{m,n}$}

 A summary of the results obtained in Ref.~\cite{CYDuals} is as follows:

  \begin{enumerate}

    \item $\CX_{m,n}$ is an abelian surface ($T^4$) fibration over $P^1$, with
      $8+M$ singular fibers, where $M$ is number of D3-branes in $T^6/\IZ_2$.

    \item The Hodge numbers of $\CX_{m,n}$ are $h^{11}=h^{21} = M+2$, where $m$,
      $n$, and $M$ are constrained by 
      \begin{equation}\label{eq:RRtadpole}
        M + 4mn = 16.
        \end{equation}
      This is the D3 charge cancellation condition $N_{\rm D3} + \int H\w F =
      N_{\rm O3}$ on $T^6/\IZ_2$ with $N_{\rm D3} = M$ and $N_{\rm O3} = 16$.

    \item The generic Mordell-Weil lattice of sections (mod torsion) of
      $\CX_{m,n}$ is $D_M$.  Here, generic means that all fibers are
      topologically I$_1\times T^2$.  This was not mentioned explicitly in
      Ref.~\cite{CYDuals}, but follows from the fact that $M$ D-branes plus an
      O-plane can coalesce to give $SO(2N)$ enhanced gauge symmetry in
      $T^6/\IZ_2$.  It also follows from consideration of D3 instantons in
      $T^6/\IZ_2$~\cite{DInstanton}.

    \item The fundamental group and isometry group of $\CX_{m,n}$ are
      \begin{equation}\label{eq:DiscreteSym}
        \pi_1 = \IZ_n\times\IZ_n\quad\text{and}\quad\MW_\tor=\IZ_m\times\IZ_m,
      \end{equation}
      corresponding to a disrete KK gauge symmetry and discrete winding gauge
      symmetry, respectively.  This follows from the fact that the flux
      parametrizes a gauging of the low energy $\CN=4$ supergravity theory of
      $T^6/\IZ_2$ (i.e., the charges coupling scalars to vectors).  For
      nonminimal flux $(m,n)\ne(1,1)$, the resulting superHiggs mechanism down
      to $\CN=2$ only partially breaks four of the $U(1)$s of $T^6/\IZ_2$,
      leaving the discrete gauge symmetry~\eqref{eq:DiscreteSym}.

    \item The polarization of the abelian fiber is $(\mbar,\nbar) =
      (m,n)/\gcd(m,n)$.  This means that the K\"ahler form on the fiber is
      proportional to a Hodge form
      \begin{equation}\label{eq:Hodge}
        \o= \nbar dy^1\w dy^2 + \mbar dy^3\w dy^4,
        \end{equation}
      a positive integer form that can be used to define a projective
      embedding.

     \item The interchange $m\leftrightarrow n$ corresponds to T-duality of the
       $T^4$ fiber.  For $m=1$, this interchange can be achieved as a quotient
       by the isometry group,\footnote{In the $m=4$ case, one can also quotient
         in two steps: $\CX_{2,2} = \CX_{4,1}/\bigl(\IZ_2\times\IZ_2\bigr)$,
         $\CX_{1,4} = \CX_{2,2}/\bigl(\IZ_2\times\IZ_2\bigr)$, where the
         $\IZ_2\times\IZ_2$ is a subgroup of $\IZ_4\times\IZ_4$ in the first
         step, and is the quotient group in the second step.}
     \begin{equation}
       \CX_{1,m}=\CX_{m,1}/\bigl(\IZ_m\times\IZ_m\bigr).
     \end{equation}

     \item In a convenient basis, the nonzero intersection numbers are 
        \begin{equation}\label{eq:NonzeroInts}
          H^2\cdot A = 2\mbar\nbar,\quad H\cdot\CE_I\cdot\CE_J = -\bar m\delta_{IJ},
          \end{equation}
        where $A$ is the abelian fiber.  These were deduced from the explicit
        harmonic forms in the approximate Calabi-Yau metric.

     \item The only nonzero intersection with the second Chern class of
        $\CX_{m,n}$ is
        \begin{equation}\label{eq:2ndChern}
          H\cdot c_2 = 8+M.
          \end{equation}
        This follows from the $F_1$ topological amplitude of $T^6/\IZ_2$, which
        to leading order is determined by the Green-Schwarz mechanism.

  \end{enumerate}

Properties 1, 5, and 6 were deduced using the existence of an approximate metric
on the type IIA Calabi-Yau manifold $\CX_{m,n}$ that is \emph{exactly dual} to
the leading order supergravity description of the type IIB $T^6/\IZ_2$
orientifold:

\begin{itemize}
   \item[9.] The approximate metric is twisted product of a Gibbons-Hawking metric and
    a $T^2$ metric,
   \begin{equation}\label{eq:ApproxMetric}
        \begin{split}
        ds^2_\text{CY} 
        &=  Z\Bigl(\frac{v_B}{\im\t} \bigl|dy^5+\t dy^6\bigr|^2 
          + {R_2}^2 \bigl(dy^2\bigr)^2\Bigr)
          + Z^{-1}{R_1}^2 \bigl(dy^1+A^1\bigr)^2\\
          &\quad + \frac{v_F}{\im\t}\bigl|\eta^3+\t\eta^4\bigr|^2,
          \qquad R_1 R_2 = (n/m) v_F,  
        \end{split}
      \end{equation}
     modulo $\IZ_2(y^{1,2,5,6})$, where $v_F$ and $v_B$ are the fiber and base
     K\"ahler moduli, respectively.  The factor $Z$ satisfies a Poisson equation
     on $T^3_{\{2,3,4\}}$.\footnote{It is the same as the warp factor of the
       $T^6/\IZ_2$ orientifold, averaged over the three T-dualized directions
       transverse to $T^3_{\{2,3,4\}}$.}

   \item[$-$] For $m,n = 0$, the first line (the Gibbons-Hawking part) approximates a
     K3 metric, and the second line is a $T^2$ metric.

   \item[$-$] For $m,n\ne0$, the two pieces are twisted:
     \begin{equation}
       \begin{split}\label{eq:dconnection}
         dA^1 &= R_1\, \star_3dZ - 2m \bigl(\eta^3\w dy^6 - \eta^4\w dy^5\bigr),\\
         d\eta^3 &= 2n dy^2\w dy^5,\quad
         d\eta^4  = 2n dy^2\w dy^6.
       \end{split}
     \end{equation}
    Since the right hand side of Eq.~\eqref{eq:dconnection} vanishes at fixed
    $y^5,y^6$, we can interpret the metric as that of a $T^4_{\{1,2,3,4\}}$
    fibration over $T^2_{\{5,6\}}/\IZ_2\cong \IP^1$.
  \end{itemize}


\subsection{Monodromy matrices for the abelian surface fibrations}
\label{Sec:CYmonodromy}

In Sec.~2, we saw that the topology of an elliptic K3 or dP$_9$ was determined
by a branch-cut-ordered set of points on $\IP^1$ and the corresponding
$SL(2,\IZ)$ monodromy matrices that determine how the coordinates and 1-forms on
$T^2$ are to be identified across branch cuts.  In the type IIB description, the
choice is interpreted as that of a collection of $(p,q)$ 7-branes.

Likewise, the topology of $\CX_{m,n}$ can be \emph{defined} by giving an ordered
set of points on the $\IP^1$ base and corresponding $SL(4,\IZ)$
monodromies\footnote{A collection of $SL(4,\IZ)$ monodromies defines a $T^4$
  fibration.  This $T^4$ fibration is an abelian surface fibration, if there
  also exists a monodromy invariant Hodge form $\omega$.  Alternatively, if
  $\omega$ is specified beforehand, the monodromy matrices must lie in the
  subgroup $Sp(\omega,\IZ)\subset SL(4,\IZ)$ that preserves $\omega$.}  acting
on $T^4$.  And, we can again determine this information explicitly via the
duality to $T^6/\IZ_2$. In App.~\ref{App:Xmonodromy}, it is shown that the $M$
D3-branes (plus orientifold planes) of $T^6/\IZ_2$ map to a collection
\begin{equation}\label{eq:CYcollection}
    \BA^M \BB_1 \BC_1 \BB_2 \BC_2 \BB_3 \BC_3 \BB_4 \BC_4
\end{equation}
of singular $T^4$ fibers of $\CX_{m,n}$.  The explicit monodromy matrices are
\begin{equation}\label{eq:XMonodI}
    K_\BA = \begin{pmatrix}
    1 & -1 & 0 & 0 \\
    0 & 1 & 0 & 0 \\
    0 & 0 & 1 & 0 \\
    0 & 0 & 0 & 1
  \end{pmatrix},
\end{equation}
from the $M$ D-branes in $T^6/\IZ_2$, and  
\begin{equation}\label{eq:XMonodII}
  \begin{split}
    K_{\BC_1} &= \begin{pmatrix}
      2 & -1 & 0 & m \\
      1 & 0 & 0 & m \\
      -n & n & 1 & -m n \\
      0 & 0 & 0 & 1
    \end{pmatrix},\\
    \noalign{\vskip1ex}
    K_{\BC_2} &= \begin{pmatrix}
      2 & -1 & 0 & 0 \\
      1 & 0 & 0 & 0 \\
      0 & 0 & 1 & 0 \\
      0 & 0 & 0 & 1
    \end{pmatrix},\\
    \noalign{\vskip1ex}
    K_{\BC_3} &= \begin{pmatrix}
      2 & -1 & -m & 0 \\
      1 & 0 & -m & 0 \\
      0 & 0 & 1 & 0 \\
      -n & n & m n & 1
    \end{pmatrix},\\
    \noalign{\vskip1ex}
    K_{\BC_4} &= \begin{pmatrix}
      2 & -1 & -m & m \\
      1 & 0 & -m & m \\
      -n & n & 1+m n & -m n \\
      -n & n & m n & 1-m n
    \end{pmatrix},
  \end{split}
  \quad
  \begin{split}
    K_{\BB_1} &= \begin{pmatrix}
      0 & -1 & 0 & -m \\
      1 & 2 & 0 & m \\
      -n & -n & 1 & -m n \\
      0 & 0 & 0 & 1
    \end{pmatrix},\\
    \noalign{\vskip1ex}
    K_{\BB_2} &= \begin{pmatrix}
      0 & -1 & 0 & 0 \\
      1 & 2 & 0 & 0 \\
      0 & 0 & 1 & 0 \\
      0 & 0 & 0 & 1
    \end{pmatrix},\\
    \noalign{\vskip1ex}
    K_{\BB_3} &= \begin{pmatrix}
      0 & -1 & m & 0 \\
      1 & 2 & -m & 0 \\
      0 & 0 & 1 & 0 \\
      -n & -n & m n & 1
    \end{pmatrix},\\
    \noalign{\vskip1ex}
    K_{\BB_4} &= \begin{pmatrix}
      0 & -1 & m & -m \\
      1 & 2 & -m & m \\
      -n & -n & 1+m n & -m n \\
      -n & -n & m n & 1-m n
    \end{pmatrix}.
  \end{split}
\end{equation}
Note that
\begin{equation}
  K_\BA = (\text{previous }K_\BA)\oplus(\text{identity})\text{ on } T^2\times T^2,
\phantom{\ +\ m,n\text{ twists.}}
\end{equation}
but
\begin{equation}
  \begin{split}
    K_{\BB_i} &= (\text{previous }K_\BB)\oplus(\text{identity})\text{ on } T^2\times T^2
    {\ +\ m,n\text{ twists,}}\\
    K_{\BC_i} &= (\text{previous }K_\BC)\oplus(\text{identity})\text{ on } T^2\times T^2
    {\ +\ m,n\text{ twists,}}
  \end{split}
\end{equation}
where the $m,n$ dependent twists in $K_{\BB_i},K_{\BC_i}$ mix $T^2_{y^1y^2}$ and
$T^2_{y^3y^4}$ and differ for $i=1,2,3,4$\@.  These monodromies uniquely
determine the topology of $\CX_{m,n}$, and preserve the Hodge
form~\eqref{eq:Hodge}.

The topology of a singular $\BA$ fiber consists of a smooth $T^2_{y^3y^4}$ times
an I$_1$ type degeneration of $T^2_{y^1y^2}$ in which the $y^1$-circle shrinks
to zero size at a point on the $y^2$-circle.\footnote{While the singular fibers
  have I$_1\times T^2$ topology, the complex structure need not respect this
  factorization.  A description of the complex structure of the singular fibers
  is given in Sec.~\ref{Sec:SurfaceS}.}  The singular locus is the smooth
$T^2_{y^3y^4}$ at the location of the singularity on $T^2_{y^1y^2}$.  The
$\BB_i$ and $\BC_i$ monodromies are related to $K_\BA$ by similarity
transformations $T K_\BA T^{-1}$, as described in App.~\ref{App:Xmonodromy}\@.
So, the singular $\BB_i$ and $\BC_i$ fibers are of the same type, except that
decomposition into singular and smooth $T^2$ differs in each case.

On the $T^4$ fibers, the analog of the $(p,q)$ 1-cycles of $H_1(T^2)$ in Sec.~2
is the group of $(p,q,r,s)$ 1-cycles of $H_1(T^4)$.  For each $\BA$, $\BB_i$ and
$\BC_i$ singular $T^4$ fiber, Table~\ref{tbl:VanishingCycle} lists the vanishing
1-cycle, the location of the singular locus, and a pair of nonvanishing 1-cycles
spanning the singular locus (which is a smooth $T^2$).  The vanishing 1-cycle
and spanning cycles of the singular locus are invariant under the monodromy
action associated to the singular fiber.  In Table~\ref{tbl:VanishingCycle}, the
spanning cycles of the smooth $T^2$ are defined only modulo the vanishing cycle,
and represent one choice out of many possible bases.  Note that in contrast to
the $T^2$ fibered case, specifying the vanishing cycle does not uniquely
determine the $SL(4,\IZ)$ monodromy.  The singular locus must also be specified.
\begin{table}[ht]
  \centering
  \begin{tabular}[ht]{cllll}
  & & & Singular locus (smooth $T^2$)\\
  Singular fiber & Vanishing cycle & Location of singular locus & \qquad spanning cycles\\
    $\BA$   & $(1,0,0,0)$   & $y^2=c$          & $(0,0,1,0)$,\quad $(0,0,0,1)$\\
    $\BB_1$ & $(1,-1,n,0)$  & $y^1+y^2+my^4=c$ & $(0,0,1,0)$,\quad $(-m,0,0,1)$\\
    $\BC_1$ & $(1,1,-n,0)$  & $y^1-y^2+my^4=c$ & $(0,0,1,0)$,\quad $(-m,0,0,1)$\\
    $\BB_2$ & $(1,-1,0,0)$  & $y^1+y^2=c$      & $(0,0,1,0)$,\quad $(0,0,0,1)$\\
    $\BC_2$ & $(1,1,0,0)$   & $y^1-y^2=c$      & $(0,0,1,0)$,\quad $(0,0,0,1)$\\
    $\BB_3$ & $(1,-1,0,n)$  & $y^1+y^2-my^3=c$ & $(m,0,1,0)$,\quad $(0,0,0,1)$\\
    $\BC_3$ & $(1,1,0,-n)$  & $y^1-y^2-my^3=c$ & $(m,0,1,0)$,\quad $(0,0,0,1)$\\
    $\BB_4$ & $(1,-1,n,n)$  & $y^1+y^2-my^3+my^4=c$ & $(m,0,1,0)$,\quad $(-m,0,0,1)$\\
    $\BC_4$ & $(1,1,-n,-n)$ & $y^1-y^2-my^3+my^4=c$ & $(m,0,1,0)$,\quad $(-m,0,0,1)$
  \end{tabular}
  \caption{Structure of the singular $\BA$, $\BB_i$, and $\BC_i$ fibers of the
    abelian surface fibration $\CX_{m,n}$.}
  \label{tbl:VanishingCycle}
\end{table}


\subsubsection{Fundamental group}

The vanishing cycles in Table~\ref{tbl:VanishingCycle} are trivial in
$H_1(\CX_{m,n},\IZ)$.  By taking linear combinations, we deduce that the cycles
$(1,0,0,0)$, $(0,1,0,0)$, $(0,0,n,0)$ and $(0,0,0,n)\in H_1(\CX_{m,n},\IZ)$ are
trivial, and generate all trivial cycles.  Thus, $(0,0,1,0)$ and $(0,0,0,1)$
generate $\IZ_n$ torsion cycles, and
\begin{equation}
  H_1(\CX_{m,n},\IZ) = \IZ_n\times\IZ_n.
\end{equation}
Provided that $\pi_1(\CX_{m,n})$ is abelian,\footnote{In general, $H_1(\CX,\IZ)$
  is the abelianization of $\pi_1(\CX)$, i.e., $\pi_1$ modulo its commutator
  subgroup.} this implies that $\pi_1 = \IZ_n\times\IZ_n$, in agreement with the
result \eqref{eq:DiscreteSym} obtained from the low energy effective field
theory of $T^6/\IZ_2$.  Indeed, the inclusion $A \hookrightarrow \CX_{m,n}$
induces a surjective map from $\pi_1(A)$ to $\pi_1(\CX_{m,n})$; that is, every
nontrivial element of $\pi_1(\CX_{m,n})$ can be deformed to lie entirely in the
abelian fiber over a single point of the base.  So, $\pi_1(\CX_{m,n})$ is
abelian.


\subsubsection{Calabi-Yau dual interpretation of $T^6/\IZ_2$ RR tadpole}

Since the base of $\CX_{m,n}$ is $\IP^1$, a loop that encloses all singular
fibers is contractible (to the point at infinity).  Therefore, as in Sec.~2, the
total monodromy must be unity.  This gives
\begin{equation}
  \begin{split}
    1 &= K_\mathrm{total}\\
    \noalign{\vskip0.5ex}
      &= K_{\BC_4}K_{\BB_4} K_{\BC_3}K_{\BB_3} K_{\BC_2}K_{\BB_2}
         K_{\BC_1}K_{\BB_1} K_\BA{}^M\\
    \noalign{\vskip0.5ex}
    &= \begin{pmatrix}
      1 & 0 &\ 0 & 0\\
        & 1 & -Q & 0\\
        &   &\ 1 & 0\\
        &   &    & 1
    \end{pmatrix},
      \quad\text{where}\quad Q = M-16+4mn,
  \end{split}
\end{equation}
so that $Q=0$.  The topological constraint $K_\mathrm{total} =1$ reproduces the
$T^6/\IZ_2$ D3 charge cancellation condition~\eqref{eq:RRtadpole}.


\subsec{Mordell-Weil lattice from junction lattice}\label{Sec:MWfromJunction}

For an abelian surface fibration, we define the junction lattice $J$ and related
lattices $J_\Loop$, $J_\eff$, $J_\Kodaira$, and $J_0$ exactly
as in Sec.~2, except that we now consider graphs on $\IP^1$ in which each
oriented string: (i)~is labeled by four charges $(p,q,r,s)$, instead of two, and
(ii)~can terminate either at a vertex or at the location of a singular
$\BX_{[p,q,r,s]}$ fiber in which a $(p,q,r,s)$ 1-cycle shrinks.

To define the equivalence classes of junctions, we let $\Bs_i$ denote an outward
oriented $(p_i,q_i,r_i,s_i)$ string emanating from $\BX_{[p_i,q_i,r_i,s_i]}$.
Then, to each string junction in standard (tree) presentation, we associate a
charge vector
  \begin{equation}
    Q = \sum_i Q_i \Bs_i,
  \end{equation}
where $Q_i\in\IZ$ is the net number of strings leaving (minus entering) the
location of the $i$th singular fiber, $\BX_{[p_i,q_i,r_i,s_i]}$.  In this case,
we treat the strings as strictly mathematical objects, in contrast to the
physical IIB strings of Sec.~2.\footnote{What carries over directly is the
  interpretation of a $(p,q,r,s)$ string as the projection to $\IP^1$ of a
  2-cycle in the abelian fibration, whose inverse image at each point
  is a $(p,q,r,s)$ 1-cycle in the abelian fiber.  So, from the
  \Mtheory\ perspective, these string graphs have an analogous physical
  interpretation to the $(p,q)$ string of type IIB:  they are M2-branes wrapped
  on $(p,q,r,s)$ cycles of the fiber of $\CX_{m,n}$.}

To apply the inner product of Sec.~2, we use the fact that an abelian surface
fibration admits a projective embedding, with corresponding Hodge form $\o$.
The Poincar\'e dual divisor class can be represented as the sum of positive
integer multiples of two $T^2$s.\footnote{In fact, one of the two integers must
  be unity (cf.~App.~\ref{App:AbelVar}.)  This is indeed the case for the
  possible values of $(\mbar,\nbar)$ of $\CX_{m,n}$.}  So, we can first
intersect with Hodge class and then use the inner
product~\eqref{eq:innerproduct} on each $T^2$.

For the abelian surfaces fibration $\CX_{m,n}$, the Hodge form $\o$ was given in
Eq.~\eqref{eq:Hodge}.  As a check, it can be readily verified that $\o$ is
invariant under the monodromies~\eqref{eq:XMonodI} and \eqref{eq:XMonodII}.  In
this case, the Poincar\'e dual divisor class can be represented by $\mbar
T^2_{y^1y^2} + \nbar T^2_{y^3y^4}$, and the charges $(p_i,q_i,r_i,s_i)$ have a
corresponding decomposition $(p_i,q_i)\oplus (r_i,s_i)$.  We define the inner
product to be
\begin{equation}\label{eq:AbelInnerProd}
  \begin{split}
  (s_i,s_i) &= -1,\\
  (s_i,s_j) &= (s_j,s_i) = \frac{\mbar}{2} \bigl(p_i q_j - p_j q_i\bigr)
                         + \frac{\nbar}{2} \bigl(r_i s_j - r_j s_i\bigr),
                         \quad i < j.
  \end{split}
\end{equation}
For the vanishing cycles $(p_i,q_i,r_i,s_i)$ listed in
Table~\ref{tbl:VanishingCycle}, the nonvanishing contribution to this inner
product comes entirely\footnote{This statement, while true, is not obvious.  For
  example, if the base were noncompact, then the inner product between a
  $(1,-1,n,0)$ string leaving a $\BB_1$ point and a string leaving a $\BB_3$,
  $\BC_3$, $\BB_4$ or $\BC_4$ point would have a contribution from the second
  term in Eq.~\eqref{eq:AbelInnerProd}.  However, it can be shown that the inner
  product between proper string junctions with no asymptotic $(p,q)$ charge
  would not.  In the compact case, the only junctions are the proper junctions.}
from the first $T^2$:
\begin{equation}\label{eq:AbelInnerProdII}
  \begin{split}
  (s_i,s_i) &= -1,\\
  (s_i,s_j) &= (s_j,s_i) = \frac{\mbar}{2} \bigl(p_i q_j - p_j q_i\bigr),
                         \quad i < j.
  \end{split}
\end{equation}
However, the $(r_i,s_i)$ data still show up in the junction lattice via the
$(p,q,r,s)$ charge conservation conditions at vertices and the terminination
conditions at the locations of singular fibers.

For the collection \eqref{eq:CYcollection}, with no coalesced 7-branes, the
resulting junction lattice is
\begin{equation}\label{eq:XmnJ}
  J= D^{\mbar -}_M\oplus J_\text{null},
  \quad\text{where}\quad J_\text{null} =
  \Bdelta_1\IZ\oplus\Bdelta_2\IZ\oplus\Bdelta_3\IZ\oplus\Bdelta_4\IZ.
\end{equation}
In the first term, the superscript indicates that the (positive definite) inner
product is $-\mbar$ times the inner product of the root lattice of $D_M$.  The
$D_M$ is generated by the proper junction lattice of $\BA^M\BB_i\BC_i$, for any
one choice of $i$, just as in the elliptically fibered case
(cf.~Refs.~\cite{DeWolfeI,DeWolfeII}).  However, unlike the case of an elliptic
K3 or dP$_9$ in Sec.~2, we do not obtain an $E_{M+1}$ lattice from the proper
junctions of $\BA^M\BB_i\BC_i\BC_j$.  The fact that $\BC_i$ and $\BC_j$ have
different $(p,q,r,s)$ for $i\ne j$ means that the analog of the $E_{M+1}$
enhancing root does not exist for this abelian fibration.

The lattice $J_\text{null}$ in Eq.~\eqref{eq:XmnJ} is the null sublattice of the
junction lattice.  A basis of generators is
\begin{align}\label{eq:XmnNullGenerators}
  \Bdelta_1 &= (0^M;\; -1,-1;\;  1,1;\; -1,-1;\; 1,1),\\
  \Bdelta_2 &= ((-1)^M;\; M-1,\, M-3;\; 5-M,\, 7-M;\; M/2-1,\, M/2-3;\; -3,-1),\\
  \Bdelta_3 &= (0^M;\; 0,0;\; 0,0;\; 1,1,\; -1,-1),\\
  \Bdelta_4 &= (0^M;\; -1,-1;\; 2,2;\; -2,-2;\; 1,1).
\end{align}
In App.~\ref{App:NullLoop}, we compute the lattice $J_\Loop$ generated by
null loop junctions.  We find
\begin{equation}\label{eq:XmnLoopLattice}
  J_\Loop = \Bdelta_1\IZ\oplus\Bdelta_2\IZ
  \oplus m\Bdelta_3\IZ\oplus m\Bdelta_4\IZ,
\end{equation}
corresponding to loops with $(p,q,r,s) = (1,0,0,0)$, $(0,1,0,0)$, $(0,0,1,0)$,
and $(0,0,0,1)$, respectively.  Therefore, the effective junction lattice is
\begin{equation}
  J_\eff = J/J_\Loop
  = D^{\mbar -}_M\oplus\Bdelta_3\IZ_m\oplus\Bdelta_4\IZ_m.
\end{equation}
Finally, $J_\Kodaira$ is trivial, since there are no coalesced branes.
Therefore, $J_0 = J_\eff$, and there is no distinction between integral and
weakly integral junctions.  We have
\begin{equation}
  \begin{split}
    \MW = \MW_0 &= J_0 = D^{\mbar -}_M\oplus\IZ_{\mbar}\oplus\IZ_{\mbar},\\
    \MW_\tor &= J_\text{null}/J_\Loop = \IZ_{\mbar}\oplus\IZ_{\mbar},
  \end{split}
\end{equation}
exactly as predicted via effective field theory considerations in
Ref.~\cite{CYDuals}.

Note that the narrow Mordell-Weil lattice $\MW_0$ has torsion.  This
distinguishes abelian surface fibrations from elliptic fibrations over $\IP^1$,
where such torsion cannot occur.  (In terms of the junction lattices,
$J_\text{null}=J_\Loop$ for elliptic surfaces.)

In fact, it is easy to identify the torsion sections explicitly.  The
sections
\begin{equation}
  \begin{split}
    \Bdelta_3 &\cong \{(y^1,y^2,y^3,y^4)=(0,0,\tfrac{1}{m},0)\text{ \ in }\CX_{m,n}\}
    \quad\text{(mod loops $\Bdelta_1,\Bdelta_2,m\Bdelta_3,m\Bdelta_4$),}\\
    \Bdelta_4 &\cong \{(y^1,y^2,y^3,y^4)=(0,0,0,\tfrac{1}{m})\text{ \ in }\CX_{m,n}\}
    \quad\text{(mod loops $\Bdelta_1,\Bdelta_2,m\Bdelta_3,m\Bdelta_4$),}\\
  \end{split}
\end{equation}
are invariant under the monodromy actions \eqref{eq:XMonodI} and
\eqref{eq:XMonodII} up to the identifications $y^i\cong y^i+1$.


\subsection{Examples with coalesced fibers}
\label{Sec:ExCoalescedFibs}

We now consider three examples of collections with coalesced fibers.  In the
first example, we assume that $M=16-4mn\ge4$ and obtain an enhancement of
$\MW_\tor(\CX_{m,n})$ from $\IZ_m\oplus\IZ_m$ to $\IZ_{2m}\oplus\IZ_m$.  In the
next two examples, we restrict to the principally polarized case $(m,n) =
(1,1)$.  We identify collections leading to $\IZ_2{}^{\oplus 4}$ and
$\IZ_4{}^{\vphantom{\oplus 2}}\oplus\IZ_2{}^{\oplus 2}$ torsion, respectively.
In Sec.~\ref{Sec:Connecting}, we describe how new abelian surface fibered
Calabi-Yau manifolds can be obtained by quotienting by these isometry groups.


\subsubsection{$\IZ_{2m}\oplus\IZ_m$ torsion, $m=1,2,3$}
\label{Sec:ZtmZmTorsion}

From the braiding operations
discussed in App.~\ref{App:Braiding}, we have
\begin{equation}\label{eq:ExIcollection}
  \begin{split}
    \BA^{M}\ \BB_1\BC_1\ \BB_2\BC_2\ \BB_3\BC_3\ \BB_4\BC_4
    &\cong
    \BA^{M-4}\ \BB_1\BC_1\ \BB_2\BC_2\ \BB_3\BC_3\BA^2\ \BB_4\BC_4\BA^2\\
    &\cong
    \BA^{M-4}\ \BB_1\BC_1\ \BB_2\BC_2\ \BB_3\BA^2\BB_3\ \BB_4\BA^2\BB_4\\
    &\cong
    \BA^{M-4}\ \BB_1\BC_1\ \BB_2\BC_2\ \BB_3{}^2\BD_3{}^2\ \BB_4{}^2\BD_4{}^2,
  \end{split}
\end{equation}
where the $\BD_i$ are defined in App.~\ref{App:Braiding} and have vanishing
cycles of the form $(0,1,*,*)$.
In the basis corresponding to the last collection of
Eq.~\eqref{eq:ExIcollection}, the generators \eqref{eq:XmnNullGenerators} of
$J_\text{null}$ become
\begin{equation}
  \begin{split}
    \Bdelta_1 &= \bigl(0^{M-4};\ -1,-1;\ 1,1;\ -1,-1,-1,-1;\ 1,1,1,1\bigr),\\
    \Bdelta_2 &= \bigl(1^{M-4};\ M-5,M-7;\ 9-M,11-M;\ (5-M/2)^2,(M/2-6)^2;\ (-1)^2,0^2\bigr),\\
    \Bdelta_3 &= \bigl(0^{M-4};\ 0,0;\ 0,0;\ m,m,m,m;\ -m,-m,-m,-m\bigr),\\
    \Bdelta_4 &= \bigl(0^{M-4};\ -m,-m;\ 2m,2m;\ -2m,-2m,-2m,-2m;\ m,m,m,m\bigr).
  \end{split}
\end{equation}
The loop junction lattice $J_\Loop$ is given by Eq.~\eqref{eq:XmnLoopLattice}.

Now, suppose that we coalesce 7-branes pairwise to obtain the collection
\begin{equation}\label{eq:ZtmZmCollection}
  \BA^{M-4}\ \BB_1\BC_1\ \BB_2\BC_2\ (\BB_3{}^2)\ (\BD_3{}^2)\ (\BB_4{}^2)\ (\BD_4{}^2).
\end{equation}
Then,
\begin{equation}
  J_\Kodaira = A^-_1\,{}^{\oplus 4},
\end{equation}
corresponding to 4 singular fibers each containing an elliptic curves of $A_1$
singularities.  In the coalesced collection \eqref{eq:ZtmZmCollection},
$\Bdelta_3/2$ becomes a weakly integral null junction, so that
\begin{equation}
  J^\weak_\text{null}=\Bdelta_1\IZ\oplus \Bdelta_2\IZ\oplus
  (\Bdelta_3/2)\IZ\oplus \Bdelta_4\IZ.
\end{equation}
and the Mordell-Weil torsion is
\begin{equation}
  \MW_\tor = J^\weak_\text{null}/J_\Loop = \Bdelta_3\IZ_{2m}\oplus\Bdelta_4\IZ_m.
\end{equation}
As in Sec.~3.4, the torsion sections can be described explicitly in terms
of the coordinates $y^i$ on the abelian fiber.  They are generated by the
sections
\begin{equation}
  (y^1,y^2,y^3,y^4) = \bigl(0,0,\tfrac{1}{2m},0\bigr)
  \text{ and }\ \bigl(0,0,0,\tfrac{1}{m}\bigr) \text{ in }\CX_{m,n}, 
\end{equation}
which are easily seen to be monodromy invariant, up to the identifications
$y^i\cong y^i+1$.


\subsubsection{$\IZ_2{}^{\oplus 2}\oplus\IZ_{2m}{}^{\oplus 2}$ torsion, $m=1,2$}
\label{Sec:Z2Zm}

In the case that $M=16-4mn\ge 8$, the isometry group of torsion sections can be
further enhanced to $\IZ_2{}^{\oplus 2}\oplus\IZ_{2m}{}^{\oplus 2}$ by
coalescing additional fibers.  Consider a coalesced collection analogous to that
of Sec.~\ref{Sec:EllipticExample}.  From the braiding operations discussed in
App.~\ref{App:Braiding}, we have
\begin{equation}\label{eq:ExIIcollection}
  \begin{split}
    \BA^M\ \BB_1\BC_1\ \BB_2\BC_2\ \BB_3\BC_3\ \BB_4\BC_4
    &\cong
    \BA^{M-8}\ \BB_1\BC_1\BA^2\ \BB_2\BC_2\BA^2\ \BB_3\BC_3\BA^2\ \BB_4\BC_4\BA^2\\
    &\cong
    \BA^{M-8}\ \BB_1\BA^2\BB_1\ \BB_2\BA^2\BB_2\ \BB_3\BA^2\BB_3\ \BB_4\BA^2\BB_4\\
    &\cong
    \BA^{M-8}\ \BB_1{}^2\BD_1{}^2\ \BB_2{}^2\BD_2{}^2\ \BB_3{}^2\BD_3{}^2\ \BB_4{}^2\BD_4{}^2.
  \end{split}
\end{equation}
In the basis corresponding to the last collection of
Eq.~\eqref{eq:ExIIcollection}, the generators \eqref{eq:XmnNullGenerators} of
$J_\text{null}$ become
\begin{equation}
  \begin{split}
    \Bdelta_1 &= \bigl(0^{M-8};\ (-1)^4;\ 1^4;\ (-1)^4;\ 1^4\bigr),\\
    \Bdelta_2 &= \bigl((-1)^{M-8};\ (M-9)^2,(M-10)^2;\ (11-M)^2,(12-M)^2;\ (M/2-5)^2,(M/2-6)^2;\ (-1)^2,0^2\bigr),\\
    \Bdelta_3 &= \bigl(0^{M-8};\ 0^4;\ 0^4;\ m^4;\ (-m)^4\bigr),\\
    \Bdelta_4 &= \bigl(0^{M-8};\ (-m)^4;\ (2m)^4;\ (-2m)^4;\ m^4\bigr).
  \end{split}
\end{equation}
The loop junction lattice $J_\Loop$ is given by Eq.~\eqref{eq:XmnLoopLattice}.

Now, suppose that we coalesce fibers pairwise to obtain the collection
\begin{equation}
  (\BA^2)\ (\BA^2)\ (\BB_1{}^2)\ (\BD_1{}^2)\ (\BB_2{}^2)\ (\BD_2{}^2)
  \ (\BB_3{}^2)\ (\BD_3{}^2)\  (\BB_4{}^2)\ (\BD_4{}^2).
\end{equation}
Then,
\begin{equation}
  J_\Kodaira = A^-_1\,{}^{\oplus 10},
\end{equation}
corresponding to 10 singular fibers, equal to the compactified Jacobian of a
genus-2 curve with an I$_2$ type degeneration, or equivalently, each with an
elliptic curve of $A_1$ singularities.  Each $\Bdelta_i/2$ becomes a weakly
integral null junction, so that
\begin{equation}
  J^\weak_\text{null}=(\Bdelta_1/2)\IZ\oplus (\Bdelta_2/2)\IZ\oplus
  (\Bdelta_3/2)\IZ\oplus (\Bdelta_4/2)\IZ.
\end{equation}
and the Mordell-Weil torsion is
\begin{equation}
  \MW_\tor = J^\weak_\text{null}/J_\Loop 
  = (\Bdelta_1/2)\IZ_2\oplus(\Bdelta_1/2)\IZ_2\oplus
  (\Bdelta_1/2)\IZ_{2m}\oplus(\Bdelta_1/2)\IZ_{2m}.
\end{equation}
The torsion sections can again be described explicitly in terms of the
coordinates $y^i$ on the abelian fiber.  They are generated by the sections
\begin{equation}
  \bigl(y^1,y^2,y^3,y^4\bigr) 
  = \bigl(\half,0,0,0\bigr),\ \bigl(0,\half,0,0\bigr),
  \ \bigl(0,0,\tfrac{1}{2m},0\bigr),\text{ and}\ \bigl(0,0,0,\tfrac{1}{2m}\bigr) 
  \text{ in }\CX_{1,1}, 
\end{equation}
which are easily seen to be monodromy invariant, up to the identifications
$y^i\cong y^i+1$.

Finally, torsion subgroups can be obtained by partially uncoalescing the
collection.  For example, if we uncoalesce
\begin{equation}\label{eq:UncoalesceChain}
  \begin{split}
    \text{the $(\BA^2)$s} 
    \quad & \Rightarrow\quad\text{we obtain \ }
    \MW_\tor = \IZ_2\oplus\IZ_{2m}{}^{\oplus 2},
    \text{ generated by $\Bdelta_1/2$, $\Bdelta_3/2$, $\Bdelta_4/2$,}\\
    \text{and $(\BB_2^2)$, $(\BD_2^2)$}
    \quad & \Rightarrow\quad\text{we obtain \ } \MW_\tor = \IZ_{2m}{}^{\oplus 2},
    \text{ generated by $\Bdelta_3/2$, $\Bdelta_4/2$,}\\
    \text{and $(\BB_3^2)$, $(\BD_3^2)$}
    \quad & \Rightarrow\quad\text{we obtain \ } \MW_\tor = \IZ_m\oplus\IZ_{2m},
    \text{ generated by $\Bdelta_3$, $\Bdelta_4/2$.}
  \end{split}
\end{equation}


\subsubsection{$\IZ_4\oplus\IZ_2{}^{\oplus 2}$ torsion and $\IZ_2{}^{\oplus 3}$ torsion}
\label{Sec:Z4}

Now focus on the principally polarized case $(m,n)=(1,1)$.  From the braiding
operations discussed in App.~\ref{App:Braiding}, we have
\begin{equation}\label{eq:ExIIIcollection}
  \begin{split}
    \BA^{12}\ \BB_1\BC_1\ \BB_2\BC_2\ \BB_3\BC_3\ \BB_4\BC_4
    &\cong
    \BA^3\BB_1\BC_1\ \BA^3\BB_2\BC_2\ \BA^3\BB_3\BC_3\ \BA^3\BB_4\BC_4\\
    &\cong
    \BD_1{}^4\BE_1\ \BD_2{}^4\BE_2\ \BD_3{}^4\BE_3\ \BD_4{}^4\BE_4,
  \end{split}
\end{equation}
where the $\BE_i$ are defined in App.~\ref{App:Braiding} and have vanishing
cycles of the form $(1,2,*,*)$.  In the basis corresponding to the last
collection of Eq.~\eqref{eq:ExIIIcollection}, the generators
\eqref{eq:XmnNullGenerators} of $J_\text{null}$ become
\begin{equation}
  \begin{split}
    \Bdelta_1 &= \bigl(1^4,-2;\ (-1)^4,2;\ 1^4,-2;\ (-1)^4,2\bigr),\\
    \Bdelta_2 &= \bigl(0^4,-1;\ (-1)^4,3;\ 0^4,-1;\ 1^4,-1\bigr),\\
    \Bdelta_3 &= \bigl(0^{10};\ (-1)^4,2;\ 1^4,-2\bigr),\\
    \Bdelta_4 &= \bigl(1^4,-2;\ 2^4,4;\ 2^4,-4;\ (-1)^4,2\bigr).
  \end{split}
\end{equation}
The loop junction lattice $J_\Loop$ is Eq.~\eqref{eq:XmnLoopLattice} with $m=1$.

Now, suppose that we coalesce quadruples of fibers to obtain the collection
\begin{equation}
  (\BD_1{}^4)\BE_1\ (\BD_2{}^4)\BE_2\ (\BD_3{}^4)\BE_3\ (\BD_4{}^4)\BE_4.
\end{equation}
Then,
\begin{equation}
  J_\Kodaira = A^-_3\,{}^{\oplus 4},
\end{equation}
corresponding to 4 singular fibers equal to the compactified Jacobian of a
genus-2 curve with an I$_4$ type degeneration, or equivalentally, 4 elliptic
curves of $A_3$ singularities.  For weakly integral null junctions, the charges
of the $\BD_i$ must be $1/4$-integral and those of the $\BE_i$ must be integral.
This gives weakly integral null junctions $\Bdelta_1/4-\Bdelta_2/2$,
$\Bdelta_3/2$, and $\Bdelta_4/2$, with
\begin{equation}
  \MW_\tor = J^\weak_\text{null}/J_\Loop = 
  (\Bdelta_1/4-\Bdelta_2/2)\IZ_4 \oplus (\Bdelta_3/2)\IZ_2 \oplus (\Bdelta_4/2)\IZ_2.
\end{equation}
In terms of the coordinates $y^i$ on the abelian fiber, the explicit torsion
sections are
\begin{equation}
  (y^1,y^2,y^3,y^4) = (1/4,1/2,0,0),\ (0,0,1/2,0),\ \text{ and }\ (0,0,0,1/2),
\end{equation}
respectively.

If instead of coalescing quadruples, we coalesce pairs of $\BD_i$ fibers,
\begin{equation}
  (\BD_1{}^2)(\BD_1{}^2)\BE_1\ (\BD_2{}^2)(\BD_2{}^2)\BE_2%
\ (\BD_3{}^2)(\BD_3{}^2)\BE_3\ (\BD_4{}^2)(\BD_4{}^2)\BE_4,
\end{equation}
then
\begin{equation}
  J_\Kodaira = A^-_1\,{}^{\oplus 8},
\end{equation}
corresponding to 8 singular fibers equal to the compactified Jacobian of a
genus-2 curve with an I$_2$ type degeneration, or equivalentally, 4 elliptic
curves of $A_1$ singularities.  The Mordell-Weil torsion is reduced from
$\IZ_4\oplus\IZ_2{}^{\oplus2}$ to $\IZ_2{}^{\oplus3}$:
\begin{equation}
  \MW_\tor = J^\weak_\text{null}/J_\Loop = 
  (\Bdelta_1/2)\IZ_2 \oplus (\Bdelta_3/2)\IZ_2 \oplus (\Bdelta_4/2)\IZ_2.
\end{equation}
In Sec.~\ref{Sec:AlgebraicConstruction}, we will reproduce this last result from
the relative Jacobian construction of $\CX_{1,1}$.


\subsection{Connections to other Calabi-Yau manifolds with nontrivial $\pi_1$}
\label{Sec:Connecting}


Let us focus on the example in Sec.~\ref{Sec:ZtmZmTorsion} in the principally
polarized case $(m,n)=(1,1)$.  The collection is
\begin{equation}
  \BA^8\ \BB_1\BC_1\ \BB_2\BC_2\ (\BB_3{}^2)\ (\BD_3{}^2)\ (\BB_4{}^2)\ (\BD_4{}^2).
\end{equation}
with
\begin{equation}
  J_\Kodaira = A^-_1\,{}^{\oplus 4}
  \quad\text{and}\quad
  \MW_\tor = J^\weak_\text{null}/J_\Loop = \IZ_2.
\end{equation}
In this case, the Calabi-Yau manifold $\CX_{1,1}$ has a $\IZ_2$ isometry.  In
terms of the coordinates $y^1,y^2,y^3,y^4$ on the abelian surface fiber, the
isometry is $y^3\mapsto y^3+1/2$.

We now quotient by this isometry and ask what resulting topology is obtained.
Since we would like preserve the condition that the fiber coordinates are
periodic modulo 1, we will rescale the $y^3$ coordinate so that $y^3_\text{new}
= 2y^3_\text{old}$.  This is implemented by conjugating all monodromy matrices:
\begin{equation}
  K_\text{old}\mapsto K_\text{new} = T K_\text{old} T^{-1},
  \quad\text{where}\quad T = \diag(1,1,2,1).
\end{equation}

The conjugation leaves $K_\BA$, $K_{\BB_2}$ and $K_{\BC_2}$ unchanged,
and maps the monodromy matrices $K_{\BC_1}$, $K_{\BB_1}$,
$K_{(\BD_3)^2}$, $K_{(\BB_3)^2}$, $K_{(\BD_4)^2}$, and $K_{(\BB_4)^2}$ to 
\begin{equation}\label{eq:QuotientMonodII}
  \begin{split}
    K_{\tilde\BC_1} &= \begin{pmatrix}
      2 & -1 & 0 & 1 \\
      1 & 0 & 0 & 1 \\
      -2 & 2 & 1 & -2 \\
      0 & 0 & 0 & 1
    \end{pmatrix},\\
    \noalign{\vskip1ex}
    K_{\tilde\BD_3} &= \begin{pmatrix}
      1 & 0 & 0 & 0 \\
      2 & 1 & -1 & 0 \\
      0 & 0 & 1 & 0 \\
      -2 & 0 & 1 & 1
    \end{pmatrix},\\
    \noalign{\vskip1ex}
    K_{\tilde\BD_4} &= \begin{pmatrix}
      1 & 0 & 0 & 0 \\
      2 & 1 & -1 & 2 \\
      -4 & 0 & 3 & -4 \\
      -2 & 0 & 1 & -1
    \end{pmatrix},
  \end{split}
  \quad
  \begin{split}
    K_{\tilde\BB_1} &= \begin{pmatrix}
      0 & -1 & 0 & -1 \\
      1 & 2 & 0 & 1 \\
      -2 & -2 & 1 & -2 \\
      0 & 0 & 0 & 1
    \end{pmatrix},\\
    \noalign{\vskip1ex}
    K_{\tilde\BB_3} &= \begin{pmatrix}
      -1 & -2 & 1 & 0 \\
      2 & 3 & -1 & 0 \\
      0 & 0 & 1 & 0 \\
      -2 & -2 & 1 & 1
    \end{pmatrix},\\
    \noalign{\vskip1ex}
    K_{\tilde\BB_4} &= \begin{pmatrix}
      -1 & -2 & 1 & -2 \\
      2 & 3 & -1 & 2 \\
      -4 & -4 & 3 & -4 \\
      -2 & -2 & 1 & -1
    \end{pmatrix},
  \end{split}
\end{equation}
respectively.  The resulting monodromy matrices are all $SL(4,\IZ)$ similar to
$K_\BA$ and therefore correspond to irreducible singular fibers of a new
collection,
\begin{equation}
  \BA^8\ \tilde\BB_1\tilde\BC_1\ \BB_2\BC_2\ \tilde\BB_3\tilde\BD_3
  \ \tilde\BB_4\tilde\BD_4.
\end{equation}
The similarity transformation is easy to see in the case of $K_{\tilde\BB_1}$
and $K_{\tilde\BC_1}$, since these are identical to the matrices $K_{\BB_1}$ and
$K_{\BC_1}$ of the $(m,n) = (1,2)$ case.  For the remaining $K_{\tilde\BX}$, an
explicit choice of matrices realizing the similarity transformation
$K_{\tilde\BX} = S_{\tilde\BX} K_\BA S_{\tilde\BX}{}^{-1}$ is
\begin{equation}\label{eq:Stildes}
  \begin{split}
    S_{\tilde\BD_3} &= \begin{pmatrix}
      0 & 0 & 1 & 0\\
      1 & 0 & 0 & 1\\
      0 & 1 & 2 & 0\\
     -1 & 0 & 0 & 0
    \end{pmatrix},\\
    \noalign{\vskip1ex}
    S_{\tilde\BD_4} &= \begin{pmatrix}
      0 & 0 &-1 & 0\\
      1 & 0 & 0 & 0\\
     -2 & 1 &-2 &-2\\
     -1 & 0 & 0 &-1
    \end{pmatrix},
  \end{split}
  \quad
  \begin{split}
    S_{\tilde\BB_3} &= \begin{pmatrix}
      1 & 0 & 0 & 1\\
     -1 & 0 &-1 &-1\\
      0 &-1 &-2 & 0\\
      1 & 0 & 0 & 0
    \end{pmatrix},\\
    \noalign{\vskip1ex}
    S_{\tilde\BB_4} &= \begin{pmatrix}
      1 & 0 & 0 & 0\\
     -1 & 0 & 1 & 0\\
      2 &-1 & 2 & 2\\
      1 & 0 & 0 & 1
    \end{pmatrix},
  \end{split}
\end{equation}
The vanishing cycles are given by $(1,-1,2,0)$ and $(1,1,-2,0)$ for
$\tilde\BB_1$ and $\tilde\BC_1$, respectively, and by the first column of the
corresponding matrix in Eq.~\eqref{eq:Stildes} for $\tilde\BB_3$, $\tilde\BD_3$,
$\tilde\BB_3$, and $\tilde\BD_3$.

From the rescaling of $y^3$, the Hodge form on the quotient (obtained from
$2\,\omega_\text{old}$) is
\begin{equation}
  \omega_\text{new} = 2 dy^1\w dy^2 + dy^3\w dy^4,
\end{equation}
of polarization $(2,1)$.  Since the $\IZ_2$ is freely acting, the quotient is
again a Calabi-Yau manifold, with trivial $\MW_\tor$ and $\pi_1=\IZ_2$.  From
these properties, we see that it is a new Calabi-Yau manifold, distinct from the
set of $\CX_{m,n}$ dual to $T^6/\IZ_2$.

In this example, the $\IZ_2$ action on a reducible (pairwise coalesced) fiber
exchanges the two components, leaving an irreducible fiber.  This is case 2
below.  More generally, there are three possibilities for the action of an
element of $\MW_\tor$ on a reducible fiber:

\begin{enumerate}
\item On each singular elliptic curve of the reducible fiber, the isometry acts
  freely by translation.  In this case, there is no change in the type of the
  reducible fiber.  The monodromy matrix factorizes into the same number of
  irreducible matrices (each similar to $K_\BA$) before and after quotienting.

\item The isometry permutes the components of the reducible fiber.  In this
  case, if there is a single orbit, then the fiber becomes irreducible after
  quotienting and the monodromy also becomes irreducible (similar to $K_\BA$).
  If there is more than one orbit, then there is one irreducible component for
  each orbit.

\item The isometry is along the vanishing cycle (i.e., each point of the
  singular locus of the reducible fiber is a fixed point of the isometry).  In
  this case, the singular fiber becomes ``more singular,'' i.e., reducible into
  more components after quotienting, and the monodromy matrix factorizes into
  more irreducible factors after quotienting than before.
\end{enumerate}


\section{Algebraic construction in the principally polarized case}
\label{Sec:AlgebraicConstruction}

In this section, we change gears and provide a second construction of the type
IIA Calabi-Yau geometry dual to $T^6/\IZ_2$, this time taking an explicit
algebro-geometric approach and focusing on the principally polarized case
$m,n=1,1$.  Following Saito \cite{Saito,SaitoKluwer,SaitoTalk}, we construct a
principally polarized abelian surface fibration over $\IP^1$ as the relative
Jacobian of a genus-2 fibered surface $S$.  We show that it satisfies all of the
required properties to be the Calabi Yau manifold $\CX_{1,1}$ described in the
previous section.  By Wall's theorem~\cite{Hubsch} and its extension due to \u
Zubr, a Calabi-Yau manifold is determined up to homotopy type by its Hodge
numbers, second Chern class, and intersection numbers~\cite{Hubsch,Tomasiello}.
We show that all of these quantities agree with those of $\CX_{1,1}$, and in
addition, compute the Mordell-Weil lattice from this perspective.

The construction begins with a pencil of \hbox{genus-2} curves---that is, with
surface $S$, fibered over $\IP^1$ via a projection map $\rho\colon S\to\IP^1$,
whose generic fibers $C_p=\rho^{-1}(p)$ are smooth curves of genus 2.

Associated to each fiber $C_p$ is its Jacobian $\CJ_{C_p}\cong T^4$, which is a
principally polarized abelian surface.  (See App.~\ref{App:CurveJac} for a
review of complex curves and their Jacobians.)  The {\it relative Jacobian}
$\CJ_{S/\IP^1}$ is an abelian surface fibration over $\IP^1$, obtained from $S$
by replacing each fiber $C_p$ with $\CJ_{C_p}$.  For the appropriate choice of
$S$, we show that $\CJ_{S/\IP^1}$ is the desired Calabi-Yau 3-fold.


\subsection{The surface $S$}
\label{Sec:SurfaceS}

Our choice of $S$ is as follows.  Let $L^2$ be a line bundle of degree $(6,2)$
over $\IP^1_{s,t}\times\IP^1_{u,v}$, and let $f(s,t;u,v)$ be a homogeneous
polynomial of degree $(6,2)$, so that $f$ defines a section of $L^2$.  Then,
$f^{1/2}$ defines a 2-fold section of the line bundle $L$ of degree $(3,1)$,
branched over the curve \hbox{$B=\{f=0\}$ in $\IP^1\times\IP^1$}. I.e., it is a
double cover of $\IP^1\times\IP^1$ branched over $B$.  This double cover is the
desired surface $S$.
\begin{figure}[ht]
  \def\BranchCurve{\mbox{\includegraphics{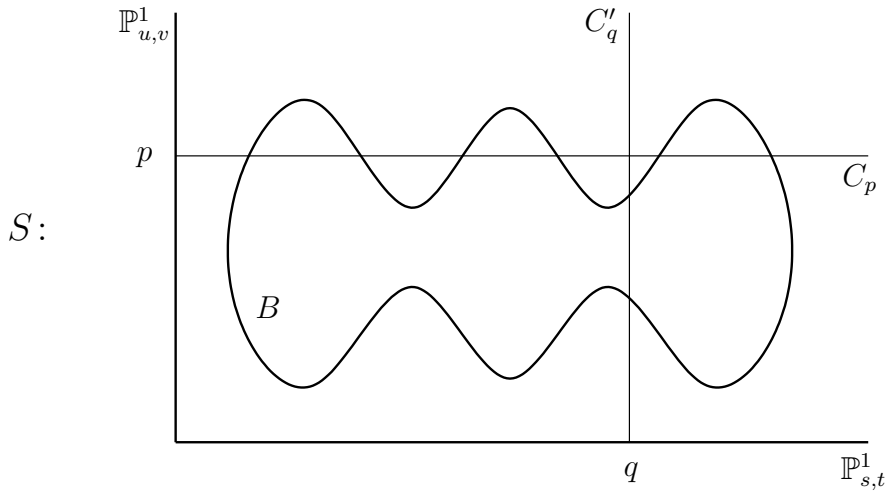}}}
  \begin{equation*}
    \begin{xy}
      \xyimport(3.8,2.5){\BranchCurve}
      (-0.15,2.40)*{\IP^1_{u,v}} ; (2.33,2.40)*{C'_q} ; (2.49,-0.15)*{q} ;
      (3.75,-0.15)*{\IP^1_{s,t}} ; (3.73,1.52)*{C_p} ; (-0.15,1.65)*{p} ;
      (0.52,0.79)*{B} ;
      (-0.75,1.25)*{\text{\Large $S\colon$}} ;
    \end{xy}
  \end{equation*}
  \caption{The surface $S$ is a double cover of $\IP^1_{s,t}\times \IP^1_{u,v}$
    branched over a degree $(6,2)$ curve~$B$.  The fiber $C_p$ over a point
    $p\in\IP^1_{u,v}$ is a double cover of $\IP^1_{u,v}$ with 6 branch points
    (genus-2), and the fiber $C'_q$ over a point $q\in\IP^1_{s,t}$ is a double
    cover of $\IP^1_{u,v}$ with 2 branch points (genus-0).}
  \label{fig:CartoonS}
\end{figure}

The surface $S$ can be viewed as a fibration in at least two ways, corresponding
to the natural projections $\rho\colon S\to\IP^1_{u,v}$ and $\rho'\colon
S\to\IP^1_{s,t}$.  In the first case, the fiber $C_p$ over a generic point
$p=[u,v]$ is a double cover of $\IP^1_{s,t}$ branched over the 6 points $[s,t]$
where $f(s,t;u,v)=0$.  In this case, $g(C_p)=2$.\footnote{Recall that an
  elliptic curve ($T^2$, with $g=1$) is a double cover of $\IP^1$ branched over
  four points.  Likewise, a genus $g\ge1$ curve is a double cover of $\IP^1$
  branched over $2g+2$ points.}  This is the desired genus-2 fibration.  In the
second case, the fiber $C'_q$ over a generic point $q=[s,t]$ is a double cover
of $\IP^1_{u,v}$ branched over 2 points, which is a $\IP^1$.


\nonnumberedsubsec{Singular fibers of $S$}

To determine the number of singular fibers of the genus-2 fibration $C\to
S\xrightarrow{\rho}\IP^1_{u,v}$, we note that
\begin{equation}\label{eq:EulerSfib}
  \begin{split}
    \chi(S) &= \chi(C)\chi(\IP^1)+n_{\rm sing}\bigl(\chi(C_{\rm sing})-\chi(C)\bigr)\\
            &= (-2)(2) + n_{\rm sing}(1) = n_{\rm sing}-4, 
  \end{split}
\end{equation}
assuming that all $n_{\rm sing}$ singular fibers are irreducible.\footnote{That
  is, we assume that $C_{\rm sing}$ is a genus-2 curve in which a single 1-cycle
  has contracted, or equivalently, an elliptic curve with 2 points identified.
  This gives a rational double point singularity of $C_{\rm sing}$.}  On the
other hand, since $S$ is the double cover of $\IP^1\times\IP^1$ ramified over
$B$,\footnote{Note that $B$ is of genus 5.  This is a special case of the result
  that $g=(\a-1)(\b-1)$ for a degree $(\a,\b)$ curve in $\IP^1\times\IP^1$
  (cf. App.~\ref{App:Genera}).}
\begin{equation}\label{eq:EulerSdbl}
  \begin{split}
    \chi(S) &= 2\chi(\IP^1\times\IP^1)-\chi(B) = 2\chi(\IP^1)^2-\chi(B)\\
            & = 2(2)^2 - (-8) = 16.
  \end{split}
\end{equation}
Equating the two expressions, we find that $n_{\rm sing}=20$.  In the same
manner, it can be shown that there are 12 singular fibers of the fibration
$C'\to S\xrightarrow{\rho'}\IP^1_{s,t}$.


\nonnumberedsubsec{Sections $\IP^1\hookrightarrow S$}

A section $\ell$ of the genus-2 fibration is a rational curve $\ell\subset S$
that projects 1-to-1 to $\IP^1_{u,v}$ and therefore intersects each fiber in
exactly one point.  There are 24 such curves.  To identify them, first consider
the projection $\rho'\colon S\to\IP_{s,t}$.  The generic fiber $C'$ is a smooth
$\IP^1$ that can be thought of as a double cover of $\IP^1_{u,v}$ ramified at
its two points of intersection with $B$.  So, it projects 2-to-1 rather than
1-to-1 $\IP^1_{u,v}$ and intersects each \hbox{genus-2} curve $C$ (i.e., locus
of fixed $[u,v]$) twice.  However, on 12 singular fibers, the two ramification
points coincide, and the fiber is a nodal $\IP^1$: it consists of two $\IP^1$s
intersecting in a point, one on each branch of the double cover.  Each of the
two $\IP^1$s projects 1-to-1 to $\IP^1_{u,v}$ and intersects each genus-2 curve
once.  This gives $2\times 12=24$ sections of the genus-2 fibration, denoted by
$\ell_I, \ell'_I$, for $I=1,\ldots,12$.  In fact, these are the only
sections. Indeed, the image in $\IP^1_{s,t} \times \IP^1_{u,v}$ of any section
is a section of $\IP^1_{s,t} \times \IP^1_{u,v}$ over $\IP^1_{u,v}$, so it must
be a copy of $\IP^1_{s,t}$. But away from the 12 reducible fibers, the inverse
image in $S$ of a copy of $\IP^1_{s,t}$ is an irreducible two-section, so it
does not contain any sections.


\nonnumberedsubsec{Cohomology of $S$}

The cohomology of $S$ is easily calculated from the Leray spectral sequence for
the projection $\pi: S \to \IP^1_{s,t}$. (See App.~\ref{App:DirIm} for
background on direct images and their relation to Leray spectral sequences.)  As
we have just seen, the generic fiber is a $\IP^1$, while precisely 12 fibers are
reducible, $\IP^1 \cup \IP^1$. We see that the derived image sheaves are:
\begin{equation}
  R^0=\IZ,\quad   R^1=0,
\end{equation}
and $R^2$ is the direct sum of $\IZ$ and 12 skyscraper sheaves supported at the
12 singular fibers.  It follows that the Leray spectral sequence degenerates at
$E^2$, and:
\begin{equation}
  h^0(S)=1,\quad   h^1(S)=0,\quad   h^2(S)=h^{1,1}(S)=2+12=14.
\end{equation}

Note that
\begin{equation}\label{eq:llprime}
  \ell_I+\ell'_I = C'\quad\hbox{in }H_2(S,\IZ),
\end{equation}
where $C'$ is the class of the generic $\IP^1$ fiber over $\IP^1_{s,t}$.
Therefore, the sections $\ell_I,\ell'_I$ of the genus-2 fibration span a 13
dimensional sublattice of $H_2(S,\IZ)$.  Adding the generic genus-2 fiber $C$
gives the full lattice $H_2(S,\IZ)$.


\subsection{The 3-fold $\CX$}
\label{Sec:3foldX}

Starting from the genus-2 fibration $C\to S\xrightarrow{\rho}\IP^1$, we define a
3-fold $\CX$ as the relative Jacobian,
\begin{equation}\label{eq:RelJac}
  \CX=\CJ_{S/\IP^1}=\Pic^0(S/\IP^1).
\end{equation}
As already mentioned, this means that $\CX$ is obtained from $S$ by replacing
the genus-2 fiber $C_p=\rho^{-1}(p)$ over each generic point $p$ with its
Jacobian abelian surface $A_p = \CJ_{C_p}$.  Singular fibers are replaced by the
compactifications of their Jacobians.  Let $\pi$ denote the corresponding
projection map.  Then, $\CX$ is an abelian surface fibration
$A\to\CX\xrightarrow{\pi}\IP^1$.

\nonnumberedsubsec{Number and type of singular fibers of $\CX$}

Like $S$, the 3-fold $\CX$ has 20 singular fibers, in agreement with the number
of singular fibers of $\CX_{1,1}$ in Sec.~2.  Each is topologically of
I$_1\times T^2$ type.  However, the complex structure does not in general
respect this factorization.  Instead, the singular fiber should be viewed as the
$\IP^1$ bundle $\IP^1\bigl(\CO(p)\oplus\CO(q)\bigr)$ over an elliptic curve $E$,
with the zero section $[0,*]$ glued to the section at infinity $[*,0]$.  Here
$p$ and $q$ are two points on $E$.\footnote{When $p=q$, the fiber factorizes as
  $E$ times I$_1$, where the I$_1$ is realized as $\IP^1$ with the points $0$
  and $\infty$ identified.  But when $p\ne q$ we instead identify
  $\infty\in\IP^1_r$ in the fiber at each point $r\in E$ with
  $\infty\in\IP^1_{r+(q-p)}$ in a \emph{different fiber} at the shifted point
  $r+(q-p)\in E$.}  This leaves a unique section $E_0\cong E_\infty$.  This
elliptic curve is the singular locus of the singular fiber.

To derive this form for the singular fibers of the $\CX$, let us first consider
the singular fibers of $S$ and then find their compactified Jacobians.  Recall
that the generic I$_1$ degeneration of an elliptic fibration can be viewed as a
$\IP^1$ with two points identified; these are the $2g+2=2$ branch points in the
presentation of $\IP^1$ as branched double cover of $\IP^1$. Likewise, a genus-2
curve $C$ is a branched double cover of $\IP^1$ with 6 branch points, so the
generic singularity is one in which 2 of these points coincide.  It can be
viewed as an elliptic curve $E$ with two points $p$ and $q$ identified
\begin{equation}
  C\cong E/\{p\sim q\}\qquad\text{(generic degeneration of the genus-2 curve $C$).}
\end{equation}
The normalization map $\nu\colon E\to C$ identifies the points $p$ and $q$.

A line bundle on $C$ pulls back via the normalization map $\nu$ to a line bundle
on $E$. In fact, specifying a line bundle on $C$ is the same as specifying a
line bundle $L$ on $E$ together with a gluing of the fibers of $L$ at $p,q$. The
set of these gluings is a copy of the group of isomorphisms from the line $C$ to
itself, which is given by the multiplicative group $C^*$. It follows that
$\Pic(C)$ is an extension of $\Pic(E)$ by $C^*$. In order to compactify, we must
allow torsion free sheaves on $C$ that are not necessarily line bundles. The
effect is to replace each fiber $C^*$ by its compactification $\IP^1$. Note what
happens when the gluing parameter $t \in C^*$ goes to $0$: the map from $L_p$ to
$L_q$, which is an isomorphism for most $t$, becomes the $0$ map as $t \to
0$. The result is the torsion free sheaf
$\nu_*\bigl(L\otimes\CO_E(-p)\bigr)$. Its fiber at the singular point has rank
$2$ rather than $1$, and consists of the direct sum of the fiber of
$L\otimes\CO_E(-p)$ at $p$ with the fiber of $L$ at $q$. On the other hand, when
the gluing parameter $t \in C^*$ goes to $\infty$, the inverse map from $L_q$ to
$L_p$, which is an isomorphism for most $t$, becomes the $0$ map as $t \to
\infty$. The result is now the torsion free sheaf
$\nu_*\bigl(L\otimes\CO_E(-q)\bigr)$. Its fiber at the singular point consists
of the direct sum of the fiber of $L$ at $p$ with the fiber of
$L\otimes\CO_E(-q)$ at $q$. In other words, the $0$ and $\infty$ sections of the
$\IP^1$ fibration over $\Pic(E)$ are glued to each other, but the gluing is not
the obvious one: it involves a shift of $\CO_E(p-q)$. This is summarized in the
following diagram:
\begin{equation}
  \begin{CD}
    0   @>>> C^*            @>>> \CJ_C         @>{\nu^*}>> \CJ_E @>>> 0\\
    @.       @V{\bigcap}VV       @VVV                      @|         @.\\
    {}  @.   \IP^1          @>>> \CJ'          @>>>        \CJ_E @.   {}\\
    @.       @.                  @VVV                      @.         @.\\
    {}  @.                  @.   \bar\CJ_C     @.          {}    @.   {}
  \end{CD}.
\end{equation}

\nonnumberedsubsec{Hodge numbers}

The number of complex structure deformations of $S$ is the choice of degree
$(6,2)$ polynomial $f(s,t;u,v)$ modulo equivalences:
\begin{equation}\label{eq:cpxmoduli}
  h^{1,1}(S) 
  = (6+1)(2+1) - (1\hbox{ overall rescaling}) - (3+3\text{ from }SL(2,\IC)^2)
  = 14.
\end{equation}
The complex structure deformations of $\CX$ are in 1-to-1 correspondence with
those of $S$.  Therefore, $h^{2,1}(\CX) = 14$.  But we have just seen that every
fiber of $\CX$ over $\IP^1$ is either an abelian surface, which is topologically
$T^4$, or it is singular, in which case it is topologically $T^2$ times a nodal
curve.  Either way, the Euler characteristics of all fibers vanish.  It follows
that the Euler characteristic of $\CX$ vanishes as well, so
$h^{1,1}=h^{2,1}=14$.  The Hodge numbers of $\CX$ agree with those $\CX_{1,1}$.

\nonnumberedsubsec{The Calabi-Yau condition}

We wish to show that the manifold $\CX$ has trivial canonical bundle.  Consider
any $\IP^1$ section of the abelian fibration $\CX$.  By the adjunction formula,
\begin{equation}\label{eq:KX}
  K_{\CX|\IP^1} = K_{\IP^1}\otimes\det(N^*_{\IP^1})
    = \CO_{\IP^1}(-2)\otimes\det(N^*_{\IP^1}),
\end{equation}
where $N^*_{\IP^1}$ is the conormal bundle to $\IP^1$ in $\CX$.  Therefore, $\CX$ is
a Calabi-Yau manifold if $\det(N^*_{\IP^1}) = \CO_{\IP^1}(2)$.  To compute
$N^*_{\IP^1}$, we note that
\begin{equation}\label{eq:DirectIm}
  N^*_{\IP^1} = \rho_* K_{S/\IP^1},
\end{equation}
where $\rho_*$ is the direct image functor\footnote{See App.~\ref{App:DirIm} for
  background on the direct image $\rho_*$ and higher direct images (derived
  functors) $R^i\rho_*$.} of the projection map $\rho\colon\ S\to\IP^1_{u,v}$.
In App.~\ref{App:ProofCY}, it is shown that this gives $N_{\IP^1} =
\CO_{\IP^1}(-1)\oplus\CO_{\IP^1}(-1)$, from which the desired result follows.


\subsection{Checks}
        
\subsubsection{Intersection numbers}
\label{Sec:IntersectionNumbers}

In App.~\ref{App:Intersections}, it is shown that to each of the 24 sections
$\ell_I,\ell'_I$, for $I=1,\ldots,12$, we can associate a \emph{theta surface}
$\Theta_I$ or $\Theta'_I\in\CX$.  The theta surfaces are embeddings of $S$ in
the Calabi-Yau threefold $\CX$.  They satisfy homology relations analogous to
those of Eq.~\ref{eq:llprime},
\begin{equation}
  \Theta_I+\Theta'_I = D\quad\text{in }H_4(\CX,\IZ),
\end{equation}
where $D$ is independent of $I$.  

The theta surfaces together with generic abelian fiber $A$ form a basis of
$H_4(\CX,\IZ)$.  Their double and triple intersections are computed in
App.~\ref{App:Intersections}.  The result for the triple intersections of theta
surfaces is
\begin{equation}\label{eq:TripleIntsNoA}
  \begin{split}
    &\Theta_I\cdot\Theta_J\cdot\Theta_K = -1,\\
    &\Theta_I\cdot\Theta_J\cdot\Theta_J=\Theta_I\cdot\Theta'_J\cdot\Theta'_J =-2,\\
    &\Theta_I\cdot\Theta_I\cdot\Theta'_I=\Theta_I\cdot\Theta_J\cdot\Theta'_J =0,\\
    &\Theta_I\cdot\Theta_I\cdot\Theta_I=-4,\\
  \end{split}
\end{equation}
for $I,J,K$ distinct, together with equations obtained from these by exchange of
$\Theta$ and $\Theta'$.  For triple intersections in which $A$ appears, we have
 $A^2=0$, and
\begin{equation}
  A\cdot\Theta_I\cdot\Theta_J = A\cdot\Theta_I\cdot\Theta'_J
  = A\cdot\Theta'_I\cdot\Theta'_J = 2,
\end{equation}
for all $I,J$, not necessarily distinct.

With the identifications,
\begin{equation}\label{eq:Theta2Hep} 
  \CE_I=(\Theta_I-\Theta'_I)/2,\quad H=(\Theta_I+\Theta'_I)/2 + A/6,
\end{equation}
these intersections precisely match the result~\eqref{eq:NonzeroInts} obtained
by classical supergravity duality in Ref.~\cite{CYDuals}.  It is important to
confirm that the integrality matches as well.  The factors of $1/2$ are exactly
as expected.  The fact that $\CE_I$ is half of an integer divisor is equivalent
to the statement in the $T^6/\IZ_2$ dual that a string stretched between a
single D3-brane and its image represents half of a root of $SO(2M)$
(cf.~Sec.~\ref{Sec:JunctionsGauge}).  The factor of 1/2 in $H$ can be understood
in a similar way. The appearance of the $A/6$ term in $H$ is more subtle and
requires a careful definition of warped volume in the $T^6/\IZ_2$ dual to
justify its appearance.  A proper treatment of this subtlety is an essential
ingredient of the analysis of duality map between D3 instantons and worldsheet
instantons under investigation in Ref.~\cite{DInstanton}, to which the reader is
referred for further details.\footnote{Note that Ref.~\cite{CYDuals} missed the
  subtlety responsible for the $A/6$ term in $H$.  There, $H$ was naively
  identified with a single theta surface, based on the fact that $A\cdot H$
  gives the homology class of $C$.  However, this intersection is preserved when
  $H$ is shifted by a multiple of $A$ or replaced by a weighted sum of theta
  surfaces of total weight 1.  Indeed, $H$ in Eq.~\eqref{eq:Theta2Hep} differs
  from a single theta surface in exactly these two ways.}


\subsubsection{Second Chern class $c_2(\CX)$}

The second Chern class $c_2(\CX)$ is the sum of the $20$ elliptic curves $E_i$
that are singular loci of the singular compactified Jacobians.  Its intersection
with a theta surface, for any of the $24$ possible choices $\Theta_I$ or
$\Theta'_I$, is $20$, and its intersection with the generic abelian fiber $A$ is
zero.  Thus, $H\cdot c_2 = 20$ for $H$, as given by Eq.~\eqref{eq:2ndChern}.

To derive $c_2$, we note that the normal bundle sequence has a very simple
modification: the vertical subsheaf of the tangent bundle is still a line
subbundle, which is still $\pi^*$ of the normal bundle to $\IP^1$ in $\CX$ (for
any $\IP^1$ section of the fibration); the horizontal piece is still
$\pi^*T_{\IP^1}$, but the map now is not surjective---instead there is a
cokernel which is the structure sheaf of the 20 elliptic curves.
\begin{equation}
  0\to N_{\IP^1}\to T_\CX\to \pi^* T_{\IP^1}
  \to \pi^* T_{\IP^1}\big|_{\bigcup_{i=1}^{20}E_i}\to 0.
\end{equation}
Here, $\Delta = \{p_1,\ldots,p_{20}\}$ is the discriminant locus of the genus-2
fibration $\rho\colon S\to\IP^1$, and the elliptic curves $E_i\in\pi^{-1}(p_i)$
are the singular loci of the degenerate fibers of $\pi\colon\CX\to\IP^1$.

Note that in the discussion of the first Chern class (Calabi-Yau condition)
above in Sec.~\ref{Sec:3foldX}, there was no special contribution from singular
fibers.  This is intuitively clear.  The $\IP^1$ sections intersect singular
fibers at smooth points, so the singular fibers should not affect $c_1$.


\subsubsection{Mordell-Weil lattice: $D_{12}^-$}
\label{Sec:MWisD}

As discussed in Sec.~2 and App.~\ref{App:AbelVar}, the Mordell-Weil group
$\MW(\CX)$ is the group of rational sections of the Abelian surface fibration
$\CX$, and the Mordell-Weil lattice $\MW(\CX)/\MW_\tor(\CX)$ is the lattice of
sections modulo torsion.  To determine this lattice, we would like to relate the
sections of $\CX$ to those of the genus-2 fibration $S$.  The sections of $\CX$
and $S$ do not quite encode the same data.  In particular, the sections of the
abelian fibration $\CX$ form a rank 12 lattice, while those of $S$ form a finite
set (just the 12 pairs $\ell_I,\ell'_I$ of Sec.~4.1).  However, given a choice
of section $\ell_0$ to play the role of zero section, we can map each of the 24
sections of $S$ to a section of $\CX$, as we now show.  This allows us to
describe the Mordell-Weil lattice of $\CX$ as a sublattice of $H_2(S,\IZ)$.

The 24 sections $\ell_I,\ell'_I\subset S$ can be thought of as elements of
$\G(\IP^1,\Pic^1(S/\IP^1))$.\footnote{See App.~\ref{App:CurveJac} for the
  definition of $\Pic^n(S/\IP^1)$ and further elaboration of this statement.}
Since $\CX=\Pic^0(S/\IP^1)$, we similarly have
\begin{equation}\label{eq:MWX}
  \MW(\CX)=\G(\IP^1,\Pic^0(S/\IP^1)).
\end{equation}
To relate the two, we note that $\Pic^1(S/\IP^1)\cong\Pic^0(S/\IP^1)$ via a
noncanonical isomorphism,
\begin{equation}\label{eq:Picisomorph}
  \Pic^1(S/\IP^1)
  \quad \xrightarrow{\otimes [\ell_0]^{-1}}\quad
  \Pic^0(S/\IP^1),
  \qquad 
  \ell \mapsto \ell-\ell_0,
\end{equation}
where $\ell_0\in\{\ell_I,\ell'_I\}$.  The isomorphism depends on the choice of
which of the 24 sections $\ell_I,\ell'_I$ maps to the zero section of $\CX$.

This isomorphism endows $\Pic^1(S/\IP^1)$ with the structure of an abelian group
and allows us to relate the Mordell-Weil lattice of $\CX$ to the N\'eron-Severi
lattice\footnote{The N\'eron-Severi lattice is roughly the same as the algebraic
  (i.e., $1,1$) part of $H_2(S,\IZ)$.  There are several closely related
  definitions of equivalence classes of divisors, for example, homological
  equivalence, linear equivalence (same line bundle) and numerical equivalence
  (same intersections).  The N\'eron-Severi lattice is the lattice of
  \emph{algebraic equivalence classes} of divisors, however, we will not need
  the technical definition of algebraic equivalence here, since on a projective
  variety, homological equivalence of divisors is equivalent to algebraic
  equivalence.  (See Ref.~\cite{GandH}, p.~462.)  In this case, N\'eron-Severi
  lattice is $\NS(S) \cong H^{1,1}(S)\cap H^2(S,\IZ) \cong \Pic(S)/\Pic^0(S)$.}
of $S$ via
\begin{equation}\label{eq:PicNS}
  \MW(\CX)\cong K^\perp\subset\NS(S),
\end{equation}
where $K$ is spanned by $\ell_0$, the generic genus-2 fiber $C$, and the
components of reducible fibers (in the case that there exist special fibers with
multiple components).  Here, the intersection pairing on $S$ is used to define
both the orthogonal complement $K^\perp$ and its height pairing.

Let us assume for simplicity that there are no reducible fibers.  Then the
Mordell-Weil lattice is just the orthogonal complement of $\ell_0$ and $C$ in
$\NS(S)$.  From the intersections
\begin{equation}\label{eq:intersections}
  \ell\cdot \ell = -1,\quad \ell\cdot C = 1,\quad C\cdot C = 0,
\end{equation}
for any section $\ell$ of $S$, the map from $\NS(S)$ to $\MW(\CX)$ is
\begin{equation}\label{eq:NStoMW}
  v\mapsto v^\perp = v-(v\cdot C)\,\ell_0 - \bigl(v\cdot(\ell_0+C)\bigr)\,C.
\end{equation}

The resulting lattice is the root lattice of $D_{12}$.  To see this, choose
$\ell_0=\ell'_{12}$ for concreteness.  Then, from
\begin{equation}\label{eq:intersectionsl}
  \ell_I\cdot \ell'_J = \delta_{IJ},\quad
  \ell_I\cdot \ell_J = \ell'_I\cdot \ell'_J = -\delta_{ij},
\end{equation}
we have
\begin{equation}\label{eq:NStoMWa}
  \begin{split}
     \ell_I  \mapsto \ell^\perp_i &= \ell_I-\ell_{12'}-C,\\
     \ell'_I \mapsto \ell'^\perp_i &= \ell'_I-\ell_{12'}-C,
  \end{split}
\end{equation}
for $i=1,\ldots,11$, and
\begin{equation}\label{eq:NStoMWb}
    \ell_{12} \mapsto \ell^\perp_{12} = \ell_{12}-\ell'_{12}-2C,
\end{equation}
with $\ell'_{12} \mapsto \ell'^\perp_{12}=0$ and $C \mapsto C^\perp=0$.

The $D_{12}$ roots are $v_I =\ell^\perp_I - \ell^\perp_{I+1}$, for $I =
1,\ldots,11$ and $v_{12} =\ell^\perp_{11}$.  In terms of the sections $\ell_I$
and $\ell'_I$,
\begin{equation}\label{eq:rootsl}
  \begin{split}
     v_I    &= \ell_I - \ell_{I+1}
             = \ell'_{I+1} - \ell'_I,\quad I = 1,\ldots, 10,\\
     v_{11} &= \ell_{11} - \ell_{12}  + C,\\
     v_{12} &= \ell_{11} - \ell'_{12} - C.
  \end{split}
\end{equation}
The roots $v_I$ generate a 12 dimensional sublattice of $\NS(S)$.  Their
intersection matrix is minus the Cartan matrix of $D_{12}$.


\subsection{Mordell-Weil torsion and connection to other CY manifolds}
\label{Sec:MWtorsionRelJac}

In the junction description, we have seen a number of examples in
Sec.~\ref{Sec:ExCoalescedFibs} with enhanced Mordell-Weil torsion from coalesced
singular fibers.  The simplest example is $\IZ_2$ torsion, which arises when 4
pairs of singular fibers of $\CX_{m,n}$ coalesce, to give 4 reducible singular
fibers (of topological type I$_2\times T^2$) and 4 elliptic curve of $A_1$
singularities of the threefold.

In the case of $\CX_{1,1}$, it should be possible to reproduce the results of
Sec.~\ref{Sec:ExCoalescedFibs} in the present description in terms of the
relative Jacobian of $S$.  In this construction, the surface $S$ is a branched
double cover of $\IP^1_{s,t}\times\IP^1_{u,v}$, so the singularity structure of
$S$, and hence of $\CX$, is completely determined by the degree $(6,2)$ branch
curve $B$.  Coalesced fibers, in the language of Sec.~\ref{Sec:ExCoalescedFibs},
arise when $B$ becomes singular in some way.  For surfaces, the correspondence
between singularities of a double cover and those of the branch curve is well
understood.  (See, for example, Ref.~\cite{Barth} Sec.~III.7.)  One way that
$B$ can develop singularities, is if it is reducible, i.e., if its defining
polynomial factorizes.  Then singularities arise from intersections of different
components of $B$.  This is not the only way in which $B$ can develop
singularities, but it will be sufficient for our purposes.

\begin{figure}[ht!]
  \def\BranchCurveZII{\mbox{\includegraphics{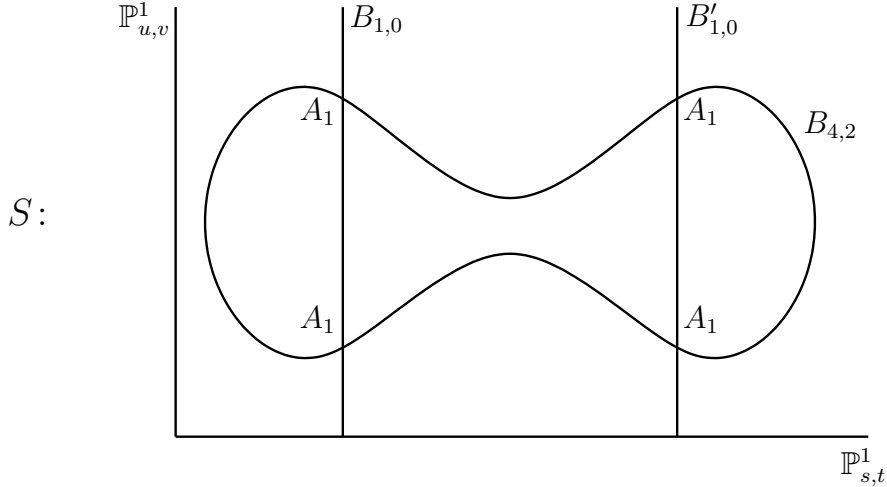}}}
  \begin{equation*}
    \begin{xy}
      \xyimport(3.8,2.4){\BranchCurveZII}
      (-0.15,2.30)*{\IP^1_{u,v}} ; (1.11,2.30)*{B_{1,0}} ; (2.93,2.30)*{B'_{1,0}} ;
      (3.75,-0.15)*{\IP^1_{s,t}} ; (3.57,1.72)*{B_{4,2}} ;
      (0.79,0.66)*{A_1} ; (0.79,1.80)*{A_1} ; 
      (2.87,0.66)*{A_1} ; (2.87,1.80)*{A_1} ; 
      (-0.75,1.25)*{\text{\Large $S\colon$}} ;
    \end{xy}
  \end{equation*}
  \caption{The surface $S$ is a double cover of $\IP^1_{s,t}\times\IP^1_{u,v}$
    branched over $B$.  When the branch curve $B$ factorizes into a degree (4,2)
    curve $B_{4,2}$ (of genus 3) and two (1,0) curves $B_{1,0}$ and $B'_{1,0}$
    (of genus 0), the difference $B_{1,0}-B'_{1,0}$ is a $\IZ_2$ torsion section
    of $S$.}
  \label{fig:Z2CartoonS}
\end{figure}
To reproduce the simplest case of $\IZ_2$ Mordell-Weil torsion, consider the
factorization $(6,2)=(4,2)+(1,0)+(1,0)$, so that the branch curve $B$ has
irreducible components $B_{4,2}$, $B_{1,0}$ and $B'_{1,0}$, with the degree of
each curve in $\IP^1_{s,t}\times\IP^1_{u,v}$ given by its subscript.  Then $B$
has an $A_1$ rational double point at each of the four points of intersection of
the components of $B$ (cf.~Fig.~\ref{fig:Z2CartoonS}), and from
Ref.~\cite{Barth} Sec.~III.7, so does the surface $S$.  Note that the $(1,0)$
curves are sections of the genus-2 fibration $S\to\IP^1_{u,v}$.  Moreover,
$2B_{1,0}$ and $2B'_{1,0}$ are each double covers of $\IP^1_{u,v}$, so each is
homologous to $C'$, the genus-0 fiber of the fibrations $S\to\IP^1_{s,t}$,
modulo irrelevant vertical components.  It follows that the difference
$2\bigl(B'_{1,0}-B'_{1,0}\bigr)$ is trivial in the Mordell-Weil group, and
$B'_{1,0}-B'_{1,0}$ is a $\IZ_2$ torsion class.  Quotienting by this isometry
gives a new Calabi-Yau manifold with Hodge numbers $h^{1,1}=h^{1,2}=10$, trivial
Mordell-Weil torsion, and fundamental group $\pi_1=\IZ_2$.  The $\IZ_2$ action
on the (resolved) reducible fibers is fixed point free, and permutes the two
components.

To obtain $\IZ_2{}^{\oplus3}$ Mordell-Weil torsion, an analogous construction
goes through with the factorization $(6,2)\to (2,2)+(1,0)+(1,0)+(1,0)+(1,0)$.
In this case the branch curve $B$ consists of an elliptic curve $B_{2,2}$
(cf.~the genus formula of App.~\ref{App:Genera}), and rational curves
$B^i_{1,0}$, for $i=1,2,3,4$, each of which is a section of the genus-2
fibration $S\to\IP^1_{u,v}$.  Again, $2B^i_{1,0}=C'$ in homology, modulo
irrelevant vertical components on the resolution, so
$\MW_\tor=\IZ_2{}^{\oplus3}$, generated by the three linearly independent
differences of the $B^i_{1,0}$.  This is the algebro-geometric description of
the Mordell-Weil torsion in the second example of Sec.~\ref{Sec:Z4}.
\begin{figure}[ht]
  \def\BranchCurveZIIcubed{\mbox{\includegraphics{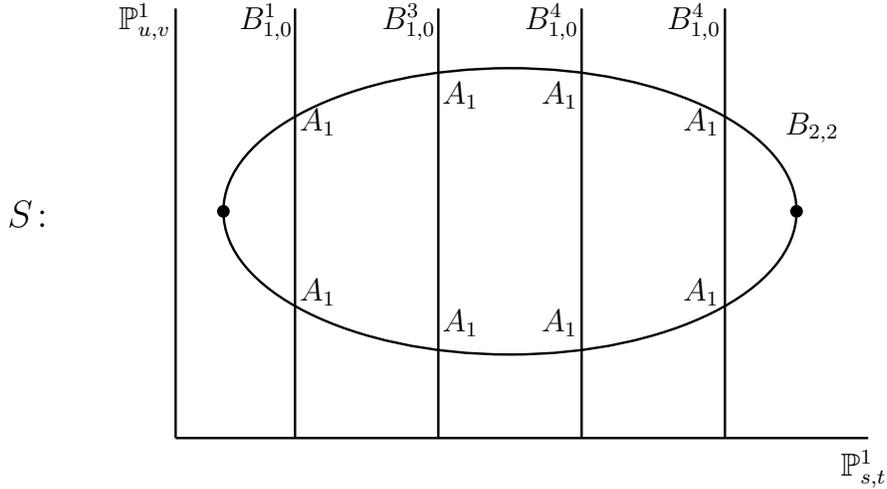}}}
  \begin{equation*}
    \begin{xy}
      \xyimport(3.8,2.4){\BranchCurveZIIcubed}
      (-0.15,2.30)*{\IP^1_{u,v}} ; 
      (0.51,2.30)*{B^1_{1,0}} ; (2.83,2.30)*{B^4_{1,0}} ;
      (1.2833,2.30)*{B^3_{1,0}} ; (2.0567,2.30)*{B^4_{1,0}} ;
      (3.75,-0.15)*{\IP^1_{s,t}} ; (3.47,1.72)*{B_{2,2}} ;
      (0.79,0.82)*{A_1} ; (0.79,1.75)*{A_1} ; 
      (1.5633,0.65)*{A_1} ; (1.5633,1.90)*{A_1} ; 
      (2.10,0.65)*{A_1} ; (2.10,1.90)*{A_1} ; 
      (2.87,0.82)*{A_1} ; (2.87,1.75)*{A_1} ; 
      (-0.75,1.25)*{\text{\Large $S\colon$}} ;
    \end{xy}
  \end{equation*}
  \caption{The surface $S$ is a double cover of $\IP^1_{s,t}\times\IP^1_{u,v}$
    branched over $B$.  When the branch curve $B$ factorizes into a degree (2,2)
    elliptic curve $B_{2,2}$ and four (1,0) rational curves $B^i_{1,0}$, the
    differences $B^i_{1,0}-B^j_{1,0}$ generate a $(\IZ_2)^{\oplus3}$ of torsion
    sections of $S$. Ramification points of the projection $B\to\IP^1_{s,t}$ are
    shown in bold.}
  \label{fig:Z2cubedCartoonS}
\end{figure}
\begin{figure}[ht!]
  \def\BranchCurveTacnode{\mbox{\includegraphics{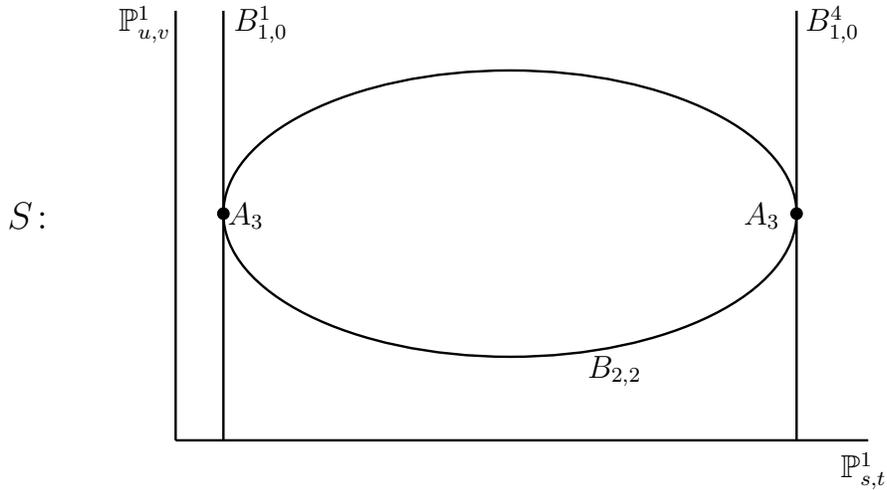}}}
  \begin{equation*}
    \begin{xy}
      \xyimport(3.8,2.4){\BranchCurveTacnode}
      (-0.15,2.30)*{\IP^1_{u,v}} ; 
      (0.48,2.30)*{B^1_{1,0}} ; (3.58,2.30)*{B^4_{1,0}} ;
      (3.75,-0.15)*{\IP^1_{s,t}} ; (2.4,0.40)*{B_{2,2}} ;
      (0.4,1.25)*{A_3} ; (3.2,1.25)*{A_3} ; 
      (-0.75,1.25)*{\text{\Large $S\colon$}} ;
    \end{xy}
  \end{equation*}
  \caption{When the rational curve $B^i_{1,0}$ of Fig.~\ref{fig:Z2cubedCartoonS}
    is placed at one of the four ramifications points of
    $B_{2,2}\to\IP^1_{s,t}$, the resulting tacnode of the branch curve yields an
    $A_3$ singularity of the surface $S$.  There are four such ramification
    points, but only two are visible in the schematic diagram above.}
  \label{fig:TacnodeCartoonS}
\end{figure}

One can ask whether a further enhancement of $\IZ_2{}^{\oplus3}$ to
$\IZ_4\oplus\IZ_2{}^{\oplus2}$ is possible, as in Sec.~\ref{Sec:Z4}, by
coalescing pairs of $A_1$ singularities of $S$ into $A_3$ singularities.  We can
indeed obtain this singularity structure by moving each $B^i_{1,0}$ to a
ramification points of the elliptic curve $B_{2,2}$, relative to the projection
$B_{2,2}\to\IP^1_{s,t}$.\footnote{Only two such points are visible in the
  cartoon of Fig.~\ref{fig:TacnodeCartoonS}, but this is a deficiency of the
  cartoon.  We know that there are really four ramification points, exactly the
  number necessary for one $B^i_{1,0}$ to be tangent at each. (Recall that an
  elliptic curve (genus 1) is the double cover of $\IP^1$ with $2g+2=4$ branch
  points.)}  Each intersection point is then a tacnode of $B$, which from
Ref.~\cite{Barth} Sec.~III.7 gives an $A_3$ singularity of $S$.\footnote{To
  apply Ref.~\cite{Barth}, we need to use the fact that the proper transform of
  a tacnode is as given in the first of the diagrams listed in Sec.~III.7.}
While this picture is tantalizing, and we have succeeded in reproducing the
desired singularity structure, we have not been able to identify a $\IZ_4$
isometry of the resulting reducible fibers of $S$.  We leave the task of
reproducing the $\IZ_4\oplus\IZ_2{}^{\oplus2}$ Mordell-Weil torsion group from
the relative Jacobian construction as an open problem.

Finally, consider the remaining factorizations of the $(6,2)$ branch curve into
a $(*,2)$ component and some number of (1,0) components:
\begin{equation}
  (5,2) + (1,0),\quad (3,2) + (1,0)^{\oplus3},
  \quad\text{and}\quad (1,2) + (1,0)^{\oplus5}.
\end{equation}
The last two factorizations give $\IZ_2{}^{\oplus2}$ and $\IZ_2{}^{\oplus4}$
Mordell-Weil torsion, respectively, as obtained in Sec.~\ref{Sec:Z2Zm} for
$m=1$.  The first factorization gives two $A_1$ singularities of $S$ and no
Mordell-Weil torsion; this is exactly the result of continuing
Eq.~\eqref{eq:UncoalesceChain} one step further, so that the only coalesced
fibers are $(\BB_1^2)$ and $(\BD_1^2)$.


\newsec{Conclusions and future directions}
\label{Sec:Conclusions}

We have seen two explicit constructions of the type IIA Calabi-Yau duals
$\CX_{m,n}$ of the type IIB $T^6/\IZ_2$ orientifold:

\begin{enumerate}
\item A monodromy and junction based description.

\item An algebro-geometric description as the relative Jacobian of a genus-2
  fibered surface~S (in the principally polarized case, $m,n = 1,1$).
\end{enumerate}

In each case, we have computed the Mordell-Weil lattice of sections, to obtain
the required $D_M$ lattice.  From a mathematical standpoint, we have shown that
the junction description provides an efficient algorithm for computing the
lattice of rational sections of an abelian surface fibration in terms of tree
graphs on the base---a generalization of F-theory string junction technology to
$T^4$ fibrations.  For the relative Jacobian construction, we have checked that
all criteria are satisfied for the application of Wall's theorem---which
classifies the threefold up to homotopy type (Hodge numbers, second Chern class,
intersection numbers)---and have reproduced the $D_{12}$ Mordell-Weil lattice
from this perspective.

\medskip
\noindent Applications to ongoing and future work are as follows:
\medskip

\emph{D-brane instantons.}  Having achieved the explicit constructions, the
rational curves of the $\CX_{m,n}$ are now well understood, which lays the
groundwork for the computation of worldsheet instanton corrections to the
$\CN=2$ prepotential.  Worldsheet instantons wrapping $\IP^1$ sections of the
$T^4$ fibration are dual to Euclidean D3-instantons in the type IIB $T^6/\IZ_2$
orientifold (and to D2-instantons in an intermediate type IIA dual).  Therefore,
they provide a duality check~\cite{DInstanton} of the modified rules for
D-brane instanton corrections in warped compactification due to
flux~\cite{AmirD3} and brane intersections~\cite{Ganor}---in similar spirit to
Ref.~\cite{CamaraDudas}.

\emph{Warped KK reduction.}  The clasical supergravity description of
$T^6/\IZ_2$ with $\CN=2$ flux is \emph{exactly} dual to the description of a
type IIA Calabi-Yau compactification in terms of an explicit, first order
approximation to the Calai-Yau metric~\eqref{eq:ApproxMetric} with known
harmonic forms~\cite{CYDuals}.  The low lying massive modes for large Calabi-Yau
base are also known.  Therefore, the known procedure for ordinary Kaluza-Klein
reduction to 4D can be re-expressed in terms of the dual variables to deduce
warped KK reduction for $T^6/\IZ_2$~\cite{WarpedKK}.  A simpler warm-up problem
applies a similar duality to deduce the warped KK reduction ansatz for the type
IIA $T^3/\IZ_2$ orientifold from the standard compactification of M-theory on a
K3 surface~\cite{WarpedKK}.

\emph{Extended SUSY breaking by topology.}  In any type II Calabi-Yau
compactification, extended supersymmetry is broken to 4D $\CN=2$ at the
compactification scale by the Calabi-Yau geometry.  In $T^6/\IZ_2$, the
quantized flux spontaneously breaks $\CN=4$ supersymmetry, at a scale
hierarchically lower than the compactification scale for large volume.  In the
dual type IIA compactification on $\CX_{m,n}$, this gives a precise sense in
which the Calabi-Yau \emph{topology} spontaneously breaks $\CN=4$ to $\CN=2$ for
large $\IP^1$ base.  The Calabi-Yau compactification, with $SU(3)$ Levi-Civita
holonomy, can be viewed as a $SU(2)$ structure compactification, the formalism
for which was worked out in Ref.~\cite{Spanjaard} (and subsequent work by the
Hamburg group, to appear).  Applying this $SU(2)$ structure formalism to the
compactification on the approximate first order metric of $\CX_{m,n}$ is an
essential step of work described in the previous paragraph, but is interesting
in its own right as a concrete example of spontaneous supersymmetry breaking by
topology.

\emph{Heterotic model building on new non simply connected manifolds.}  The
Calabi-Yau duals $\CX_{m,n}$ have $\pi_1=\IZ_n\times\IZ_2$ for $n=1,2,3,4$.
Moreover, at special points in moduli space, these Calabi-Yau manifolds develop
enhanced isometry groups (cf.~Secs.~\ref{Sec:ExCoalescedFibs}
and~\ref{Sec:MWtorsionRelJac}), the quotients by which yield new Calabi-Yau
manifolds with other fundamental groups.  Since few Calabi-Yau manifolds with
nontrivial $\pi_1$ are known~\cite{GrossPheno}, these constructions are
mathematically interesting.\footnote{From a mathematical standpoint these
  manifolds are also interesting in that Calabi-Yau manifolds with $T^4$
  fibrations do not arise as hypersurfaces in 4D toric varieties, so only a few
  examples exist compared a much larger class of known K3 fibrations.}  In
addition, they furnish a new class of compactification manifolds for Wilson line
breaking of GUT groups in heterotic models~\cite{DonagiSaito,SaitoTalk}.

\medskip
\noindent Finally, let us point out two additional connections to recent work:
\medskip

\emph{D(imensional) duality.}  Ref.~\cite{Dduality} considered compactifications
that interpolated between a compactification in the critical dimension on a
Riemann surface $C$ of genus $g$ and a supercritical compactification on its
Jacobian torus $\CJ_C$ with a timelike linear dilaton.  One might expect similar
compactifications to exist, connecting a subcritical compactification on $C$ to
an asymptotic region with compactified on $\CJ_C$ with an asymptotically
constant linear dilaton.  Such compactifications (fibered over a $\IP^1$ base)
provide a context in which not only Calabi-Yau $\CX_{1,1}$, but also the
auxilliary surface $S$ of Sec.~\ref{Sec:AlgebraicConstruction}, is part of the
physical compactification geometry.

\emph{T-fold compactifications.}  T-fold compactifications are nongeometric
compactifications similar to compactifications on $T^n$ fibrations, except that
the transition functions lie in the T-duality group $O(n,n;\IZ)$ rather than its
geometric subgroup $GL(n;\IZ)$~\cite{Hull}.  For $n=3$, the component of the
T-duality group continuously connected to the identity is $SO(3,3;\IZ)_+\cong
SL(4;\IZ)$.  So, in this case an $n=3$ T-fold encodes the same data as $T^4$
fibration over the same base~\cite{McGreevy}.  Thus, the collections of
$SL(4,\IZ)$ monodromies defining the Calabi-Yau manifolds $\CX_{m,n}$ in
Sec.~\ref{Sec:DualityMap} can alternatively be taken to define T-fold
compactifications.  In fact, we have already seen this as part of the duality
map described in Sec.~\ref{Sec:DualityMap}.  The connection between the $T^4$
fibration and $n=3$ T-fold compactification is that \Mtheory\ on the former is
type IIA compactified on the latter.  Thus, starting from M-theory on
$\CX_{m,n}$, For one choice of \Mtheory\ circle, we recover a type IIA T-fold
compactification that happens to be purely geometric---the intermediate D6/O6
orientifold with flux in the duality chain~2 of Sec.~\ref{Sec:DualityMap}.  For
other choices of the M-theory circle, the T-fold is expected to be nongeometric.


\bigskip\centerline{\bf Acknowledgements}\nobreak\medskip\nobreak

We are grateful to P.~Argyres, K.~Becker, V.~Bouchard, V.~Braun, O.~DeWolfe,
D.~Freed, K.~Hori, M.-H.~Saito and G.~Segal for conversations.  The research of
R.D.  is supported by NSF grants DMS 0612992 and Research and Training Grant DMS
0636606.  The work of P.G.  and M.S.  is supported in part by the DOE under
contract DE-FG02-95ER40893, the National Science Foundation under Grant
No.~PHY99-07949, the National Science and Engineering Council (NSERC) of Canada,
and by a start-up grant at Bryn Mawr College.  P.G. thanks ETH Zurich, the CERN
TH Division, Perimeter Institute, and the University of British Columbia for
hospitality.  M.S. thanks the University of Toronto, the Aspen Center for
Physics, and the CERN TH divison for hospitality during the course of this work,
as well as the University of Pennsylvania for continued hospitality.
\pagebreak

\appendix


\section{Braiding operations and monodromy matrices}
\label{App:Braiding}


\subsection{Elliptic fiber}

As described at the end of Sec.~\ref{Sec:MonodBraid}, when a $(p,q)$ 7-brane is
transported across the branch cut of another 7-branes, its $(p,q)$ type changes.
The reason is simple: Consider a $(p,q)$ string ending on a $(p,q)$ 7-brane, and
transport the 7-brane through a branch cut.  From Eq.~\eqref{eq:pqMonodromy}, we
know that the $(p,q)$ charge of the string transforms when it crosses the branch
cut.  Therefore, the charge of 7-brane on which it ends must transform in the
same way.  We will use the symbol $\cong$ to denote equivalences under such
$\emph{braiding operations}$, i.e., under brane motions of type IIB string
theory or motions of singular fibers of elliptic fibrations. The basic relation
is
\begin{equation}
  \BX_{\Bz_1} \BX_{\Bz_1} \cong \BX_{\Bz_2}\BX_{\Bz_1'},
\end{equation}
via motion of the brane or singular fiber $\BX_{\Bz_1}$ counterclockwise
through the branch cut of $\BX_{\Bz_2}$.  Here, $\Bz_i = \binom{p_i}{q_i}$, and
\begin{equation}
  \Bz_1' = K_{[p_1,q_1]}\Bz_1.
\end{equation}
The transformation of the monodromy matrix follows by writing
\begin{equation}
  K_{\Bz_2}K_{\Bz_1} = K_{\Bz_1'}K_{\Bz_2},
\end{equation}
from which we deduce that
\begin{equation}
  K_{\Bz_1'} = K_{\Bz_2}K_{\Bz_1}K_{\Bz_2}^{-1}.
\end{equation}
The monodromy matrix of $\BX_{\Bz_1}$ is conjugated by that of the branch cut it
crosses.

As examples of braiding, we now explain how to realize the $SO(2N)$ enhancement
from $N<4$ D7-branes at an O7-plane, in terms of nonperturbative description of
the type IIB $T^2/\IZ_2$ orientifold, with 24 $(p,q)$ 7-branes (F-theory on K3,
with 24 I$_1$ fibers).  It is not immediately apparent how the enhancement
occurs in the nonperturbative description, since the collection $\BA^N\BB\BC$
cannot be coalesced for $N<4$.  We provide explicit brane motions that make the
$D_1 \cong A_1$, $D_2\cong A_1\oplus A_1$ and $D_3\cong A_3$ lattices manifest
on subcollections of $(p,q)$ 7-branes.

In addition to $\BA$, $\BB$ and $\BC$ defined in Sec.~\ref{Sec:MonodBraid}, it
is also convenient below to define $\BD = \BX_{0,1}$ and $\BE=\BX_{1,2}$, with
monodromy matrices
\begin{equation}
  K_\BD = \begin{pmatrix}
    1 & 0\\
    1 & 1
    \end{pmatrix}
  \quad\text{and}\quad
  K_\BE = \begin{pmatrix}
    3 & -1\\
    4 & -1
    \end{pmatrix},
\end{equation}
respectively.  Note the following useful braiding relations,
\begin{equation}
  \begin{split}
    \BD\BA &\cong \BA\BB \cong \BB\BD,\\
    \BD\BC &\cong \BC\BA \cong \BA\BD,
  \end{split}\quad
  \begin{split}
    \BC\BD &\cong \BD\BE,\\
    \BB\BC &\cong \BC\BX_{3,1},
  \end{split}
\end{equation}
which can be combined to give
\begin{equation}
  \BC\BA^2 \cong \BA^2\BB\quad\text{and}\quad \BA\BB\BC\cong \BB\BC\BA.
\end{equation}
Most of these relations are used below.

For $N=1$, we have
\begin{equation}
  \BA\BB\BC\cong \BB\BC\BA\cong \BB\BA\BD\cong \BA\BD^2\cong \BC^2\BA.
\end{equation}
Coalescing the $\BD^2$ or $\BC^2$ in the last two expressions gives $A_1\cong
D_1$ (I$_2$ reducible fiber).

For $N=2$, we have
\begin{equation}
  \BA^2\BB\BC\cong \BB\BC\BA^2\cong \BB\BA^2\BB\cong \BB^2\BD^2.
\end{equation}
Coalescing the $\BB^2$ and $\BD^2$ gives $A_1\oplus A_1\cong D_2$ (I$_2$ fiber at
two different points on $\IP^1$).

For $N=3$, we have
\begin{equation}
  \begin{split}
    \BA^3\BB\BC &\cong \BA\BB\BC\BA^2\cong \BA\BB\BA^2\BB\cong \BA\BB^2\BD^2\\
    &\cong \BD\BA\BB\BD^2\cong \BD\BB\BD^3\cong \BD^4\BE.
  \end{split}
\end{equation}
Coalescing the $\BD^4$ gives $A_3\cong D_3$ (I$_4$ reducible fiber).

For $N\ge 4$, $\BA^N \BB\BC$ can be coalesced to give a $D_N$ singularity
(I$^*_{N-4}$ fiber) of the elliptic surface.


\subsection{Abelian surface fiber}

For the abelian surface fibrations $\CX_{m,n}$ studied in this paper, we
similarly define $\BX_{[p,q],i}$ fibers by the monodromy matrices
\begin{equation}
  \begin{split}
    K_{[p,q],i} &= T_i K_{[p,q],2}T_i^{-1}\quad\text{for $i=1,3,4$,}\\
    K_{[p,q],2} &= K_{[p,q]}\oplus I_{2\times2}
    = \begin{pmatrix}
      K_{p,q} &\\ & I_{2\times2}
    \end{pmatrix},\\
  \end{split}
\end{equation}
where the matrices $T_i(m,n)$ are given in Eq.~\eqref{eq:AbelTMatrices} below.
The vanishing cycles of the $\BX_{[p,q],i}$ fibers are of the form $(p,q,*,*)$.
Explicitly, for $i=2$ the vanishing cycle is $(p,q,0,0)$, while for $i\ne2$ it
is $(p,q,0,0)\,T_i^{\rm T}$.  Defining $\BA_i$, $\BB_i$, $\BC_i$, $\BD_i$ and
$\BE_i$ in this way, we find that $K_{\BA_i}$ is independent of $i$, so we drop
the subscript.  Then, by similarity transformation (using $T_i$), the braiding
relations for elliptic fibrations in the previous subsection all remain valid
for abelian surface fibrations, provided that fixed $i$ is used throughout each
relation.


\section{Complex tori, abelian varieties, and the Mordell-Weil lattice}
\label{App:AbelVar}

First, note the following definitions \cite{GandH,MWJunction}:

\begin{itemize}
  \item {\it Complex torus.} Starting from the vector space $\IC^g$, and a
    discrete lattice $\L\subset C^g$ of maximal rank $2g$, we define a complex
    torus by the quotient $T^{2g} = \IC^g/\L$.  For example, a complex $T^2$ is
    the quotient of $\IC$ by $\L = \IZ+\t\IZ$.

\item {\it Group law.} Addition of vectors endows $\IC^g$ with the structure of
  an abelian group, such that $\L$ is a subgroup, and $T^{2g}$ a quotient group.
  The group law $T^{2g}\times T^{2g}\to T^{2g}$ is just addition of points
  modulo $\L$.

\item {\it Abelian variety.} A complex torus is called an abelian variety if it
  admits an embedding in projective space.  For a $T^2$, this embedding is
  always possible and the abelian variety is called an elliptic curve.

\item {\it Weierstrass model.}  An elliptic curve $E$ over the complex numbers
  is given by the Weierstrass model,
  \[ z y^2 = 4x^3 - g_2 x z^2 - g_3 z^3,\qquad g_2,g_3,\in\IC, \]
  where $[x,y,z]\in\IP^2$.  In this case, the explicit map from points on the
  complex torus $t\in \IC^2/\bigl(\IZ+\t\IZ\bigr)$ to points on the elliptic
  curve is $[x,y,z] = [\wp(t),\wp'(t),1]$, where $\wp$ is the Weierstrass
  $\wp$-function.  The complex modulus $\tau$ is determined by the Weierstass
  $j$-function.

\item {\it Rational points.} When the coeffients of the defining polynomials of
  an abelian variety $A$ over $\IC$ are rational numbers, we can consider the
  subgroup $A(\IQ)$ of rational points on $A$---those points whose projective
  coordinates are also rational.  (For example if $g_2,g_3\in\IQ$ in the
  Weierstrass model, the rational points are solutions with $x,y\in\IQ$.)

\item {\it Mordell's theorem.}  The group of rational points of an abelian
  variety is known as the Mordell-Weil group.  Mordell's theorem states that
  this group is finitely generated:
  \[ A(\IQ) = \IZ^{\oplus r}\oplus A(\IQ)_\tor,\quad
  \text{where}\quad A(\IQ)_\tor = \oplus_{i=1}^{\dim A}\,\IZ_{m_i},\]
  for positive integers $m_i$.

\item {\it Mordell-Weil lattice.}  An abelian variety $A$ over $\IC$ together
  with a \emph{symmetric divisor}, determines a canonical \emph{height function}
  on the points of $A$, and a corresponding inner product.  This inner product
  gives $A(\IQ)/A(\IQ)_\tor$ the structure of a lattice, known as the
  Mordell-Weil lattice.  (For further details, including the definitions of
  height function and symmetric divisor, which will not concern us here, see
  Ref.~\cite{HeightPairing}.)

\end{itemize}
We can also work over a general number field $k$.\footnote{A \emph{number field}
  is a finite extension of $\IQ$.}  In this case, an abelian variety is a
defined to be a projective variety---the simultaneous solution to polynomial
equations in $k$-projective space---together with the additional structure of an
abelian group.  In the discussion of complex tori, we obtain one such
interpretation for each embedding of $k$ in $\IC$.  The remaining definitions
can be carried over to the more general case, and Mordell's theorem remains
valid.

One reason to consider this generalization is that it allows us to consider an
abelian fibration
\[\pi\colon \CX\to \CB\text{ with abelian fiber }A \]
to be an abelian variety in its own right:

\begin{itemize}
  
  \item {\it Rational functions.}  Given a field $k$, let $K[x]$ denote the ring of
    polynomial functions of $x$ with coefficients in $k$. The field of rational
    functions $K(x)$ consists of functions that are ratios of polynomials in
    $K[x]$.

  \item {\it Abelian fibration over $\CB$.}  Consider an abelian variety over
    the function field $K=k(\CB)$: the field of rational functions on the base
    $\CB$.  The defining polynomials are functions on $\CB$.  By evaluating
    these functions, we obtain an abelian variety over $k$ at each point of
    $\CB$, i.e., an abelian fibration.  A single rational point over $K$ gives a
    rational point of the fiber over every point of the base, i.e., a section.
    The Mordell-Weil group is thus the group of rational sections of the
    fibration modulo constant sections.
\end{itemize}
For example, in the Weierstrass model, if we take $g_2(w),g_3(w)\in\IC(w)$ to be
rational functions on $\IP^1$, then the rational points are those solutions
$[x(w),y(w),1]$ such that $x$ and $y$ are also rational functions on $\IP^1$.
For each point $w\in\IP^1$, the Weierstrass model gives a single elliptic curve
over $\IC$.  Each rational point $[x(w),y(w),1]$ gives a section: a map from
$\IP^1$ to the elliptic fibration.  For this reason, we generally use the
terminology \emph{rational section} rather than rational point.

Given a complex torus $\IC^g/\L\cong T^{2g}$, it is natural to ask under what
conditions the torus is an abelian variety, i.e., can be embedded in complex
projective space.  This leads to the following definitions:

\begin{itemize}

\item \emph{Hodge form.}  A complex torus is an abelian variety if there exists
  a Hodge form $\o$: a closed positive form of type (1,1) representing an
  integer cohomology class.  (See Ref.~\cite{GandH}, p.~302.).

\item \emph{Polarization.}  When such a form exists, the cohomology class $[\o]$
  is called a polarization, and can be represented by an invariant form
\begin{equation}
  \o = \sum_{i=1}^g\delta_idx^i\w dy^i
  \quad\text{with}\quad \delta_i \bigm| \delta_{i+1},
\end{equation}
in terms of coordinates $x^i\cong x^i+1$ and $y^i\cong y^i+1$ dual to a suitably
chosen integer basis of the lattice $\L$.  The positive integers $\d_i$ are
called the \emph{elementary divisors of the polarization}.  The class $[\o]$ is
a \emph{principal polarization} if $\d_i=1$ for all $i$~\cite{GandH}.
\end{itemize}


\section{Mordell-Weil height pairing from intersections}
\label{App:MWfromInt}

As mentioned in the previous section, the Mordell-Weil group modulo torsion
$\MW/\MW_\tor$ of an abelian fibration $\CX$ can be given the structure of a
lattice.  The lattice inner product is determined by the homological
intersection pairing on a surface in $\CX$.  However, this first requires a map
from $\MW(\CX)$ to $H_2(\CX)$.  Here, we describe the map and resulting inner
product, including a slight subtlety regarding the integrality of the map when
$\CX$ has reducible fibers (related to the weakly integral junctions of
Sec.~\ref{Sec:WeaklyIntegral}).


\subsection{Elliptic fibration over a curve}

The theory of the Mordell-Weil lattice of an elliptic surface $\pi\colon\CX\to
\CB$ was given by Shioda in Refs.~\cite{ShiodaI,ShiodaII}.  We follow
Ref.~\cite{ShiodaII} closely.  Let $P$ and $Q$ denote rational sections of the
elliptic fibration, $O$ the zero section, and $[P]$, $[Q]$, $[O]$ the
corresponding divisor classes in $\NS(\CX)$.  Let $F$ denote the class of the
generic fiber.  We assume that there is at least one singular fiber.

\nonnumberedsubsec{Irreducible fibers}

In the case that the singular fibers are all irreducible, the inner product
$\langle P,Q\rangle$ on the Mordell-Weil lattice of $\CX$ is simply the
intersection pairing
\begin{equation}\label{eq:MWinnerprod}
  \langle P,Q\rangle = [P]^\perp\cdot [Q]^\perp,
\end{equation}
where $[P]^\perp = [P] - a[O]- bF\in H_2(S,\IZ)$, with $a$ and $b$ chosen so
that $[P]^\perp\cdot[O] = [P]^\perp\cdot F = 0$.  We can compute $a$ and $b$
explicitly using
\begin{equation}
  \s^2 = -\chi,\quad \s\cdot F = 1,\quad\text{and}\quad F^2 = 0. 
\end{equation}
Here, $\s$ is the class of any section and $\chi$ is the arithmetic
genus\footnote{The arithmetic genus is defined by $\chi(\CX) = h^{2,0}-h^{1,0}+
  h^{0,0}$.  For $\CX$ a dP$_9$, this gives $\chi=1$ and for a K3 surface,
  $\chi=2$.}  of $\CX$.  The result is $a=1$ and $b= [P]\cdot[O] + \chi$.  This
defines an embedding of the Mordell-Weil lattice,
\begin{equation}\label{eq:MWembeddingI}
  \varphi\colon \MW/\MW_\tor \hookrightarrow H_2(\CX,\IZ),
  \quad\text{such that}\quad
  P\mapsto [P]^\perp.
\end{equation}

\nonnumberedsubsec{Reducible fibers}

When the elliptic fibration has reducible fibers, a similar story holds.
However, in this case, the best we can do is to embed the Mordell-Weil lattice
in $H_2(\CX,\IQ)$.  Let $F_p = \pi^{-1}(p)$ denote the fiber over a point
$p\in\CB$, and
\begin{equation}
  R = \{p\in\CB \mid F_p\text{ is reducible}\}.
\end{equation}
Then, for each $p\in\CB$, we can decompose $F_p$ into irreducible components
$F_{p,i}$.  For simplicity, we assume that each component appears with
multiplicity 1, so that
\begin{equation}
  F_p = \sum_{i=0}^{m_p-1}F_{p,i}.
\end{equation}
We label the components so that $F_{p,0}$ is the unique component that
intersects the zero section.  Let $T$ denote the subgroup of $\NS(\CX)$
generated by components of fibers and $[O]$.  Under the assumption that the
elliptic surface $\CX$ has at least one singular fiber, both $\NS(\CX)$ and $T$
are torsion-free.  The Mordell-Weil group is the quotient
\begin{equation}
  \MW(\CX) = \NS(\CX)/T.
\end{equation}

To define the Mordell-Weil lattice, we seek an embedding $\varphi\colon
\MW/\MW_\tor\hookrightarrow H_2(\CX,\IQ)$.  We can proceed as before, this time
requiring that $[P]^\perp$ in the inner product~\eqref{eq:MWinnerprod} lie in
the orthogonal complement of $[O]$, $F$, and all components $F_{p,i}$ of
reducible fibers.  The analog of Eq.~\eqref{eq:MWembeddingI} is
\begin{equation}\label{eq:MWembeddingII}
  \varphi\colon \MW/\MW_\tor \hookrightarrow H_2(\CX,\IQ),
  \quad\text{taking}\quad
  P\mapsto [P]^\perp,
\end{equation}
where
\begin{equation}
  [P]^\perp = [P] - [O] - ([P]\cdot[O] + \chi)F
  -\sum_{p\in R} \bigl(F_{p,1},\ldots,F_{p,m_{p-1}}\bigr)\, A_p^{-1}
  \begin{pmatrix}
    [P]\cdot F_{p,1}\\
    \vdots\\
    [P]\cdot F_{p,m_{p-1}}
  \end{pmatrix}.
\end{equation}
Here, $A_p$ is the intersection matrix of the extra ($i\ne0$) components of the
reducible fiber $F_p$,
\begin{equation}
  A_p = F_{p,i}\cdot F_{p,j},
  \quad\text{where}\quad 1\le i,j\le m_p-1.
\end{equation}
Since $A_p^{-1}$ is not integral, $[P]^\perp$ lies in $\NS(\CX)_\IQ =
\NS(\CX)\otimes\IQ$ but not in $\NS(\CX)$.

In fact, we can be more explicit about how close to integrality $\im\varphi$ is.
Define the \emph{essential sublattice of $\NS(\CX)$} to be $L = T^\perp$, and let
$\ell = \lcm\{m_p\mid p\in R\}$.  Then,
\begin{equation}
  \im\varphi \subset \frac{1}{\ell}L.
\end{equation}
This is the geometric analog of the weak integrality condition on string
junctions in Sec.~\ref{Sec:WeaklyIntegral}.

Shioda goes on to show that the narrow Mordell-Weil lattice is
\begin{equation}
  \MW_0\cong L,
\end{equation}
and that 
\begin{equation}\label{eq:MWvsLdual}
  \MW/\MW_\tor \subset L^*,
\end{equation}
where $L^*$ is the lattice dual to $L$:
\begin{equation}
  L^* = \{x\in L\otimes\IQ \mid \langle x,y\rangle\in\IZ\text{ for all }y\in L\}.
\end{equation}

\nonnumberedsubsec{The unimodular case}

When $NS(\CX)$ is unimodular, Eq.~\eqref{eq:MWvsLdual} becomes an equality.
This is the case, for example, for $\CX=\text{dP}_9$.  Using this result, Shioda
also gives a useful expression for the torsion subgroup of the Mordell-Weil
group.  It is
\begin{equation}
  \MW_\tor \cong T/T',
\end{equation}
where $T' = (T\otimes\IQ) \cap \NS(\CX)$.


\subsection{Abelian surface fibration over a curve}

The description of the Mordell-Weil lattice of an abelian surface fibration, or
higher dimensional abelian fibration, is very similar to that for an elliptic
fibration.  Here, we sketch the new ingredients necessary to define the height
pairing in the general case.

On a smooth compact $n$-dimensional complex manifold $\CX$, any class $c \in
H^2(\CX,\IR)$ gives a map
\begin{equation}
  H^a(\CX,\IR) \to H^{2n-a}(\CX,\IR),
\end{equation}
given by cup product with the $(n-a)$th power of $c$.  When $c$ is a K\"ahler
class, and in particular when $c$ is the Chern class $c_1(L)$ of an ample line
bundle $L$ on a smooth projective variety, the Hard Lefschetz theorem
(cf.~Ref.~\cite{GandH}, p.~122) says that this map is an isomorphism for $a$
between $0$ and $n$.  (The real coefficients can be replaced throughout by the
rationals, but the result fails over the integers.) We apply this in our case,
with $n=3$, $a=2$, where $L$ is an ample theta divisor.  This allows us to
identify $H_2(X)=H^4(X)$ with $H^2(X)$. The natural height pairing on the
Mordell-Weil lattice is determined in terms of this identification.


\section{Monodromy matrices for the abelian fibration $\CX_{m,n}$}
\label{App:Xmonodromy}

First let us fix conventions in the simpler elliptically fibered case.  In
Sec.~\ref{Sec:MonodBraid}, the monodromy matrices $K$ were defined to act on
vectors $\binom{p}{q}$, that is, on the components of the homology class ${\bf
  z} = p\alpha+q\beta$ relative to a basis $\{\alpha,\beta\}$ of $H_1(T^2,\IZ)$.
Let $y^i\cong y^i+1$ denote coordinates on the $T^2$.  Then, a convenient basis
for $H^1(T^2,\IZ)$ is $\{[dy^i]\}$, and a corresponding basis for $H_1(T^2,\IZ)$
is $\alpha = [S^1_1]$, $\beta=[S^1_2]$, where the circles $S^1_i$ are chosen so
that
\begin{equation}
  \int_{S^1_i} dy^i = \delta^i_j.
\end{equation}
In this basis, a class $[\omega]\in H^1(T^2,\IZ)$ can be represented by $\omega
= \omega_idy^i$ and a class $[{\bf z}]\in H_1(T^2,\IZ)$ by ${\bf z} = z^i
S^1_i$.  Then, $\binom{z^1}{z^2} = \binom{p}{q}$, and by a slight abuse of
notation we simply write ${\bf z} = \binom{p}{q}$ as in
Sec.~\ref{Sec:MonodBraid}.

When a branch cut is crossed in the positive (counterclockwise) direction about
the branch point, the components and basis elements transform as
\begin{subequations}\label{eq:Kaction}
\begin{align}
  z^i &\mapsto K^i{}_j z^j, & S^1_i &\mapsto S^1_j\bigl(K^{-1}\bigr)^j{}_i,\\
  \o_i &\mapsto \o_j \bigl(K^{-1}\bigr)^j{}_i, & dy^i &\mapsto K^i{}_j dy^j.
\end{align} 
\end{subequations}
These definitions all carry over to the case of $T^4$ fiber except that
$i=1,2,3,4$ and we write ${\bf z}^T = (p,q,r,s)$, by the same slight abuse of
notation.

\nonnumberedsubsec{Warm-up: $\CN=4$ case, ${\rm K3}\times T^2$}

Before describing the monodromy matrices for the abelian surface fibration
$\CX_{m,n}$, it is useful to consider the $\CN=4$ case.  In the absence of flux,
the IIA dual of the type IIB $T^6/\IZ_2$ orientifold is a compactification on
K3$\times T^2$, which we can think of as a $T^4$ fibration over $\IP^1$ in which
a $T^2\subset T^4$ trivially factorizes.  Let $y^1,y^2$ denote the coordinates
on the nontrivial $T^2$ fiber of K3, $y^3,y^4$ coordinates on the trivial $T^2$,
and $y^5,y^6$ coordinates on the $\IP^1$ base.\footnote{Compared to
  Ref.~\cite{CYDuals}, we have $(y^1,y^2,y^3,y^4,y^5,y^6)_\text{here} =
  (x^{10},-x^8,x^4,x^5,x^6,x^7)_\text{there}$.}  The monodromy matrices for K3
were given in Sec.~\ref{Sec:MonodBraid}.  The collection of singular fibers is
$\BA^{16}\BB\BC\BB\BC\BB\BC\BB\BC$, with matrices $K_\BA$, $K_\BB$ and $K_\BC$
given by Eq.~\eqref{eq:Kmonod}.  To obtain the corresponding monodromy matrices
on the $T^4$ fibration K3$\times T^2$, we simply tensor with the identity matrix
in the $y^3y^4$ block:
\begin{equation}\label{eq:KABC}
  K_\BA = \begin{pmatrix}
    1 & -1 & 0 & 0 \\
    0 & 1 & 0 & 0 \\
    0 & 0 & 1 & 0 \\
    0 & 0 & 0 & 1
  \end{pmatrix},\quad
  K_\BC = \begin{pmatrix}
    2 & -1 & 0 & 0 \\
    1 & 0 & 0 & 0 \\
    0 & 0 & 1 & 0 \\
    0 & 0 & 0 & 1
  \end{pmatrix},\quad
  K_\BB = \begin{pmatrix}
    0 & -1 & 0 & 0 \\
    1 & 2 & 0 & 0 \\
    0 & 0 & 1 & 0 \\
    0 & 0 & 0 & 1
  \end{pmatrix}.
\end{equation}
In the classical supergravity duality (cf.~end of Sec.~3.1), only the
combined monodromy of the pair $\BO=\BB\BC$ is visible, with monodromy
\begin{equation}\label{eq:KO}
  K_\BO=K_\BC K_\BB = \begin{pmatrix}
    -1 & 4 & 0 & 0 \\
    0 & -1 & 0 & 0 \\
    0 & 0 & 1 & 0 \\
    0 & 0 & 0 & 1
  \end{pmatrix}.
\end{equation}
The fact that $K_\BA$ and $K_\BO$ have zeros in the $2,3,4$ components of the
first column reflects the duality origin of K3$\times T^2$ in the \Mtheory\ lift
from the $T^3/\IZ_2\times T^3$ orientifold: The $y^1$ direction is the
\Mtheory\ circle, and in the perturbative lift, this circle is fibered over the
other directions.

\nonnumberedsubsec{$\CN=2$ case, $\CX_{m,n}$}

In the $\CN=2$ case, the manifold $\CX_{m,n}$ arises as follows.  We first
T-dualize the IIB $T^6/\IZ_2$ orientifold along a $T^3$ to obtain a IIA D6/O6
orientifold.  The orientifold is not quite $T^3/\IZ_2\times T^3$, since the IIB
NS flux dualizes to twists of the $T^3\times T^3$ topology.  Instead, the
orientifold is $\CY_n(y^2,y^3,y^4,y^5,y^6)/\IZ_2\times S^1$, where $\CY_n$ is an
$S^1_3\times S^1_4$ fibration over $T^3_{\{2,5,6\}}$.  The global
1-forms on $\CY_n$ are $dy^2$, $\eta^3$, $\eta^4$, $dy^5$, and $dy^6$, with
\begin{equation}
  d\eta^3 = 2n dy^2\w dy^5,\quad d\eta^4 = 2n dy^2\w dy^6.
\end{equation}
Up to a choice of coordinate gauge (equivalent to a gauge
choice for the NS $B$-field in $T^6/\IZ_2$), we can take
\begin{equation}\label{eq:IIAtwist}
  \eta^3 = dy^3 + 2n y^2 dy^5,\quad \eta^4 = dy^4 + 2n dy^2.
\end{equation}
The $\IZ_2$ involution is $(-1)^{F_L}\Omega\CI_3$, where $F_L$ is left moving
fermion number, $\Omega$ is worldsheet parity, and $\CI_3$ is the inversion
$\CI_3\colon (y^2,y^5,y^6)\mapsto -(y^2,y^5,y^6)$, which acts on the 1-forms as
\begin{equation}
  \CI^*_3\colon (dy^2,\eta^3,\eta^4,dy^5,dy^6)\mapsto (-dy^2,\eta^3,\eta^4,-dy^5,-dy^6).
\end{equation}

This type IIA orientifold lifts to \Mtheory\ on $\CX_{m,n}\times S^1$.
Compactifying on the $S^1$ factor then gives the type IIA Calabi-Yau
compactification on $\CX_{m,n}$.  In the classical supergravity description of
the lift, $\CX_{m,n}$ is obtained by fibering the \Mtheory\ circle over
$\CY_{m,n}$ and then quotienting by the $\IZ_2$ involution $\CI_4\colon
(\eta^1,y^2,y^5,y^6)\mapsto -(\eta^1,y^2,y^5,y^6)$.  Here, $\eta^1$ is the
1-form along the \Mtheory\ circle, and satisfies
\begin{equation}
  \begin{split}
    d\eta^1 &= F^\text{IIA orientifold}_{(2)}/\bigl(2\pi\sqrt{\a'}\bigr)\\ 
    &= -2m\eta^3\w dy^6 + 2m \eta^4\w dy^5 + (\text{warp factor dependence}).
  \end{split}
\end{equation}
This leading order description falls short of the exact description of
$\CX_{m,n}$, since it ignores KK modes around the M-theory circle, and breaks
the $U(1)$ isometry only by the explicit $\IZ_2$.  Nevertheless, it contains
sufficient information to parametrize the exact description.

In summary, at the level of this description, the steps to construct $\CX_{m,n}$
are
\begin{enumerate}
\item Fiber the $S^1_3$ and $S^1_4$ circles over $T^3_{\{2,5,6\}}$,

\item Fiber the $S^1_1$ circle over the resulting manifold $\CY_n$,

\item Quotient by $\CI_4$, which inverts the  $1,2,4,5$ directions. 
\end{enumerate}
To make the abelian surface fibration manifest, step~1 can be equivalently
described as
\begin{enumerate}
  \item[$1'$.] Fiber the torus $T^3_{\{2,3,4\}}$ over $T^2_{\{5,6\}}$,
\end{enumerate}
Then, the abelian surface fibration $\CX_{m,n}$ can be understood as the
$T^4_{\{1,2,3,4\}}$ fibration over $T^2_{\{5,6\}}$, quotiented by $\CI_4$.  The
base of the resulting $T^4$ fibration is $T^2_{\{5,6\}}/\IZ_2\cong \IP^1$.

The collection of singular fibers visible in this description is $\BA^M
\BO_1\BO_2\BO_3\BO_4$, where the the locations of the $\BO_i$ on $\IP^1\cong
T^2_{\{5,6\}}/\IZ_2$ are the $\IZ_2$ fixed points $p_i$, with $(y^5,y^6)$
coordinates
\begin{equation}
  p_1 = (1/2,0),\quad p_2 = (0,0),\quad p_3 = (0,1/2),\quad p_4 = (1/2,1/2).
\end{equation}
With the appropriate coordinate gauge choice for $y^1$, the monodromies $K_\BA$
and $K_{\BO_i}$ are the same as those for K3$\times T^2$.  That is, $K_{\BO_2} =
K_\BO$ of Eq.~\eqref{eq:KO}.  However, the remaining monodromies $K_{\BO_i}$
differ from $K_\BO$.  We deduce these monodromies as follows.

In the basis $\eta^1, dy^2, \eta^3, \eta^4$ (restricted to the $T^4$ fiber), the
monodromies $K_{\BO_i}$ at all four fixed points on the base are simply
\begin{equation}
  \begin{pmatrix}
    -1 & 0 & 0 & 0 \\
    0 & -1 & 0 & 0 \\
    0 & 0 & 1 & 0 \\
    0 & 0 & 0 & 1
  \end{pmatrix}
\end{equation}
from the $\IZ_2$ action on the fiber.  However, we seek the monodromies in the
coordinate basis $dy^1, dy^2, dy^3, dy^4$ instead.\footnote{Recall that the
  homology \emph{components} $z^i = (p,q,r,s)$ and cohomology \emph{basis}
  transform in the same way under monodromy transformations
  (cf.~Eq.~\eqref{eq:Kaction}).}  Since the fixed points $p_i$ are locally
equivalent to one another, the monodromy matrices $K_{\BO_i}$ must be related by
similarity transformation,
\begin{equation}\label{eq:AbelTMatrices}
  K_{\BO_i} = T_i K_{\BO_2} T_i^{-1},\quad
  \text{where}\quad S_i\in\SL(4,\IZ).
\end{equation}
To deduce the transformation matrices $T_i$ we first use the definition
\eqref{eq:IIAtwist} to determine the lower $3\times3$ block.  From
\begin{align*}
  \eta^3\big|_{y^5 = \ha} &= dy^3-ndy^2, & \eta^3\big|_{y^5=0} = dy^3,\\
  \eta^4\big|_{y^6 = \ha} &= dy^4-ndy^2, & \eta^4\big|_{y^6=0} = dy^4,
\end{align*}
we obtain the lower $3\times3$ blocks of the following matrices:
\begin{equation}\label{eq:simT}
  T_1 = \begin{pmatrix}
    1 & 0 & 0 & -m\\
    0 & 1 & 0 & 0\\
    0 & -n & 1 & 0\\
    0 & 0 & 0 & 1
    \end{pmatrix},\quad
 T_3 = \begin{pmatrix}
   1 & 0 & m & 0\\
   0 & 1 & 0 & 0\\
   0 & 0 & 1 & 0\\
   0 & -n & 0 & 1
 \end{pmatrix},\quad
T_4 = \begin{pmatrix}
  1 & 0 & m & -m\\
  0 & 1 & 0 & 0\\
  0 & -n & 1 & 0\\
  0 & -n & 0 & 1
\end{pmatrix},
\end{equation}
with $T_2$ equal to the $4\times4$ identity.  The zeros in the first column
follow from the fact that the \Mtheory\ circle is fibered over $\CY_{n}$ in our
construction.  The first row can be determined by a careful analysis of the
connection for the M-theory circle fibration, however, a simpler route is to
note that the similarity transformations $T_i$ must leave the Hodge form
\eqref{eq:Hodge} invariant.  This determines the first row except for the second
component.  Finally, this component is required to vanish so that the $T_i^{-1}$
are also integral and $T_i\in SL(4,\IZ)$.

Eqs.~\eqref{eq:KO} and~\eqref{eq:simT} together give
\begin{equation}\label{eq:KOi}
  \begin{split}
    K_{\BO_1} &= \begin{pmatrix}
      -1 & -4 & 0 & -2m\\
      0 & -1 & 0 & 0\\
      0 & 2n & 1 & 0\\
      0 & 0 & 0 & 1
    \end{pmatrix},\\
    \noalign{\vskip1ex}
    K_{\BO_3} &= \begin{pmatrix}
      -1 & -4 & 2m & 0\\
      0 & -1 & 0 & 0\\
      0 & 0 & 1 & 0\\
      0 & 2n & 0 & 1
    \end{pmatrix},
  \end{split}
  \qquad
  \begin{split}
    K_{\BO_2} &= \begin{pmatrix}
      -1 & -4 & 0 & 0\\
      0 & -1 & 0 & 0\\
      0 & 0 & 1 & 0\\
      0 & 0 & 0 & 1
    \end{pmatrix},\\
    \noalign{\vskip1ex}
    K_{\BO_4} &= \begin{pmatrix}
      -1 & -4 & 2m & -2m\\
      0 & -1 & 0 & 0\\
      0 & 2n & 1 & 0\\
      0 & 2n & 0 & 1
    \end{pmatrix}.
  \end{split}
\end{equation}
While the classical supergravity duality does not resolve $\BO_i = \BB_i\BC_i$
into its two constituents, we know that $K_{\BO_2} = K_{\BO}$ of
Eq.~\eqref{eq:KO} can be factored into the two monodromies $K_\BB$ and $K_\BC$
of Eq.~\eqref{eq:KABC}, each related to $K_\BA$ via similarity transformation.
Consequently, the $K_{\BO_i}$ factorize as
\begin{equation}
  K_{\BO_i} = K_{\BC_i} K_{\BB_i}, \quad\text{where}\quad
  K_{\BB_i} = T_i K_\BB T_i^{-1}\quad\text{and}\quad
  K_{\BC_i} = T_i K_\BC T_i^{-1}.
\end{equation}
This gives the matrices quoted in Sec.~\ref{Sec:CYmonodromy}.  The factorization is
unique up to braiding and overall $SL(4,\IZ)$ conjugation.


\section{Null loop junctions of $\CX_{m,n}$}
\label{App:NullLoop}

To determine the junction lattice vectors~\eqref{eq:XmnLoopLattice} of the null
loop junctions of $\CX_{m,n}$, we first transform the loop junctions to standard
presentation, and then read off the number of strings emanating from each $\BA$,
$\BB_i$ and $\BC_i$ point on $\IP^1$.  To transform to standard presentation, we
push the lower half of the loop through each branch point in succession, from
left to right, applying the Hanany-Witten effect at each step
(cf.~Fig.~\ref{fig:HananyWitten}).  Once this has been done, the original loop
is contractible to a point, leaving just the new Hanany-Witten strings
intersecting at this point.  The discontinuity in the $(p,q,r,s)$ charge of the
original segment of string across a branch cut of $\BX_i$ is equal to the charge
$\Bz_i$ of the new string \emph{grown} via the Hanany-Witten effect.
\begin{figure}[ht]
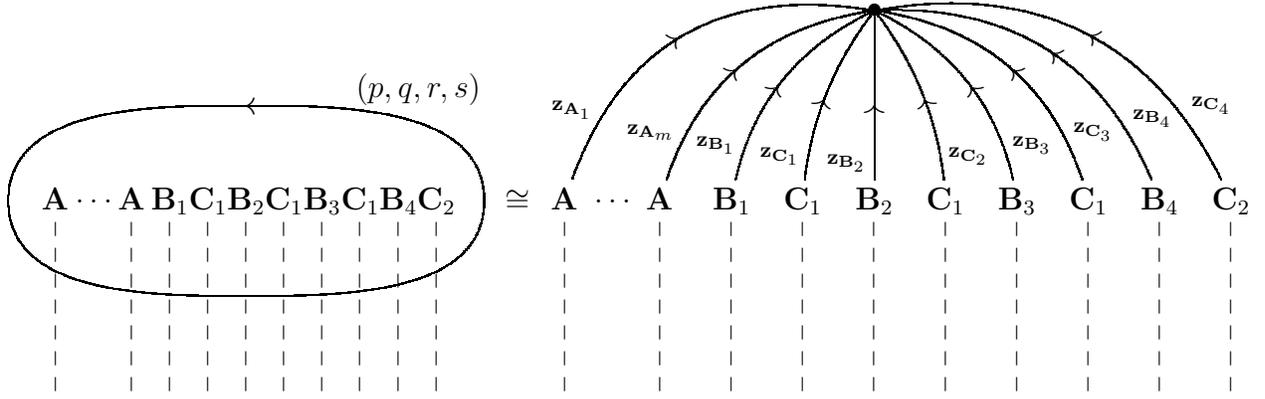

  \begin{equation*}
    \hskip-1.3em
    \xygraph{
      !~:{@{-}}
      !~-{@{--}}
      {\BA_{\vphantom1}}="A1"-[dd] [r(0.4)uu]{\dots}="dots"  [r(0.4)]
      {\BA_{\vphantom1}}="AM"-[dd] [r(0.4)uu] 
      {\BB_1}="B1"-[dd] [r(0.4)uu]{\BC_1}="C1"-[dd] [r(0.4)uu]
      {\BB_2}="B2"-[dd] [r(0.4)uu]{\BC_1}="C2"-[dd] [r(0.4)uu]
      {\BB_3}="B3"-[dd] [r(0.4)uu]{\BC_1}="C3"-[dd] [r(0.4)uu]
      {\BB_4}="B4"-[dd] [r(0.4)uu]{\BC_2}="C4"-[dd]
      ,"B2"[u] ="N"
      ,"A1"[l(0.5)] ="W"
      ,"B2"[d] ="S"
      ,"C4"[r(0.5)] ="E"
      ,"N":@(l,u)"W" :@(d,l)"S" :@(r,d)"E" :@{->}@(u,r)"N"_{\strut\displaystyle(p,q,r,s)}
      ,"E" [r(0.35)]{\cong} [r(0.5)]
      !~:{@{-}|@{>}}
      !~-{@{--}}
      {\BA_{\vphantom1}}="AA1"-[dd] [r(0.5)uu]{\dots}="dots"  [r(0.5)]
      {\BA_{\vphantom1}}="AAM"-[dd] [r(0.75)uu] 
      {\BB_1}="BB1"-[dd] [r(0.75)uu]{\BC_1}="CC1"-[dd] [r(0.75)uu]
      {\BB_2}="BB2"-[dd] [r(0.75)uu]{\BC_1}="CC2"-[dd] [r(0.75)uu]
      {\BB_3}="BB3"-[dd] [r(0.75)uu]{\BC_1}="CC3"-[dd] [r(0.75)uu]
      {\BB_4}="BB4"-[dd] [r(0.75)uu]{\BC_2}="CC4"-[dd]
      ,"BB2"[uu] ="Vertex"
      {\bullet}
      ,"AA1":@/^2.5pc/"Vertex"^(0.2){\Bz_{\BA_1}}
      ,"AAM":@/^1.5pc/"Vertex"^(0.2){\Bz_{\BA_m}}
      ,"BB1":@/^1.0pc/"Vertex"^(0.2){\Bz_{\BB_1}}
      ,"CC1":@/^0.5pc/"Vertex"^(0.2){\Bz_{\BC_1}}
      ,"BB2":"Vertex"^(0.2){\Bz_{\BB_2}}
      ,"CC2":@/_0.5pc/"Vertex"_(0.2){\Bz_{\BC_2}}
      ,"BB3":@/_1.0pc/"Vertex"_(0.2){\Bz_{\BB_3}}
      ,"CC3":@/_1.5pc/"Vertex"_(0.2){\Bz_{\BC_3}}
      ,"BB4":@/_2.0pc/"Vertex"_(0.2){\Bz_{\BB_4}}
      ,"CC4":@/_2.5pc/"Vertex"_(0.2){\Bz_{\BC_4}}
    }
  \end{equation*}
  \caption{A loop junction of $\CX_{m,n}$ transformed to standard presentation.}
  \label{fig:XmnLoopsAsTrees}
\end{figure}

Consider the $(p,q,r,s) = (1,0,0,0)$ loop.  After crossing each successive
branch cut in the counterclockwise direction, the new $(p,q,r,s)$ charge is
determined by multiplication by the monodromy matrices~\eqref{eq:XMonodI} and
\eqref{eq:XMonodII}.  For the loop on the LHS of Fig.~\ref{fig:XmnLoopsAsTrees},
this gives
{\small
\begin{equation}\label{eq:SuccessiveCharges}
  \begin{split}
    \BA\colon &\ \text{no change},\\
    \BB_1\colon &
    \begin{pmatrix}
      0 &-1 & 0 &-m\\
      1 & 2 & 0 & m\\
     -n &-n & 1 &-mn\\
      0 & 0 & 0 & 1
      \end{pmatrix}
    \begin{pmatrix}
      1 \\ 0 \\ 0\\ 0
    \end{pmatrix}
    =
    \begin{pmatrix}
      0 \\ 1 \\ -n\\ 0
    \end{pmatrix},\\
    \BC_1\colon &
    \begin{pmatrix}
      2 &-1 & 0 & m\\
      1 & 0 & 0 & m\\
     -n & n & 1 &-mn\\
      0 & 0 & 0 & 1
      \end{pmatrix}
    \begin{pmatrix}
      0 \\ 1 \\-n\\ 0
    \end{pmatrix}
    =
    \begin{pmatrix}
     -1 \\ 0 \\ 0 \\ 0
    \end{pmatrix},\\
    \BB_2\colon &
    \begin{pmatrix}
      0 &-1 & 0 & 0\\
      1 & 2 & 0 & 0\\
      0 & 0 & 1 & 0\\
      0 & 0 & 0 & 1
      \end{pmatrix}
    \begin{pmatrix}
     -1 \\ 0 \\ 0\\ 0
    \end{pmatrix}
    =
    \begin{pmatrix}
      0 \\-1 \\ 0\\ 0
    \end{pmatrix},\\
    \BC_2\colon &
    \begin{pmatrix}
      2 &-1 & 0 & 0\\
      1 & 0 & 0 & 0\\
      0 & 0 & 1 & 0\\
      0 & 0 & 0 & 1
      \end{pmatrix}
    \begin{pmatrix}
      0 \\-1 \\ 0\\ 0
    \end{pmatrix}
    =
    \begin{pmatrix}
      1 \\ 0 \\ 0\\ 0
    \end{pmatrix},
  \end{split}\quad
  \begin{split}
    \phantom{\BA\colon} &\phantom{\text{no change},}\\
    \BB_3\colon &
    \begin{pmatrix}
      0 &-1 & m & 0\\
      1 & 2 &-m & 0\\
      0 & 0 & 1 & 0\\
     -n &-n & mn& 1
      \end{pmatrix}
    \begin{pmatrix}
      1 \\ 0 \\ 0\\ 0
    \end{pmatrix}
    =
    \begin{pmatrix}
      0 \\ 1 \\ 0\\-n
    \end{pmatrix},\\
    \BC_3\colon &
    \begin{pmatrix}
      2 &-1 &-m & 0\\
      1 & 0 &-m & 0\\
      0 & 0 & 1 & 0\\
     -n & n & mn& 1
      \end{pmatrix}
    \begin{pmatrix}
      0 \\ 1 \\ 0\\-n
    \end{pmatrix}
    =
    \begin{pmatrix}
     -1 \\ 0 \\ 0\\ 0
    \end{pmatrix},\\
    \BB_4\colon &
    \begin{pmatrix}
      0 &-1 & m    &-m\\
      1 & 2 &-m    & m\\
     -n &-n & 1+mn &-mn\\
     -n &-n & mn   & 1-mn
      \end{pmatrix}
    \begin{pmatrix}
     -1 \\ 0 \\ 0\\ 0
    \end{pmatrix}
    =
    \begin{pmatrix}
      0 \\-1 \\ n\\ n
    \end{pmatrix},\\
    \BC_4\colon &
    \begin{pmatrix}
      2 &-1 &-m    & m\\
      1 & 0 &-m    & m\\
     -n & n & 1+mn &-mn\\
     -n & n & mn   & 1-mn
      \end{pmatrix}
    \begin{pmatrix}
      0 \\-1 \\ n\\ n
    \end{pmatrix}
    =
    \begin{pmatrix}
      1 \\ 0 \\ 0\\ 0
    \end{pmatrix}.
  \end{split}
\end{equation}
} 
The charges $\Bz_{\BX_i}$ in the tree junction on the RHS of
Fig.~\ref{fig:XmnLoopsAsTrees}, are the differences between succesive
$(p,q,r,s)$ charges in Eq.~\eqref{eq:SuccessiveCharges}:
\begin{equation}\label{eq:zXmn}
  \begin{split}
    \Bz_{\BA_i} &= 0,\quad
    \Bz_{\BB_1}  = -\begin{pmatrix} 1\\ -1\\  n\\  0\end{pmatrix},\quad
    \Bz_{\BC_1}  = -\begin{pmatrix} 1\\  1\\ -n\\  0\end{pmatrix},\quad
    \Bz_{\BB_2}  =  \begin{pmatrix} 1\\ -1\\  0\\  0\end{pmatrix},\quad
    \Bz_{\BC_2}  =  \begin{pmatrix} 1\\  1\\  0\\  0\end{pmatrix},\\
    \Bz_{\BB_3} &= -\begin{pmatrix} 1\\ -1\\  0\\  n\end{pmatrix},\quad
    \Bz_{\BC_3}  = -\begin{pmatrix} 1\\  1\\  0\\ -n\end{pmatrix},\quad
    \Bz_{\BB_4}  =  \begin{pmatrix} 1\\ -1\\  n\\  n\end{pmatrix},\quad
    \Bz_{\BC_4}  =  \begin{pmatrix} 1\\  1\\ -n\\ -n\end{pmatrix}.
  \end{split}
\end{equation}
Note that $\Bz_{\BX_i}\propto (p,q,r,s)_{\BX_i}$, where $(p,q,r,s)_{\BX_i}$ is
the vanishing 1-cycle of $\BX_i$. The factors of proportionality give the the
components of the junction lattice vector.  Reading off the coefficients from
Eq.~\eqref{eq:zXmn}, we see that the lattice vector of the $(1,0,0,0)$ loop is
\begin{equation}
  \Bdelta_1 = (0^M;\ -1,-1;\ 1,1;\ -1,-1;\ 1,1),
\end{equation}
as claimed in Sec.~\ref{Sec:MWfromJunction}.  In the same way, the $(0,1,0,0)$,
$(0,0,1,0)$ and $(0,0,0,1)$ loops of $\CX_{m,n}$ give junction lattice vectors
$\Bdelta_2$, $m\Bdelta_3$ and $m\Bdelta_4$, respectively, where the $\Bdelta_i$
are defined in Eq.~\eqref{eq:XmnNullGenerators}.


\section{Complex curves and their Jacobians}
\label{App:CurveJac}

In this Appendix, we provide background on complex curves and their Jacobian
varieties, and on the relation of the Jacobian variety to divisors, line
bundles, and the Picard group.  This review largely follows Ref.~\cite{GandH}.
The broad picture to keep in mind is that in algebraic geometry, the Jacobian
$\CJ_C$ of a genus-$g$ curve $C$ is a $T^{2g}$ analogous to what a physicist
would call the moduli space of $U(1)$ Wilson lines on $C$.

A genus-$g$ curve has a fundamental group $\pi_1(C)$ with $2g$ generators $a_i$
and $b_i$, for $i=1,\ldots g$, subject to the relation
\begin{equation}\label{eq:CgRelation}
  \bigl(a_1 b_1 a_1{}^{-1} b_1{}^{-1}\bigr) \bigl(a_2 b_2 a_2{}^{-1} b_2{}^{-1}\bigr)
  \cdots \bigl(a_g b_g a_g{}^{-1} b_g{}^{-1}\bigr)=1.
\end{equation}

The homology group $H_1(C,\IZ)$ is the abelianization of $\pi_1(C)$, that is,
$\pi_1(C)$ modulo its commutator subgroup $\{h_1h_2h_1{}^{-1}h_2{}^{-1} \mid
h_1,h_2\in \pi_1(C)\}$.  Thus,
\begin{equation}
  H_1(C,\IZ) = \IZ^{2g},
\end{equation}
generated by $2g$ linearly independent 1-cycles $\gamma_i$.  In a canonical
basis of $A$-cycles and $B$-cycles, we choose
\begin{equation}
  \gamma_i = A_i\quad\text{and}\quad\gamma_{g+i}=B_i\quad
  \text{for}\quad i=1,\ldots,g,
\end{equation}
with
\begin{equation}
  A_i\cap B_j = \d_{ij}, \quad A_i\cap A_j = B_i\cap B_j = 0.
\end{equation}

So far, we have taken a topological perspective.  However, it is also natural to
take a holomorphic perspective.  The cohomology group
\begin{equation}
  H^0(C,\Omega^1) = \IC^g
\end{equation}
is generated by $g$ holomorphic 1-forms $\omega_1,\ldots,\omega_g$.  By
integration, we then have $2g$ period vectors
\begin{equation}
  \Pi_i = \begin{pmatrix}
    \int_{\gamma_i}\omega_1\\
    \noalign{\vskip2pt}
    \int_{\gamma_i}\omega_2\\
    \vdots\\
    \int_{\gamma_i}\omega_g
  \end{pmatrix},\quad\text{for}\quad i=1,\ldots,2g,
\end{equation}
and a $g\times 2g$ period matrix $\Pi =
\bigl(\Pi_1\ \Pi_2\ \cdots\ \Pi_{2g}\bigr)$.  By $SL(3,\IC)_L\times SL(6,\IZ)_R$
change of basis, the period matrix can be put in standard form
$\Pi=\bigl(I\mid\t\bigr)$, where $I$ is the $g\times g$ identity matrix and $\t$
is a symmetric $g\times g$ matrix.

The Jacobian $\CJ_C$ of a genus-$g$ curve $C$ is the complex torus
\begin{equation}
  \IC^g/\L \cong T^{2g},
\end{equation}
where $\L$ is the lattice generated by the period vectors $\Pi_i\in\IC^g$.  It
is a principally polarized abelian variety, as defined in
App.~\ref{App:AbelVar}, and conversely, any smooth principally polarized abelian
variety is the Jacobian of some complex curve \cite{GandH}.  If $x^i$ and $y^i$
denote the coordinates on $\IC^g$ relative to the lattice basis, then the
holomorphic 1-forms are the familiar $dx+\t dy$ linear combinations
\begin{equation}
  \omega_i = \delta_{ij}dx^j + \tau_{ij} dy^j.
\end{equation}
For a generic complex torus, $\t$ need not have any special symmetry properties,
however the existence of a polarization implies that $\tau$ is symmetric.

Given a base point $p_0\in C$, the \emph{Abel-Jacobi map} takes points on $C$ to
points on its Jacobian variety $\CJ_C$:
\begin{equation}
  \mu\colon C\to\CJ_C,\qquad
  \mu(p) = \begin{pmatrix}
    \int_{p_0}^p\omega_1\\
    \noalign{\vskip2pt}
    \int_{p_0}^p\omega_2\\
    \vdots\\
    \int_{p_0}^p\omega_g
  \end{pmatrix}
  \quad\text{mod $\L$.}
\end{equation}

To connect this discussion to that in Sec.~\ref{Sec:MWisD} and
App.~\ref{App:Intersections}, we now reinterpret the Abel-Jacobi map in terms of
line bundles on $C$.  The \emph{Picard group} $\Pic(C)$ is the space of
holomorphic line bundles on $C$.\footnote{In sheaf theoretic terms, $\Pic(C) =
  H^1(C,\CO^*_C)$.}  To each divisor $D$ of $C$ (i.e., to each linear
combination of points of $C$), is associated a line bundle $[D]$, and to each
line bundle $L$ is associated its first Chern class $c_1(L)$.  If we write $D =
\sum a_i p_i$ and $L=[D]$, then $c_1(L)$ counts the net number of points of $D$,
called the \emph{degree} of $L$:
\begin{equation}
  \deg L = \int_C c_1(L) = \sum a_i.
\end{equation}
We use a superscript $\Pic^d(C)\subset \Pic(C)$ to denote the subspace of line
bundles of degree $d$.  Note that the degree of a line bundle is multiplicative,
$\Pic^m(C)\times\Pic^n(C)\to\Pic^{m+n}(C)$, so that only $\Pic^0(C)$ has a
canonical group structure.  In terms of $U(1)$ Wilson lines, $\Pic^0(C)$ is the
space of flat connections on $C$, that is, of holomorphic potentials $A$ such
that the field strength $F_{i\jbar}$ vanishes.  This makes it clear that for
curves of nonzero genus, a line bundle is \emph{not} uniquely determined by its
first Chern class.  Finally, it can be shown that each divisor $D=\sum a_ip_i$
defines a meromorphic section $(D)$ of the line bundle $[D]=\otimes
[p_i]^{a_i}$, with a zero of degree $a_i$ at $p_i$ for each $a_i>0$ and a pole
of degree $-a_i$ at $p_i$ for each $a_i<0$ \cite{GandH}.\footnote{In
App.~\ref{App:DirIm}, we correspondingly associate two sheaves to a divisor
$D=\sum a_ip_i$.  $\CO(D)$ is the sheaf of meromorphic sections of $[D]$ with
poles of degree $\le a_i$ at $p_i$ for each $a_i<0$; and $\CO(-D)$ is the sheaf
of meromorphic sections of $[D]$ with zeros of degree $\ge a_i$ at $p_i$ for
each $a_i>0$.}

In this description, the Jacobian variety of $C$ is defined to be the space of
degree zero line bundles $\CJ_C=\Pic^0(C)$.  Note that each $\Pic^d(C)$ is
noncanonically isomorphic to $\Pic^0(C)$, so it is also a $T^{2g}$.  For
example,
\begin{equation}\label{eq:Pic1toPic0}
  \Pic^1(C)\xrightarrow{\cong}\Pic^0(C),
\end{equation}
by tensoring with a line bundle of degree $-1$, and similarly,
$\Pic^d(C)\cong\Pic^0(C)$ by tensoring with the $d$th power of this line bundle.

The Abel-Jacobi map directly follows. A point (divisor) $p_0$ defines a map
\eqref{eq:Pic1toPic0} via tensoring with the line bundle $[p_0]^{-1}=[-p_0]$ of
degree $-1$.  Since any point $p\in C$ defines an element $[p]\in\Pic^1(C)$, we
obtain a map from $C$ to $\CJ_C=\Pic^0(C)$:
\begin{equation}
  \mu(p)=[p-p_0].
\end{equation}


\section{Genera of curves in $\IP^1\times\IP^1$}
\label{App:Genera}

Consider a degree $(\a,\b)$ curve $B_{\a,\b}\subset\IP^1\times\IP^1$.  Let $x,y$
denote the hyperplane classes of the respective $\IP^1$ factors, with
$x^2=y^2=0$.  By the adjunction formula,
\begin{equation}\label{eq:adjuction}
  \begin{split}
    c(T_{B_{\a,\b}}) &= c(\IP^1\times\IP^1)/c(N_{B_{\a,\b}})\\
    &= (1+x)^2(1+y)^2/(1+\a x+\b y)\\
    &= 1+(2-\a)x+(2-\b)y+\cdots,
  \end{split}
\end{equation}
so,
\begin{equation}\label{eq:chiab}
  \begin{split}
    \chi(B_{\a,\b})
      &= \int_{B_{\a,\b}}c_1(T_{B_{\a,\b}})
       = \int_{\IP^1\times\IP^1}c_1(T_{B_{\a,\b}})\w c_1(N_{B_{\a,\b}})\\
      &= 2 - 2(\a-1)(\b-1).
  \end{split}
\end{equation}
On the other hand, $\chi = 2-2g$ for a genus $g$ curve.  Therefore,
\begin{equation}\label{eq:genusab}
  g=(\a-1)(\b-1).
\end{equation}


\section{Direct image functor}
\label{App:DirIm}

Given a continuous map of topological spaces $f\colon X\to Y$, the direct image
functor, used in the proof of the Calabi-Yau condition in
Eq.~\eqref{eq:DirectIm} and App.~\ref{App:ProofCY}, provides a map from sheaves
on $X$ to sheaves on $Y$.  The purpose of this appendix is to provide background
on the direct image functor and higher direct images.  For completeness, we
begin with the definition of a sheaf.  For a more complete discussion, the
reader is referred to Secs.~0.3, 1.1, and 3.5 of Ref.~\cite{GandH}, which we
follow closely.

A \emph{presheaf} $\CF$ on a topological space $X$ assigns a set $\CF(U)$ to
each open set $U\subset X$, as well as a restriction map $r_{U,V}\colon
\CF(U)\to\CF(V)$ to each pair $U\subset V$.  The restriction map is required to
satisfy $r_{W,U} = r_{V,U}\circ r_{W,V}$ for $U\subset V\subset W$.  We will
assume below that the $\CF(U)$ are abelian groups.

Then, $\CF$ is a \emph{sheaf}, provided that the following conditions are
satisfied.

\begin{enumerate}
 \item Sections $\s_1\in\CF(U)$ and $\s_2\in\CF(V)$ whose restrictions agree
   over the intersection $U\cap V$ are the restrictions of a unique section over
   the union $U\cup V$:
   \begin{equation}
     \s_1\big|_{U\cap V} = \s_2\big|_{U\cap V}
     \quad\Rightarrow\quad
     \exists{\rho\in\CF(U\cup V)} \text{ such that }
     \rho\big|_U = \s_1\text{ and }\rho\big|_V = \s_2.
   \end{equation}
 \item If $\s\in\CF(U\cup V)$ and $\s\big|_U = \s\big|_V = 0$, then $\s=0$.

\end{enumerate}

\noindent Common sheaves that we might consider are:
\begin{description}
\item[{\it The locally constant sheaves\/ $\IZ$, $\IQ$, $\IR$, $\IC$,}] with
  $\IZ(U)=$ additive group of locally constant \hbox{$\IZ$-valued} functions on
  $U$, and similar definitions for $\IQ$, $\IR$, and $\IC$.
\item[{\it The structure sheaf\/ $\CO$,}] with $\CO(U) =$ additive group of
  holomorphic functions on $U$.
\item[{\it The sheaf\/ $\CO^*$,}] with $\CO^*(U) =$ multiplicative group of
  nowhere vanishing holomorphic functions on $U$.
\item[{\it The sheaf\/ $\CO(D)$.}] Given a divisor $D=\sum a_iV_i$ in terms of
  irreducible hypersurfaces $V_i$, and its corresponding line bundle $[D]$, we
  define $\CO(D) =$ additive group of meromorphic sections of $[D]$, with poles
  of order $\le a_i$ on $V_i$.
\item[{\it The sheaf\/ $\O^p$,}] with $\O^p(U) =$ additive group of holomorphic
  $p$-forms on $U$.
\end{description}

Maps of sheaves of abelian groups are group homomorphisms compatible with the
sheaf conditions on the restriction maps.  \v Cech cohomology gives a definition
of the cohomology of sheaves.  We refer the reader to Ref.~\cite{GandH} for a
description of \v Cech cohomology.  For our purposes, it should suffice to note
that (i) such a cohomology can be defined, and (ii) it agrees with de Rham and
singular cohomology for the locally constant sheaves $\IR$ and $\IZ$,
respectively.  Likewise, the \v Cech cohomology of $\O^p$ agrees with Dolbeault
cohomology:
\begin{equation}
  H^q(X,\O^p) \cong H^{p,q}_{\bar\partial}.
\end{equation}

Given a continuous map $f\colon X\to Y$ of topological spaces and a sheaf
$\CF\colon U\to\CF(U)$, a natural map of sheaves is the \emph{direct image
  functor} $f_*$, defined by
\begin{equation}
  f_*\CF\colon V\to \CF(f^{-1}(V)),
\end{equation}
for $V$ an open set in $Y$.  Likewise, we define the \emph{higher direct images}
$R^qf_*$ by
\begin{equation}
  R^q f_*\CF\colon V\to H^q(f^{-1}(V),\CF).
\end{equation}
The `$R\,$' stands for \emph{right derived functor}.\footnote{The direct image
  functor is left exact, but not necessarily right exact.  However, under mild
  assumptions, there is a canonical way to extend the sequence to the right to
  form a long exact sequence, by appending ``right derived functors'' of $\CF$.
  These right derived functors are precisely the higher direct images.}

For a fibration $F\to X\xrightarrow{f}\CB$ with generic fiber $F$, and
$U\subset\CB$ an open set of the base, the open set $f^{-1}(U)\cong U\times F$
contains the entire fiber over each point of $U$.  Thus, to first approximation
$R^qf_*(\IQ)$ is the constant sheaf $H^q(F,\IQ)$---the $q$th cohomology group
along the fiber (for contractible $U$).  We then need to modify this
approximation to take into account the monodromy action of the fundamental group
of $\CB$ on the cycles in $H^q(F,\IQ)$.

This definition via a first approximation followed by corrections can be made
precise in terms of a \emph{Leray spectral sequence} $\{E^{p,q}_r\}$ with
$E_\infty\Rightarrow H^*(X,\CF)$ whose second step is
\begin{equation}
  E^{p,q}_2 = H^p(\CB,R^qf_*\CF),
\end{equation}
as explained on p.~462--468 of Ref.~\cite{GandH}.  For example, in de Rham
cohomology, we have
\begin{equation}
  E^{p,q}_2 = H^p_\text{DR}(\CB,H^q_\text{DR}(F)),
\end{equation}
where the right hand side is defined by viewing $H^q_\text{DR}(F)$ as a vector
bundle over $\CB$ associated to a representation of the fundamental group.

In general, the cohomology rings $H^*(X,\CF)$ and $H^*(\CB,R^*f_*\CF)$ need not
agree, however, under further restrictions this is the case.  For $\CF=\IQ$, and
$X$ and $B$ compact and K\"ahler, the sequence $\{E^{p,q}_r\}$ can be shown to
converge at the second step, so that
\begin{equation}
  E_2\cong E_\infty\quad\text{and}\quad
  H^*(X,\IQ)\cong H^*(\CB,R^*f_*\IQ).
\end{equation}
In this case, the $H^p(\CB,R^qf_*\IQ)$ give a filtration of $H^*(X,\IQ)$
according to base and fiber degree, much like the Dolbeault cohomology gives a
filtration of the de Rham cohomolog according to holomorphic and antiholomorphic
degree.


\section{Proof of the Calabi-Yau condition}
\label{App:ProofCY}

To evaluate Eq.~\eqref{eq:DirectIm}, it is convenient to factorize the
projection map $\rho\colon S\to\IP^1_{u,v}$ as $\rho = f\circ\varphi$, where
$\varphi$ is the double cover $\varphi\colon\ S\to
\IP^1_{s,t}\times\IP^1_{u,v}$, and we define $f$ and $g$ to be the projection
maps $f\colon\ \IP^1_{s,t}\times\IP^1_{u,v}\to \IP^1_{u,v}$ and
$g\colon\ \IP^1_{s,t}\times\IP^1_{u,v}\to \IP^1_{s,t}$.  Then,
\begin{equation}\label{eq:Rfunctor}
  \rho_* K_{S/\IP^1} = \bigl(R^1\rho_*\CO_S\bigr)^*
  = \bigl(R^1 f_*\varphi_*\CO_S\bigr)^*,
\end{equation}
where we have used Serre duality on the genus-2 fibers in the first equality.
For a double cover $\varphi\colon\ A\to B$, we have the general
result\footnote{For example, for the double cover
  $\varphi\colon\ \IP^1\to\IP^1$, $z\mapsto z^2$, we have branch points at $0$
  and $\infty$, and $\varphi_*\CO_{\IP^1} = \CO_{\IP^1}\oplus\CO_{\IP^1}(-1)$.
  For $\varphi\colon\ E\to\IP^1$, with $E$ an elliptic curve, we have four
  branch points, and $\varphi_*\CO_E = \CO_{\IP^1}\oplus\CO_{\IP^1}(-2)$.}
\[ \varphi_*\CO_A = \CO_B\oplus\CO_B(-\half\bigl(\hbox{branch locus})\bigr). \]
In our case, this gives $\varphi_*\CO_S = \CO_{\IP^1\times\IP^1} \oplus
\CO_{\IP^1\times\IP^1}(-3,-1)$, from which
\begin{equation}\label{eq:PiKStar}
  R^i\rho_*\CO_S
  = R^i f_*\CO_{\IP^1\times\IP^1}\oplus R^i f_*\CO_{\IP^1\times\IP^1}(-3,-1).
\end{equation}
The first term is
\begin{equation}\label{eq:RfFirst}
  R^i f_*\CO_{\IP^1\times\IP^1} = \biggl\{
  \begin{matrix}\CO_{\IP^1} & i = 0,\\ 0 & i=1,\end{matrix}\biggr.
\end{equation}
and the second is
\begin{equation}\label{eq:RfSecond}
  \def\phz{{\vphantom{0}}}
  \begin{split}
    R^i f_*\CO_{\IP^1\times\IP^1}(-3,-1)
    &= R^i f_*\bigl(f^*\CO_{\IP^1}(-1)\otimes g^*\CO_{\IP^1}(-3)\bigr)\\
    &= \CO_{\IP^1}(-1)\otimes R^i f_* g^* \CO_{\IP^1}(-3)\\
    &= \CO_{\IP^1}(-1)\otimes g_0^* R^i f^\phz_{0\,*} \CO_{\IP^1}(-3)\\
    &= \CO_{\IP^1}(-1)\otimes H^i\bigl(\IP^1,\CO(-3)\bigr)\\
    &= \biggl\{
       \begin{matrix}
         0 & i = 0,\\
         \CO_{\IP^1}(-1)\oplus\CO_{\IP^1}(-1) & i = 1.
       \end{matrix}\biggr.
  \end{split}
\end{equation}
Here, $f_0$ and $g_0$ are maps from $\IP^1$ to a point $p$, so that we have the
following commutative diagram:
\begin{equation}\label{eq:fgzero}
  \begin{CD}
    \IP^1_{s,t}\times\IP^1_{u,v}  @>{f}>>     \IP^1_{u,v}\\
    @VV{g}V                                   @VV{g_0}V\\
    \IP^1_{s,t}                   @>{f_0}>>   p
  \end{CD}
\end{equation}
In the last line of Eq.~\eqref{eq:RfSecond}, we have used $H^0(\IP^1,\CO(-3)) =
0$, and by Serre duality\footnote{For $E$ a vector bundle and $\dim M = n$,
  Serre duality states that $H^p(M,E)=H^{n-p}(M,K\otimes E^*)^*$.}
$H^1\bigl(\IP^1,\CO(-3)\bigr) = H^{0}\bigl((\IP^1,\CO(1)\bigr))^* =
\IC\oplus\IC$\@.  Combining results \eqref{eq:DirectIm} through
\eqref{eq:RfSecond}, we obtain $N_{\IP^1} =
\CO_{\IP^1}(-1)\oplus\CO_{\IP^1}(-1)$, as claimed.  The Calabi-Yau condition
follows.


\section{Intersections of theta surfaces of $\CX$}
\label{App:Intersections}

In this appendix, we define the theta surfaces $\Theta_I,\Theta'_I$ of
Sec.~\ref{Sec:IntersectionNumbers}, and compute their double and triple
intersections in $\CX$, as well as their intersections with the abelian surface
fiber $A$.

An outline of the construction of the theta surfaces is as follows.  The genus-2
fibration $S\to\IP^1$ has $2\times 12 =24$ sections $\ell_I,\ell'_I$.  Each
defines an isomorphism
\begin{equation}
  \Pic^1(S/\IP^1)\xrightarrow{\cong}\Pic^0(S/\IP^1)=\CX.
\end{equation}
Once one such isomorphism has been chosen, each section $\ell_I$ or $\ell'_I$ of
$S$ defines a corresponding theta surface $\Theta_I$ or $\Theta'_I$,
respectively, embedding $S$ in the Calabi-Yau manifold $\CX$.

The correspondence is most easily understood fiberwise.  Therefore, we first
focus on a single genus-2 curve $C$.  In this case, each point $p$ on the curve
defines a theta divisor $C_p$ of the Jacobian abelian surface $A$, which is an
embedding of $C$ in $A$.  We compute the intersections of these $C_p$.  Then, we
fiber this construction over $\IP^1$.  This replaces $C$ by $S$, $p$ by a
section $\ell$ of $S$, and $A$ by the relative Jacobian $\CX$ of $S$.  The
double intersections in $A$ become curves in $\CX$, and we use this intermediate
result to compute the desired triple intersections in $\CX$.

\nonnumberedsubsec{Warm-up: intersections of curves in an abelian surface}

Let us first review the map between points of a complex curve and theta divisors
of its Jacobian, and then use this map to compute intersections the theta
divisors in the \hbox{genus-2} case, where the Jacobian is an abelian surface.
For background on complex curves $C$, their Jacobians $\CJ_C$, and spaces of
degree $d$ line bundles $\Pic^d(C)$, the reader is referred to
App.~\ref{App:CurveJac}.

Let $C$ be any curve of genus $g$\@.  $\Pic^{g-1}(C)$ has a canonical
$\Theta$-divisor,
\begin{equation}
  \Theta = \{ L\in \Pic^{g-1}(C) \mid h^0(L) > 0 \}.
\end{equation}
More generally, for all $d\ge 0$, consider the variety
\begin{equation}
  W_d = \{ L\in \Pic^d(C) \mid h^0(L) > 0 \}.
\end{equation}
Its dimension is $\min(d,g)$.  When $d=g-1$, $W_d$ is a divisor of $\Pic^d(C)$.
Otherwise it is a subvariety of other dimension.  Now, given a line bundle
$L\in\Pic^1(C)$, we have an isomorphism $\Pic^d(C)\cong\Pic^0(C)$, realized by
tensoring with $L^{-\otimes d}$, so that $W_d$ can also be viewed as a
subvariety of $\Pic^0(C)$.  Therefore, a line bundle $L\in \Pic^{g-1}(C)$
determines a theta divisor $\Theta_L \subset \Pic^0(C)$.  In the case of
interest $g=2$, the result is as follows:\medskip

\noindent\emph{A line bundle $L\in\Pic^1(C)$ determines a theta divisor
  $\Theta_L\subset\Pic^0(C)$, embedding the genus-2 curve $C$ in its Jacobian
  surface $\CJ_C=\Pic^0(C)$.}\medskip

In the genus-2 case, intersections of theta divisors in the abelian surface
$A=\Pic^0(C)$ are computed as follows.  Let us focus on line bundles $L =
\CO(p)$, where $p$ is a point on the genus-2 curve $C$, and denote the
corresponding theta divisors (embeddings $C\hookrightarrow A$) by $C_p$.  Given
two points $p_1,p_2\in C$, we have
\begin{equation}
  \begin{split}
    C_{p_1}\cap_A C_{p_2} &= \{ L\in \Pic^0(C) \mid
    h^0(L\otimes\CO(p_1))>0,\ h^0(L\otimes\CO(p_2))>0 \}\\
    &= \CO(-p_1)\otimes \{q\in C \mid h^0\bigl(\CO(q + p_2 - p_1)\bigr)>0 \}
    \qquad \bigl(\CO(q)\approx L\otimes \CO(p_1)\bigr)\\
    &= \CO(-p_1)\otimes\{\CO(p_1),\CO(p_2')\},\\
    &= \{ \CO(0), \CO(p'_2 - p_1) \}\quad\text{for}\quad p_1\ne p_2.
  \end{split}
\end{equation}
Here, $p'=\imath(p) = h-p$, where $\imath$ denotes the hyperelliptic involution
on the genus-2 curve (sending one branch to the other) and $h$ is independent of
$p$.  For $p_1=p_2$, the intersection is the whole curve $C_{p_1}=C_{p_2}$.  We
write this result simply as
\begin{equation}\label{eq:CdotC}
      C_{p_1}\cap_A C_{p_2} =  \left\{\begin{matrix}
      \{0, p'_2 - p_1\} & p_1\ne p_2,\\
      \,C_{p_1}\hfill & p_1 = p_2. \end{matrix}\right.
\end{equation}
Note that $p'_2-p_1 = p'_1-p_2$, so that the intersection is symmetric.  For
$p_1=p'_2$, the intersection consists of the single point $0\in A$, counted with
multiplicity 2 for the homological intersection.

\nonnumberedsubsec{Intersections of pairs of surfaces in the threefold $\CX$}

Next, consider the genus-2 fibration $S$ and its relative Jacobian $\CX =
\Pic^0(S/\IP^1)$.  In this case, the analog of the theta curve $C_p$ associated
to each point $p\in C$ is a family of genus-2 curves $C_\ell$ associated to each
section $\ell$ of the genus-2 fibration.  In the same way that each $p$
determined a line bundle $\CO(p)\in\Pic^1(C)$ above, each section $\ell$ now
determines a section of $\Pic^1(S/\IP^1)$.

Let $\Theta_I, \Theta'_I$ denote the total space associated to the family
$C_{\ell_I},C_{\ell'_I}$, respectively.  Each of these \emph{theta surfaces} is
an embedding of $S$ in the Calabi-Yau threefold $\CX$.  From the previous
result~\eqref{eq:CdotC}, the intersections of pairs of distinct theta surfaces
are
\begin{equation}\label{eq:DblInts}
  \begin{split}
    \Theta_I\cdot\Theta_J &= \s_0 + \s_{\ell'_I-\ell_J},\quad
    \Theta'_I\cdot\Theta'_J = \s_0 + \s_{\ell_I-\ell'_J},\\
    \Theta'_I\cdot\Theta_J &= \s_0 + \s_{\ell_I-\ell_J},\quad
    \Theta_I\cdot\Theta_I' = 2\s_0 + C_{\ell_I\cap\ell'_I}.
  \end{split}
\end{equation}
Here, $L\to \s_L$ is the isomorphism identifying degree zero line bundles in
$\Pic^0(S/\IP^1)$ with sections of the abelian fibered threefold $\CX$.  Note
that $\ell'_I-\ell_J = \ell'_J-\ell_I$ as a consequence of
Eq.~\eqref{eq:llprime}.  The curve $C_{\ell_I\cap\ell'_I}$ is the common genus-2
fiber of $\Theta_I$ and $\Theta'_I$.  For self intersections, we have
\begin{equation}
  \Theta_I\cdot\Theta_I = c_1(K_{\Theta_I}).
\end{equation}
Recall that $C$ is the generic genus-2 fiber of the projection $S\to\IP^1_{u,v}$
and $C'$ is the generic genus-0 fiber of the projection $S\to\IP^1_{s,t}$.
Since $\Theta_I\cong S$, the relevant fact for applying the last result to the
computation of triple intersections is
\begin{equation}
  c_1(K_S) = C' - C.
\end{equation}
This result follows from the double cover formula $K_S =
\pi^*(K_{\IP^1\times\IP^1}\otimes L)$ (cf.~Lemma.~17.1 in Ref.~\cite{Barth}).
Here, $L^2$ is defined by the branch curve $B$ of the double cover
$S\to\IP^1\times\IP^1$ as in Sec.~\ref{Sec:SurfaceS}:
\begin{equation}
  L^2 = \CO_{\IP^1\times\IP^1}(B) = \CO(6,2).
\end{equation}
This gives
\begin{equation}
  \begin{split}
    K_S &= \pi^*\bigl(K_{\IP^1\times\IP^1}\otimes\CO(3,1)\bigr),\\
    c_1(K_S) &= -2(C+C') + (3C' + C) = C'-C.
  \end{split}
\end{equation}

\nonnumberedsubsec{Triple intersections of divisors of $\CX$}

Finally, the triple intersections of divisors in $\CX$ can obtained as double
intersections of curves in surfaces. For example, for $I,J,K$ distinct,
\begin{equation}\label{eq:TrpIntIJK}
  \begin{split}
    \Theta_I\cdot\Theta_J\cdot\Theta_K &= 
    (\Theta_I\cdot\Theta_J)\cdot_{\Theta_J}(\Theta_J\cdot\Theta_K)\\ 
    &= (\s_0+\s_{\ell'_I-\ell_J})\cdot_{\Theta_J}(\s_0+\s_{\ell'_K-\ell_J})\\ 
    &\cong (\ell_J+\ell'_I)\cdot_S(\ell_J+\ell'_K)\\ 
    &= -1.
  \end{split}
\end{equation}
Here, we have used the fact that $\s_{\ell-\ell_J}$ maps to $\ell\in S$ under
the isomorphism $\Theta_J\to S$.  The remaining triple intersections of theta
surfaces are
\begin{equation}\label{eq:TrpIntList}
  \begin{split}
    &\Theta_I\cdot\Theta_J\cdot\Theta_J=\Theta_I\cdot\Theta'_J\cdot\Theta'_J =-2,\\
    &\Theta_I\cdot\Theta_I\cdot\Theta'_I=\Theta_I\cdot\Theta_J\cdot\Theta'_J =0,\\
    &\Theta_I\cdot\Theta_I\cdot\Theta_I=-4,\\
  \end{split}
\end{equation}
together with equations obtained from these by exchange of $\Theta$ and
$\Theta'$.  The computation is analogous to the previous one.  Using
Eq.~\eqref{eq:DblInts}, it is possible to confirm that the result is independent
of choice of which of the three theta surfaces is used to perform the double
intersection.

We now turn to intersections involving the generic abelian surface fiber $A$. In
this case, $A^2=0$, and
\begin{equation}
  A\cdot\Theta_I\cdot\Theta_J = A\cdot\Theta_I\cdot\Theta'_J 
  = A\cdot\Theta'_I\cdot\Theta'_J = 2,
\end{equation}
for any $I,J$, not necessarily distinct.  This is most easily proven from the
intersection of curves in the abelian fiber $A$.  For example,
\begin{equation}\label{eq:TrpsThTh2}
  \begin{split}
    A\cdot\Theta_I\cdot\Theta_J
    &= (A\cdot\Theta_I)\cdot_A(A\cdot\Theta_J)\\ 
    &\cong C\cdot_A C=2,
  \end{split}
\end{equation}
as desired.  (In an abelian surface, the self-intersection of a genus-$g$
curve is $2g-2$.)  

The same result is obtained if the intersections are performed in a theta
surface.  Let $C_I$ denote the genus-2 fiber of $\Theta_I\cong S$.  Then, for
example, for $I\ne J$,
\begin{equation}\label{eq:TrpsThTh}
  \begin{split}
    A\cdot\Theta_I\cdot\Theta_J
    &= (A\cdot\Theta_I)\cdot_{\Theta_I}(\Theta_I\cdot\Theta_J)\\
    &= C_I\cdot_{\Theta_I}(\s_0+\s_{\ell'_J-\ell_I})\\
    &\cong C\cdot_S(\ell_I+\ell'_J)=2,\\
    A\cdot\Theta_I\cdot\Theta_I
    &= (A\cdot\Theta_I)\cdot_{\Theta_I}(\Theta_I\cap\Theta_I)\\
    &= C\cdot_{\Theta_I} c_1(K_{\Theta_I})\\
    &\cong C\cdot_S (C'-C)=2.
  \end{split}
\end{equation}
In the last step, we have used the fact that the genus-2 fiber $C$ of $S$
satisfies $C^2=0$ and $C\cdot C' = C\cdot(\ell_K+\ell'_K) = 2$.


\end{document}